\documentclass[traditabstract]{aa} 
\usepackage{graphicx}
\usepackage{lscape}
\usepackage{longtable}
\usepackage{natbib,twoopt}
 \bibpunct{(}{)}{;}{a}{}{,}    
\usepackage{graphics}
\usepackage{color}
\usepackage{ulem} 
\usepackage{txfonts}
\begin{document}
   \title{Primordial $^4$He abundance: a determination based on  
the largest sample of H~{\sc ii} regions with a methodology tested on model H~{\sc ii} regions 
\thanks{Based on observations
collected at the European Southern Observatory, Chile, programs
073.B-0283(A), 081.C-0113(A), 65.N-0642(A), 68.B-0310(A), 69.C-0203(A), 69.D-0174(A), 
70.B-0717(A), 70.C-0008(A), 71.B-0055(A).}
\thanks{Based on observations at the Kitt Peak National Observatory, 
National Optical Astronomical Observatory, operated by the Association of 
Universities for Research in Astronomy, Inc., under contract with the National 
Science Foundation.}
\thanks{Tables \ref{taba1} and \ref{taba2} are available 
only in the electronic edition.}
}

   \author{Y. I. Izotov \inst{1,2,3},
          G. Stasi\'nska \inst{1}
          \and
          N. G. Guseva \inst{3,2}}
\offprints{Y. I. Izotov, izotov@mao.kiev.ua}

   \institute{LUTH, Observatoire de Meudon, F-92195 Meudon Cedex, France
         \and
                     Main Astronomical Observatory,
                     Ukrainian National Academy of Sciences,
                     Zabolotnoho 27, Kyiv 03680,  Ukraine
         \and
                     Max-Planck-Institut f\"ur Radioastronomie, 
                     Auf dem H\"ugel 69, 53121 Bonn, Germany
             }

   \date{Received  accepted }

  \abstract{
We verified the validity of the empirical method  to derive the $^4$He 
abundance used in our previous papers
by applying it to  CLOUDY (v13.01) models.  Using newly published  He~{\sc i} 
emissivities, for which we present convenient fits as well as the output CLOUDY 
case B hydrogen and He~{\sc i} line intensities, we found that the empirical 
method is able to reproduce the input CLOUDY $^4$He abundance with an accuracy
of better than 1\%. 
The CLOUDY output data also allowed us to derive the non-recombination 
contribution to the intensities of the strongest Balmer hydrogen 
H$\alpha$, H$\beta$, H$\gamma$, and H$\delta$ emission lines and the
ionisation correction factors for He. With these improvements we used our 
updated empirical method to derive the $^4$He abundances and to test
corrections for several systematic effects in a sample of 1610 
spectra of low-metallicity extragalactic H~{\sc ii} regions, the largest sample
used so far. From this sample we extracted a subsample of 111
H~{\sc ii} regions with  H$\beta$ equivalent width 
EW(H$\beta$) $\geq$ 150\AA, with excitation parameter
$x$ = O$^{2+}$/O $\geq$ 0.8, and with  helium mass 
fraction $Y$ derived with an accuracy better than 3\%. With this subsample we derived the primordial 
$^4$He mass fraction $Y_{\rm p}$ = 0.254 $\pm$ 0.003 from linear regression 
$Y$ -- O/H. The derived value of $Y_{\rm p}$ is higher at the 68\% 
confidence level (CL) than that 
predicted by the standard big bang nucleosynthesis (SBBN) model, possibly
implying the existence of different types of neutrino species in addition to 
the three known types of active neutrinos. Using the most recently derived 
primordial abundances D/H = (2.60$\pm$0.12)$\times$10$^{-5}$ 
and $Y_{\rm p}$ = 0.254 $\pm$ 0.003 and 
the $\chi^2$ technique, we found that
the best agreement between abundances of these light elements is achieved
in a cosmological model with baryon mass density $\Omega_{\rm b} h^2$ = 
0.0234$\pm$0.0019 (68\% CL) and an effective number of the
neutrino species $N_{\rm eff}$ = 3.51$\pm$0.35 (68\% CL).
}
\titlerunning{Primordial $^4$He abundance}
\authorrunning{Y. I. Izotov et al.}
   \maketitle
%

\section{Introduction}\label{intro}

In the standard theory of big bang nucleosynthesis (SBBN), given the 
number of light neutrino species, the abundances of light elements 
D, $^3$He, $^4$He and $^7$Li depend only on one cosmological parameter, the   
baryon-to-photon number ratio $\eta$, which is related 
to the baryon density parameter $\Omega_{\rm b}$, the present ratio of the 
baryon mass density to the critical density of the Universe, by
the expression 10$^{10}$$\eta$ = 273.9 $\Omega_{\rm b}h^2$,
where $h$~=~$H_0$/100~km~s$^{-1}$~Mpc$^{-1}$ and 
$H_0$ is the present value of the Hubble parameter \citep{St05,St12}. 

Because of the strong dependence of the D/H abundance ratio on $\Omega_{\rm b}h^2$, 
deuterium is the best-suited light element for determining the baryon mass
fraction. Its abundance can accurately be  measured in high-redshift 
low-metallicity QSO 
Ly$\alpha$ absorption systems. Although the data are still scarce -- there 
are only ten absorption systems for which such a D/H measurement has 
been carried out \citep{PC12} -- the measurements appear to converge to a mean 
primordial value D/H~$\sim$~(2.5 - 2.9)~$\times$~10$^{-5}$,
which corresponds to $\Omega_{\rm b}h^2$~$\sim$~0.0222 - 0.0223 
\citep{Io09,N12,PC12}.
This estimate of $\Omega_{\rm b}h^2$ agrees excellently with the value 
of 0.0221 - 0.0222 obtained from studies of the fluctuations of the cosmic 
microwave background (CMB) with {\sl WMAP} and {\sl Planck} \citep{K11,A13}.

While deuterium is sufficient to derive the baryonic mass density from BBN, 
accurate measurements of the 
primordial abundances of at least two different relic elements are required 
to verify the consistency of SBBN. The primordial abundance of $^4$He 
can in principle be derived accurately from observations of 
the helium and hydrogen emission lines from low-metallicity blue compact dwarf 
(BCD) galaxies, which have undergone little chemical evolution. 
Several groups have used this technique to derive the primordial  
$^4$He mass fraction $Y_{\rm p}$. The most recent determinations of $Y_{\rm p}$ have resulted
in high values of 0.2565~$\pm$~0.0060 (2$\sigma$) \citep{IT10} and 
0.2534~$\pm$~0.0083 (1$\sigma$)
\citep{A12}, implying deviations from the SBBN and existence of additional 
types of neutrino species such as sterile neutrinos. However, taking
into account large statistical and systematic errors in the $Y_{\rm p}$ determination,
one can conclude that these latest determinations of $Y_{\rm p}$ are broadly
(at the 3$\sigma$ level) consistent with the prediction of SBBN,
$Y_{\rm p}$ = 0.2477~$\pm$~0.0001 \citep{A13}.
Although $^4$He is not a sensitive baryometer ($Y_{\rm p}$ depends only 
logarithmically on the baryon density), its primordial abundance depends much 
more sensitively on the expansion rate of the Universe and thus on any small 
deviation from SBBN, than that of the other primordial light elements 
\citep{St06}. 

However, to detect small deviations from SBBN and make cosmological inferences,
$Y_{\rm p}$ has to be determined to a level of accuracy
of less than one percent. This is not an easy task. While it is relatively 
straightforward to derive the helium abundance
in an H~{\sc ii} region with an accuracy of 10
percent if the spectrum is adequate, gaining a factor of several in accuracy
requires many conditions to be met.
First, the observational data have to be of good quality and
should constitute a large sample to reduce statistical uncertainties
in the determination of $Y_{\rm p}$. This has been the concern of studies 
conducted for instance by \citet{ITL94,ITL97},
\citet{IT98,IT04}, who obtained 93 high signal-to-noise
spectra of low-metallicity extragalactic H~{\sc ii} regions,
which includes a total of 86 H~{\sc ii} regions in 77 galaxies 
\citep[the HeBCD sample, see ][]{IT04}. 
Furthermore, \citet{I09,I11b} and \citet{G11}
collected a sample of 75 spectra of low-metallicity H~{\sc ii} regions in
different galaxies observed with the Very Large Telescope (VLT) 
(VLT sample). Finally,
1442 high-quality spectra of low-metallicity H~{\sc ii} regions were 
extracted from the Sloan Digital Sky Survey (SDSS) (SDSS sample). 
They were selected 
from the SDSS Data Release 7 (DR7), as were those with the [O~{\sc iii}]
$\lambda$4363 emission line measured with an accuracy better than 25\%. 
Additionally, all strongest He~{\sc i} emission lines in the optical range,
$\lambda$3889, $\lambda$4471, $\lambda$5876, $\lambda$6678, and $\lambda$7065,
had to be present in the spectrum and measured with good accuracy. Part of
the SDSS sample containing $\sim$ 800 galaxies with the highest
H$\beta$ luminosities, was discussed
by \citet{I11b}. In total, the HeBCD, VLT and SDSS samples
together constitute by far the largest sample of high-quality spectra
assembled to investigate the problem of the primordial helium abundance. 
Because of greatly increased samples it  turns out 
that the accuracy of the determination of the primordial $^4$He
abundance is limited at present more by systematic uncertainties and biases
than by statistical errors.   

Different empirical methods have been used to derive physical conditions and 
element abundances and to convert the observed He~{\sc i} 
emission line intensities to a $^4$He abundance 
\citep[e.g. ][]{ITL94,ITL97,IT10,Pe07,A10,A11,A12}. All of them use fits of 
different processes 
\citep[e.g. ][]{I06}, including the fits of
He~{\sc i} and H emissivities and different effects that result in
observed line intensities that are different from their recombination values. 
There are many known effects we need to correct for  
to transform the observed He~{\sc i} line intensities into a $^4$He abundance. 
Neglecting or misestimating them may lead 
to systematic errors in the $Y_{\rm p}$ determination that are larger than the 
statistical errors.
These effects are (1) reddening, (2) underlying stellar
absorption in the He~{\sc i} lines, (3) collisional 
excitation of the He~{\sc i} lines that result in intensities 
different from their recombination values, 
(4) fluorescence of the He~{\sc i} lines, which also result in intensities 
different from their 
recombination values, (5) collisional excitation of the hydrogen lines, 
(6) possible departures from case B in the emissivities of H and He~{\sc i} 
lines, (7) the temperature 
structure of the H~{\sc ii} region, and (8) its ionisation structure. 
The role of each of these effects is discussed for instance by
\citet{I07}. 

Most of the systematic effects are  taken into account in the most recent
papers \citep[e.g. ][]{I07,IT10,Pe07,A10,A12}. 
However, one important problem remains.  While corrections for many effects 
were obtained with photoionised H~{\sc ii} region models 
\citep[e.g. with CLOUDY, ][]{F98,F13}, the \textit{overall} procedure for 
determining the $^4$He abundance has never been tested on photoionisation 
models. Since a photoionisation code such as CLOUDY takes into account all the
processes affecting the He~{\sc i} line intensities (radiation transfer, ionisation 
and recombination processes, heating and cooling,
collisional and fluorescence excitation, etc.), it produces in principle
model H~{\sc ii} regions that should be close to real  H~{\sc ii} 
regions. Of course, the gas density distributions, the distribution of the
ionising stars and their spectral energy distributions are simplified with 
respect to reality, but how could one trust a method that does not recover
the correct helium abundance used to compute these models when treating the 
models in the same way as observations are treated? Of course, such an 
exercise needs to be made using exactly the same atomic ingredients as are 
implied in CLOUDY calculations, since the objective is to judge the method 
itself and, if necessary, improve it.

In the present paper we test the empirical method developed in a number of our 
papers \citep[e.g. ][]{ITL94,ITL97,I07,IT10} using a 
grid of photoionised H~{\sc ii} region models calculated with
version v13.01 of the CLOUDY code \citep{F13}. 
In Sect. \ref{cloudy} we describe our  grid of CLOUDY photoionisation models.
The fits of new emissivities for 32 He~{\sc i} emission lines tabulated
by \citet{P13} are discussed in Sect. \ref{Heemiss}. Ionisation correction 
factors for He and their fits are obtained in Sect. \ref{icfHe}. We
consider the non-recombination excitation of hydrogen in Sect. \ref{hydro}.
In Sect. \ref{compar} we examine how well our empirical method recovers the $^4$He 
abundance with which the photoionisation models were constructed, and update 
it to improve the results. 
In Sect. \ref{real} we discuss the
additional systematic effects that appear in real H~{\sc ii} regions,
describe our sample of low-metallicity H~{\sc ii} regions, apply our updated 
empirical method to determine the $^4$He abundance in these objects and justify
the linear regressions used for the primordial $^4$He abundance determination.
In Sect. \ref{primo} we present the linear regressions $Y$ -- O/H and derive
the primordial $^4$He mass fraction $Y_{\rm p}$. Cosmological implications of 
the derived $Y_{\rm p}$ are discussed in Sect. \ref{cosmo}. In particular, we
derive the effective number of neutrino species $N_{\rm eff}$. Section
\ref{summary} summarises our main results. 

\section{Grid of photoionisation CLOUDY models}\label{cloudy}

Using version v13.01 of CLOUDY code \citep{F13}, we calculated a grid
of 576 homogeneous spherical ionisation-bounded H~{\sc ii} region models 
with parameters shown in Table \ref{tab1},
which cover the entire range of parameters in real 
low-metallicity H~{\sc ii} regions used for the $^4$He abundance determination.
In particular, the range of oxygen abundances is 12~+~log~O/H~=~7.3~--~8.3.
The abundances of other heavy elements relative to oxygen are kept 
constant corresponding to typical values obtained for low-metallicity
emission-line galaxies \citep[e.g. ][]{I06}. The $^4$He abundance in all
models, however, is the same with $Y$~=~0.254. We also included dust, scaling it
according to the oxygen abundance. The characteristics adopted for the dust 
are those offered by CLOUDY as ``Orion nebula dust''. For all the models 
we used the "iterate" option.

We adopted three values of the number of ionising photons Q and the shape
of the ionising radiation spectrum corresponding to the Starburst99
model with the ages of 1.0, 2.0, 3.5, and 4.0 Myr and different metallicities 
Thus, for 12+logO/H = 7.3 and 7.6 we adopted Starburst99
models with a heavy-element mass fraction $Z$ = 0.001, for those
with 12+logO/H = 8.0 models with $Z$ = 0.004, and for those 
with 12+logO/H = 8.3 models with $Z$ = 0.008.
All Starburst99 models were calculated with the stellar tracks from 
\citet{M94} and the \citet{HM98} and \citet{P01} stellar atmosphere set. 
We also varied the log of the volume-filling factor $f$ between $-$0.5 
and $-$2.0 to obtain CLOUDY models with different ionisation parameters. 
For the subsequent analysis, out of the
576 H~{\sc ii} region models we selected the 363 that had a 
volume-averaged ionisation parameter of log $U$ between $-$3 and $-$2. This 
range of log $U$ is typical for the real 
high-excitation low-metallicity H~{\sc ii} regions used for the $^4$He
abundance determination.

Our range of the number density $N_{\rm e}$ = 10 -- 10$^3$ cm$^{-3}$, 
which was kept constant along
the H~{\sc ii} region radius, covers the whole range expected for the 
extragalactic H~{\sc ii} regions used for the $^4$He abundance determination.

Additionally, to study the effect of density inhomogeneities, we 
calculated two sets of models, each consisting of the 192 H~{\sc ii} 
region models with parameters from Table \ref{tab1} (excluding $N_{\rm e}$). 
The density in the first
set of inhomogeneous models has a Gaussian distribution along the radius $r$ 
according to
\begin{equation}
N_{\rm e}(r)=N_{\rm e}(0)\exp\left[-\frac{r^2}{(30\,{\rm pc})^2}\right], \label{gauss}
\end{equation}
where $N_{\rm e}(0)$ = 10$^3$ cm$^{-3}$. 
The volume-averaged log $U$ in 86 H~{\sc ii} region models out of 
192 H~{\sc ii} region models with Gaussian density distribution is in 
the range $-$3 - $-2$. 

The density in the second set of 192 inhomogeneous models was varied
periodically with radius in the range $N_{\rm e}$= 10 -- 10$^2$ cm$^{-3}$.
A volume-averaged log $U$ in 117 H~{\sc ii} region models out of 
192 H~{\sc ii} region models with periodic density distribution is in 
the range $-$3 - $-2$.
Thus, the total number of the models we used
for our analysis is 363 + 86 + 117 = 566.

In our comparison, we used CLOUDY-calculated emission-line intensities as the 
input parameters for our empirical method.

\begin{table}
\caption{Input parameters for the grid of the photoionised H~{\sc ii}
region models \label{tab1}}
\begin{tabular}{lc} \hline \hline
Parameter                  & Value   \\ \hline
log $Q$(H)$^{\rm a}$             &  52, 53, 54 \\
Starburst age, Myr         &  1.0, 2.0, 3.5, 4.0    \\
$N_{\rm e}$$^{\rm b}$                  &  10, 10$^2$, 10$^3$, var \\
log $f$$^{\rm c}$                & $-$0.5, $-$1.0, $-$1.5, $-$2.0 \\
Oxygen abundance 12+logO/H & 7.3, 7.6, 8.0, 8.3 \\
$^4$He mass fraction $Y$       &  0.254   \\ \hline
\end{tabular}

$^{\rm a}$log of the number of ionising photons in units s$^{-1}$.

$^{\rm b}$The electron number density in units cm$^{-3}$. The electron
number density in models labelled ``var'' is varied along the
radius according to Eq.~\ref{gauss}.

$^{\rm c}$log of volume-filling factor.

\end{table}

\section{He~{\sc i} emissivities}\label{Heemiss}

\subsection{Emissivity fits}\label{fits}

In our empirical method we used the latest set of He~{\sc i} emissivities 
tabulated by \citet{P13} for a wide range of the electron temperature $T_{\rm e}$
and the electron number density $N_{\rm e}$. Similar to \citet{P07}, we first fitted
tabulated data for 32 He~{\sc i} emission lines in the
low-density limit with negligible collisional excitation,
\begin{equation}
\frac{4\pi j_\lambda}{N_{\rm e}N_{{\rm He}^+}} =
\left[a+b(\ln T_{\rm e})^2+c\ln T_{\rm e}+\frac{d}{\ln T_{\rm e}}\right]T^{-1}_e \label{eqa1}
\times 10^{-25},
\end{equation}
where the emissivity $4\pi j_{\lambda}/N_{\rm e}N_{{\rm He}^+}$ is in 
ergs cm$^3$ s$^{-1}$. The coefficients
of the fits for the electron temperature range $T_{\rm e}$ = 5000 -- 25000K are 
shown in Table \ref{taba1} (available only in the electronic edition).
These fits reproduce tabulated data for low $N_{\rm e}$ = 10 cm$^{-3}$ 
with an accuracy of better than 0.1\%. 

\begin{figure}
   \centering
   \includegraphics[width=7cm,angle=-90]{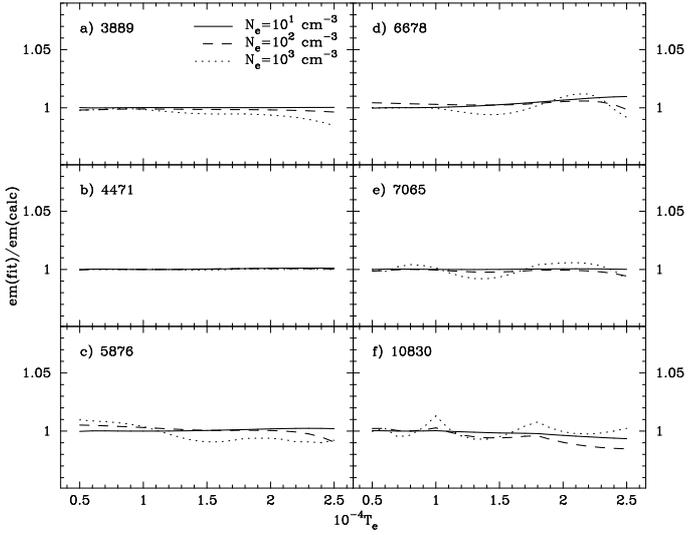}
      \caption{Comparison of calculated and fitted emissivities
of the strongest He~{\sc i} emission lines for three values of
the electron number densities $N_{\rm e}$ = 10, 10$^2$, and 10$^3$ cm$^{-3}$.
Collisional excitation is taken into account.}
         \label{fig1}
   \end{figure}

To fit the ratio of collisional to recombination excitation of He~{\sc i} 
emission lines we used the equation \citep{KF95}
\begin{equation}
\frac{C}{R} = \left(1+\frac{3552t^{-0.55}_{\rm e}}{N_{\rm e}}\right)^{-1}
\sum_i a_it^{b_i}_{\rm e} \exp\left(\frac{c_i}{t_{\rm e}}\right), \label{eqa2}
\end{equation}
where $t_{\rm e}$ is $T_{\rm e}/10000$, and $i$ is an index that varies from 1 to 9.
The coefficients of the fits for 32 He~{\sc i} emission
lines are given in Table \ref{taba2} 
(available only in the electronic edition). We note, however, that we use nine 
terms in Eq. \ref{eqa2}, while \citet{P07} used six terms at most.
To find the total emissivity of a given line, we simply multiplied the result 
obtained from Eq. \ref{eqa1} by the quantity $1 + C/R$ obtained in 
Eq. \ref{eqa2}.
In Fig. \ref{fig1} we compare our fits of emissivities including collisional
excitation for some of the brightest He~{\sc i} emission lines for the electron number 
densities $N_{\rm e}$ = 10, 10$^2$, and 10$^3$ cm$^{-3}$ and the entire range 
of electron temperatures $T_{\rm e}$ with those tabulated by \citet{P13}. It is
seen that the accuracy of our fits is similar to or better 
than 1\% for the electron temperatures $T_{\rm e}$ = 10000 -- 20000K and 
entire range of $N_{\rm e}$, which are the ranges of 
$T_{\rm e}$ and $N_{\rm e}$ in H~{\sc ii} regions used for the $^4$He abundance
determination.

\subsection{Calculating the He~{\sc i}
emission-line intensities in CLOUDY}\label{problem}

\begin{figure}
   \centering
   \includegraphics[width=6cm,angle=-90]{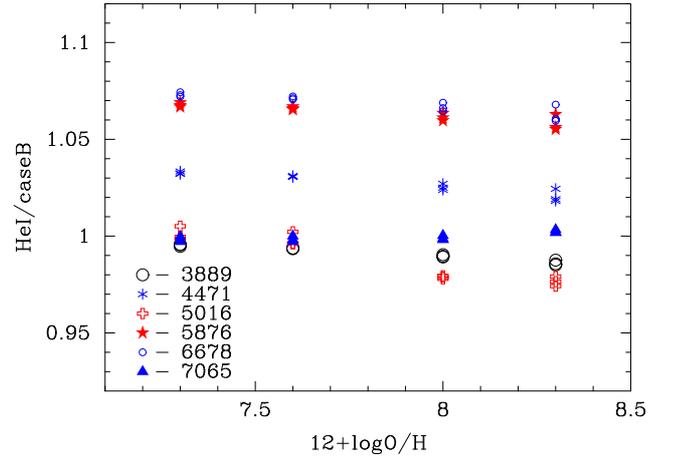}
      \caption{Ratio of the CLOUDY-calculated intensities
with all processes included
to CLOUDY case B intensities for several brightest He~{\sc i} emission
lines as a function of the oxygen abundance for the models with
low electron number density $N_{\rm e}$ = 10 cm$^{-3}$. For clarity only 
models with log $Q$(H) = 53 are shown.}
         \label{fig2}
   \end{figure}

Now we analyse the calculation of He~{\sc i} line intensities in CLOUDY with 
respect to the He~{\sc i} emissivities.
CLOUDY outputs include
H and He~{\sc i} emission-line intensities calculated under different
assumptions: 1) case A; 2) case A, including collisional contribution;
3) case B; 4) case B, including collisional contribution, etc.
One of the CLOUDY outputs are the 
He~{\sc i} line intensities calculated taking into account all
detailed physics regarding collisional transitions, radiative transfer, etc.
This is the  ``predicted line intensity with all processes included''
in the CLOUDY output. One would expect that for a
low-density ionisation-bounded H~{\sc ii} region
these calculated He~{\sc i} emission-line intensities would be close to the
CLOUDY output recombination case B intensities. 
In Fig. \ref{fig2}  we show the ratios of CLOUDY line intensities calculated
with taking into account all processes
to CLOUDY case B intensities for some important He~{\sc i} 
emission lines in the
models with the electron number density $N_{\rm e}$ = 10 cm$^{-3}$. The
collisional excitation of He~{\sc i} emission lines at this low $N_{\rm e}$ is 
very low. It is highest for the $\lambda$7065 emission line and does not 
exceed 1.5\% of the recombination intensity. 

The only known mechanism that may cause line intensities to deviate from their
recombination 
values in Fig. \ref{fig2} is the fluorescent excitation due to the
non-negligible optical depth for the He~{\sc i} $\lambda$3889 emission line.
Two emission lines, He~{\sc i} $\lambda$3889 and $\lambda$7065, are the most 
sensitive to the fluorescent excitation. However, the optical depth of the 
He~{\sc i} $\lambda$3889
emission line in the models considered in Fig. \ref{fig2} is small, $\la$ 0.1. 
Therefore, the effect of fluorescent excitation is low in the considered 
models with $N_{\rm e}$ = 10 cm$^{-3}$. It is seen in Fig. \ref{fig2}
that the intensities of He~{\sc i} $\lambda$3889 and $\lambda$7065
emission lines with ``all processes included'' are close to the case B
intensities.

Similarly, the calculated intensity of the singlet He~{\sc i} $\lambda$5016
emission line is close to the case B value, indicating that
the considered models have high optical depths in the resonance transitions
from the ground level of the singlet He~{\sc i} state 
(e.g., the optical depth of the $\lambda$584
line in the CLOUDY output is $\ga$ 10$^5$), closely corresponding to the
case B.

On the other hand, the intensities of two important lines, 
$\lambda$5876 and $\lambda$6678, calculated with ``all processes included''
are higher by $\sim$ 6\% -- 7\% than the case B values, which is difficult to
understand, because both lines are less affected by collisional and
fluorescent excitation than the  $\lambda$3889 and $\lambda$7065
lines. The same effect to a lesser extent is also present for the  
$\lambda$4471 line. It is likely that there is a problem in the 
computation 
of the He~{\sc i} emission line intensities in version v13.01 of the 
CLOUDY code.
Therefore, we did not use the CLOUDY He~{\sc i} intensities with 
``all processes included'' in our subsequent analysis for all He~{\sc i} 
emission lines. Instead, we adopted the CLOUDY He~{\sc i} case B intensities 
enhanced by collisions.

   \begin{figure}
   \centering
   \includegraphics[width=6cm,angle=-90]{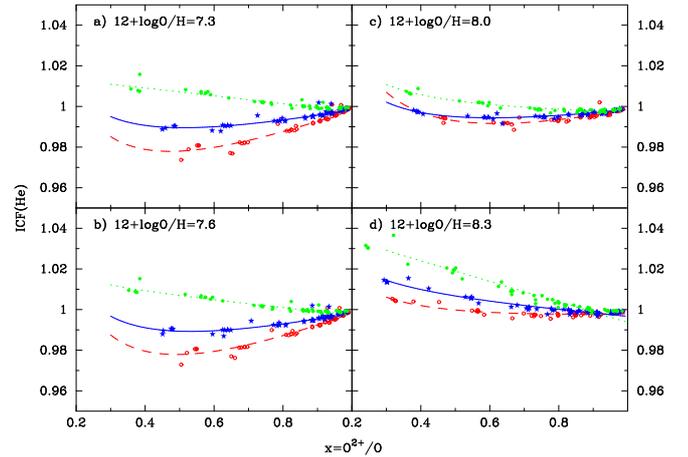}
      \caption{Ionisation correction factors $ICF$(He) vs.
excitation parameter $x$ = O$^{2+}$/O from the CLOUDY models with various 
oxygen abundances 12+logO/H and starburst ages. Red dashed, blue solid,
and green dotted lines correspond to the starburst ages of 1.0-2.0, 3.5, 
4.0 Myr, respectively. Symbols are CLOUDY-modelled data.}
         \label{fig3}
   \end{figure}

\section{Ionisation correction factors for He}\label{icfHe}

For our comparison of the empirical method with the CLOUDY models
we also needed to take into account the ionisation structure of the H~{\sc ii}
region in the empirical method. The He$^+$ zone can be slightly larger or 
slightly smaller than
the H$^+$ zone, depending on the spectral energy distribution
of the ionising radiation. This
effect was taken into account in our previous work \citep[e.g. ][]{I07} by
introducing the ionisation correction factor $ICF$(He) as a function
of the excitation parameter $x$=O$^{2+}$/(O$^+$+O$^{2+}$).
The CLOUDY output  allows us to obtain the exact value of $ICF$(He), 
which is equal to $x$(H$^+$)/($x$(He$^+$)+$x$(He$^{2+}$)), where $x$ stands for
volume-integrated ionic fractions.  
We fitted these $ICF$(He) (colour symbols 
in Fig. \ref{fig3})
for four values of 12 + logO/H = 7.3, 7.6, 8.0 and 8.3 
and for four values of the starburst age 1.0, 2.0, 3.5 and 4.0 Myr by the 
expressions (we note that expressions for 1.0 Myr and 2.0 Myr are
identical)\\

\vspace{0.2cm}

\noindent a) for 12 + logO/H = 7.3

\begin{equation}
ICF(1.0, 2.0~{\rm Myr})=0.07289 \times x+0.90898+0.01628/x, \label{icf7.3_2.0}
\end{equation}
\begin{equation}
ICF(3.5~{\rm Myr})=0.03663 \times x+0.95204+0.00959/x, \label{icf7.3_3.5}
\end{equation}
\begin{equation}
ICF(4.0~{\rm Myr})=-0.01986 \times x+1.01744-0.00017/x, \label{icf7.3_4.0}
\end{equation}

\vspace{0.2cm}

\noindent b) for 12 + logO/H = 7.6

\begin{equation}
ICF(1.0, 2.0~{\rm Myr})=0.07844 \times x+0.90076+0.01896/x, \label{icf7.6_2.0}
\end{equation}
\begin{equation}
ICF(3.5~{\rm Myr})=0.04144 \times x+0.94511+0.01176/x, \label{icf7.6_3.5}
\end{equation}
\begin{equation}
ICF(4.0~{\rm Myr})=-0.01640 \times x+1.01232+0.00144/x, \label{icf7.6_4.0}
\end{equation}

\vspace{0.2cm}

\noindent c) for 12 + logO/H = 8.0

\begin{equation}
ICF(1.0, 2.0~{\rm Myr})=0.04555 \times x+0.93531+0.01740/x, \label{icf8.0_2.0}
\end{equation}
\begin{equation}
ICF(3.5~{\rm Myr})=0.02508 \times x+0.96411+0.00915/x, \label{icf8.0_3.5}
\end{equation}
\begin{equation}
ICF(4.0~{\rm Myr})=0.00158 \times x+0.98968+0.00620/x, \label{icf8.0_4.0}
\end{equation}

\vspace{0.2cm}

\noindent d) for 12 + logO/H = 8.3

\begin{equation}
ICF(1.0, 2.0~{\rm Myr})=0.00812 \times x+0.98390+0.00593/x, \label{icf8.3_2.0}
\end{equation}
\begin{equation}
ICF(3.5~{\rm Myr})=-0.01228 \times x+1.00508+0.00390/x, \label{icf8.3_3.5}
\end{equation}
\begin{equation}
ICF(4.0~{\rm Myr})=-0.05020 \times x+1.04383+0.00011/x, \label{icf8.3_4.0}
\end{equation}
which are applicable for $x$ $\ga$ 0.4, 
corresponding to the case for 
high-excitation H~{\sc ii} regions used for the $^4$He abundance determination.
These fits are shown in Fig. \ref{fig3} by red dashed, blue solid and
green dotted lines for starburst ages of 1.0-2.0, 3.5, and 4.0 Myr, 
respectively. $ICF$s are lower for models
with lower oxygen abundance 12 + logO/H = 7.3 and harder ionising radiation.
They are higher for the highest metallicity models with
12~+~logO/H~=~8.3, because the spectral energy distribution
of the ionising radiation is softer. 
$ICF$s derived from Eqs. \ref{icf7.3_2.0} -- \ref{icf8.3_4.0}, which we use
below, are close to those used by e.g. \citet{I07}.
We found that $ICF$ values are not sensitive to the assumption on the 
density distribution. Ionisation correction factors $ICF$(He) for all CLOUDY 
models (colour symbols in Fig. \ref{fig3}) are reproduced by the same fits for 
the homogeneous and inhomogeneous models.

\section{Non-recombination excitation of hydrogen}\label{hydro}

All element abundances in H~{\sc ii} regions are commonly derived relative to 
that of hydrogen.
In particular, the $^4$He abundance is derived from the ratio of the recombination
He~{\sc i} line intensity and the recombination intensity of 
the hydrogen H$\beta$
emission line. Additionally, the observed Balmer decrement in real
H~{\sc ii} regions is used for dust-reddening corrections. 
Therefore, prior to the reddening correction and abundance determination, 
hydrogen line intensities should be corrected for  
collisional and fluorescent excitation, which cause them to deviate 
from the recombination 
values. Much work has been done in previous studies to take these effects
into account \citep[see discussion and references in ][]{I07,IT10}.

CLOUDY outputs allow one to obtain the fraction of non-recombination contribution
($C+F$)/$I$ to intensities of hydrogen lines, where $C$ and $F$ are
collisional and fluorescent contribution to the intensity, and $I$ 
is the intensity. In particular, this contribution
for the H$\beta$ emission line is defined as a ratio 
[$I$(H$\beta$) -- caseB(H$\beta$)]/$I$(H$\beta$), where $I$(H$\beta$) is 
the H$\beta$ intensity calculated with ``all processes included'' and 
caseB(H$\beta$) is the case B recombination value. Similar expressions can
be applied for other hydrogen lines. We used our grid of CLOUDY models to
fit ($C+F$)/$I$ for the four strongest Balmer hydrogen H$\alpha$, H$\beta$,
H$\gamma$, and H$\delta$ emission lines as follows:\\

\vspace{0.2cm}

\noindent a) for the ages of 1.0 Myr and 2.0 Myr

\begin{eqnarray}
\frac{C+F}{I}({\rm H}\alpha)&=&-48.3963+19.0348A({\rm O})-2.4792A({\rm O})^2 \nonumber \\
                            & &+0.1071A({\rm O})^3,
\label{collHa_2.0}
\end{eqnarray}
\begin{eqnarray}
\frac{C+F}{I}({\rm H}\beta)&=&-28.3279+11.1685A({\rm O})-1.4571A({\rm O})^2 \nonumber \\
                            & &+0.0629A({\rm O})^3,
\label{collHb_2.0}
\end{eqnarray}
\begin{eqnarray}
\frac{C+F}{I}({\rm H}\gamma)&=&-24.3327+9.6403A({\rm O})-1.2631A({\rm O})^2 \nonumber \\
                            & &+0.0548A({\rm O})^3,
\label{collHg_2.0}
\end{eqnarray}
\begin{eqnarray}
\frac{C+F}{I}({\rm H}\delta)&=&-8.6082+3.5564A({\rm O})-0.4814A({\rm O})^2 \nonumber \\
                            & &+0.0215A({\rm O})^3,
\label{collHd_2.0}
\end{eqnarray}

\vspace{0.2cm}

\noindent b) for the age of 3.5 Myr

\begin{eqnarray}
\frac{C+F}{I}({\rm H}\alpha)&=&-14.0001+5.7342A({\rm O})-0.7696A({\rm O})^2 \nonumber \\
                            & &+0.0340A({\rm O})^3,
\label{collHa_3.5}
\end{eqnarray}
\begin{eqnarray}
\frac{C+F}{I}({\rm H}\beta)&=&-7.6283+3.1765A({\rm O})-0.4315A({\rm O})^2 \nonumber \\
                            & &+0.0192A({\rm O})^3,
\label{collHb_3.5}
\end{eqnarray}
\begin{eqnarray}
\frac{C+F}{I}({\rm H}\gamma)&=&3.2205-1.0250A({\rm O})+0.1099A({\rm O})^2 \nonumber \\
                            & &-0.0040A({\rm O})^3,
\label{collHg_3.5}
\end{eqnarray}
\begin{eqnarray}
\frac{C+F}{I}({\rm H}\delta)&=&20.6912-7.7845A({\rm O})+0.9793A({\rm O})^2 \nonumber \\
                            & &-0.0411A({\rm O})^3,
\label{collHd_3.5}
\end{eqnarray}

\vspace{0.2cm}

\noindent c) for the age of 4.0 Myr

\begin{eqnarray}
\frac{C+F}{I}({\rm H}\alpha)&=&22.4758-8.4465A({\rm O})+1.0630A({\rm O})^2\\ \nonumber
                            & &-0.0447A({\rm O})^3,
\label{collHa_4.0}
\end{eqnarray}
\begin{eqnarray}
\frac{C+F}{I}({\rm H}\beta)&=&21.3317-8.0875A({\rm O})+1.0252A({\rm O})^2 \nonumber \\
                            & &-0.0434A({\rm O})^3,
\label{collHb_4.0}
\end{eqnarray}
\begin{eqnarray}
\frac{C+F}{I}({\rm H}\gamma)&=&33.8930-12.9456A({\rm O})+1.6505A({\rm O})^2 \nonumber \\
                            & &-0.0702A({\rm O})^3,
\label{collHg_4.0}
\end{eqnarray}
\begin{eqnarray}
\frac{C+F}{I}({\rm H}\delta)&=&50.5361-19.3833A({\rm O})+2.4787A({\rm O})^2 \nonumber \\
                            & &-0.1056A({\rm O})^3,
\label{collHd_4.0}
\end{eqnarray}
where $A$(O) = 12 + log O/H. 

\begin{figure}
   \centering
   \includegraphics[width=6cm,angle=-90]{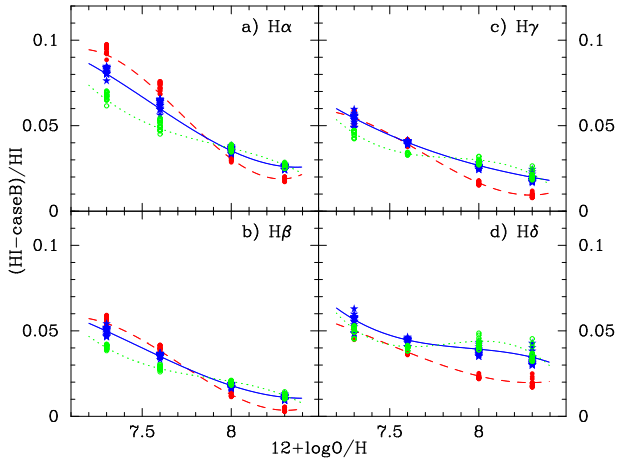}
      \caption{Non-recombination contribution of Balmer
hydrogen line intensities as a function of the oxygen abundance.
The red dashed, blue solid, and green dotted lines show fits for starburst
ages of 2.0, 3.5, and 4.0 Myr, respectively. The dependences for starburst
age of 1.0 Myr are identical to those for 2.0 Myr.
Symbols are CLOUDY-modelled data.}
         \label{fig4}
   \end{figure}

These fits are shown in Fig. \ref{fig4} 
by red dashed, blue solid, and green dotted lines for models with starburst
ages of 2.0, 3.5, and 4.0 Myr, respectively.
It is seen from the figure that
the non-recombination contribution in low-metallicity H~{\sc ii} regions
can be as high as 9\% and 6\% for the H$\alpha$ and other hydrogen lines,
respectively. Neglecting this effect would result in appreciable 
underestimation of the $^4$He/H abundance ratio. 
Recently, \citet{L09} considered
collisional excitation of hydrogen lines based on calculations with an
earlier version of the CLOUDY code and probably with earlier set of the atomic
data for hydrogen. She found that collisional contribution could be as high
as 5\% for the H$\alpha$ emission line, or about twice as low as
the non-recombination contribution in Fig. \ref{fig4}. 
In the following,  we do not specify the collisional contribution and use the 
CLOUDY output values that include all processes that cause the hydrogen line 
intensities to deviate from the recombination
values. In particular, fluorescent excitation may play a role, as was
suggested by \citet{L09a}. To derive the helium abundance we used 
the non-recombination contribution to the 
hydrogen line intensities obtained in the present paper.
 
\section{Comparing the CLOUDY input  and empirically derived values of
the $^4$He abundance}\label{compar}

Now we examine how well our empirical method described in  
\citet{IT04} and \citet{I07} for determining the $^4$He abundance 
recovers the input $^4$He abundance in CLOUDY models.
For this we first updated our empirical code so that its atomic 
ingredients were compatible with those used in CLOUDY: we incorporated the 
He~{\sc i} emissivities from \citet{P13}, the new data on the 
non-recombination excitation of hydrogen emission lines, and the new values 
of the ionisation correction factors for He obtained 
in the previous sections. Then we applied our updated empirical code 
to our 566 CLOUDY models, treating the CLOUDY output line intensities as if 
they were observed ones. 

We used
hydrogen line intensities calculated with CLOUDY, which include the
non-recombination contribution. 
Furthermore, we used CLOUDY case B He~{\sc i} line intensities enhanced only 
by collisions. The latter was done because of the
problems with CLOUDY He~{\sc i} line intensities for
``all processes included'' that we discussed in Sect. \ref{problem}.
Therefore, the fluorescent excitation of He~{\sc i} 
emission lines in the empirical method was set to zero when comparing 
with CLOUDY He {\sc i} case B emission-line intensities. However, we note that
in real H {\sc ii} regions discussed below fluorescent excitation
is taken into account to correct the observed intensities of He {\sc i}
emission lines. The CLOUDY forbidden line intensities of heavy elements
were used to derive the electron temperature and heavy element abundances.

Then, we derived the $^4$He$^+$ abundance $y^+_i$ = $^4$He$^+_i$/H$^+$ from a 
number of strongest He~{\sc i} emission lines.

The weighted mean of the $y^+_i$, 
$y^+_{\rm wm}$, is defined by
\begin{equation}
y^+_{\rm wm}=\frac{\sum_i^n{y^+_i/\sigma^2(y^+_i)}}
{\sum_i^n{1/\sigma^2(y^+_i)}}\label{eq2},
\end{equation}
where $y^+_i$ is the $^4$He$^+$ abundance derived from the intensity of the 
He~{\sc i}
emission line labelled $i$, and $\sigma(y^+_i)$ is the statistical error
of $y^+_i$. 

We applied the 
Monte Carlo procedure described in \citet{IT04} and \citet{I07}, 
randomly varying the electron temperature $T_{\rm e}$(He$^+$) and the electron number
density $N_{\rm e}$(He$^+$) within a specified range, to minimise the quantity
\begin{equation}
\chi^2=\sum_i^n\frac{(y^+_i-y^+_{\rm wm})^2}{\sigma^2(y^+_i)}\label{eq1}.
\end{equation}
The resulting $y^+_{\rm wm}$ is the value representing the empirically derived 
$^4$He$^+$ abundance in each model.

In our comparison analysis we used 
$y^+_i$ successively with equal weights and with weights proportional
to the He~{\sc i} emission line intensity. The latter case is appropriate
to the observational data because the intensities of weaker emission lines 
are more uncertain and therefore should be considered with lower weights.

Additionally, in cases with the nebular He~{\sc ii} $\lambda$4686 
emission line, we added the abundance of doubly ionised 
helium $y^{2+}$ $\equiv$ He$^{2+}$/H$^+$ to $y^+$. Although the He$^{2+}$ zone
is hotter than the He$^{+}$ zone, we 
adopted $T_{\rm e}$(He$^{2+}$) = $T_{\rm e}$(He$^{+}$).
The last assumption has only a minor effect on the $y$ value, because
 $y$$^{2+}$ is small ($\leq$ 0.5\% of $y^+$) in CLOUDY models.

The total $^4$He abundance $y$ is obtained from the expression 
$y$~=~$ICF$(He)$\times$($y^+$+$y$$^{2+}$) and is converted to the
$^4$He mass fraction using equation
\begin{equation}
Y=\frac{4y(1-Z)}{1+4y}, \label{eq:Y}
\end{equation}
where $Z$ = $B$$\times$O/H is the heavy-element mass fraction.
The coefficient $B$ depends on O/$Z$, where O is the oxygen mass fraction.
\citet{M92} derived O/$Z$ = 0.66 and 0.41 for $Z$ = 0.001 and 0.02, 
respectively. The latter value is close to the most recent one in the Sun 
\citep[0.43 using the abundances from ][]{A09}. Adopting 
the \citet{M92} values, we obtain $B$ = 18.2 and
27.7 for $Z$ = 0.001 and 0.02, respectively. In our calculations  for
every H~{\sc ii} region we used a value of $B$ that
linearly scales with $A$(O) (=~12~+~log O/H) as follows:
\begin{equation}
B = 8.64A({\rm O})-47.44.\label{B}
\end{equation} 
For comparison, \citet{P92} and \citet{I07}, for example, used constant
$B$ of 20 and 18.2, respectively.
We note, however, that neglecting the $B$ -- $A$(O) dependence would
result in tiny uncertainties in $Y$, not exceeding 0.1\% in the whole range of
$A$(O) considered in this paper.

   \begin{figure*}
   \centering
  \hbox{ \includegraphics[width=4cm,angle=-90]{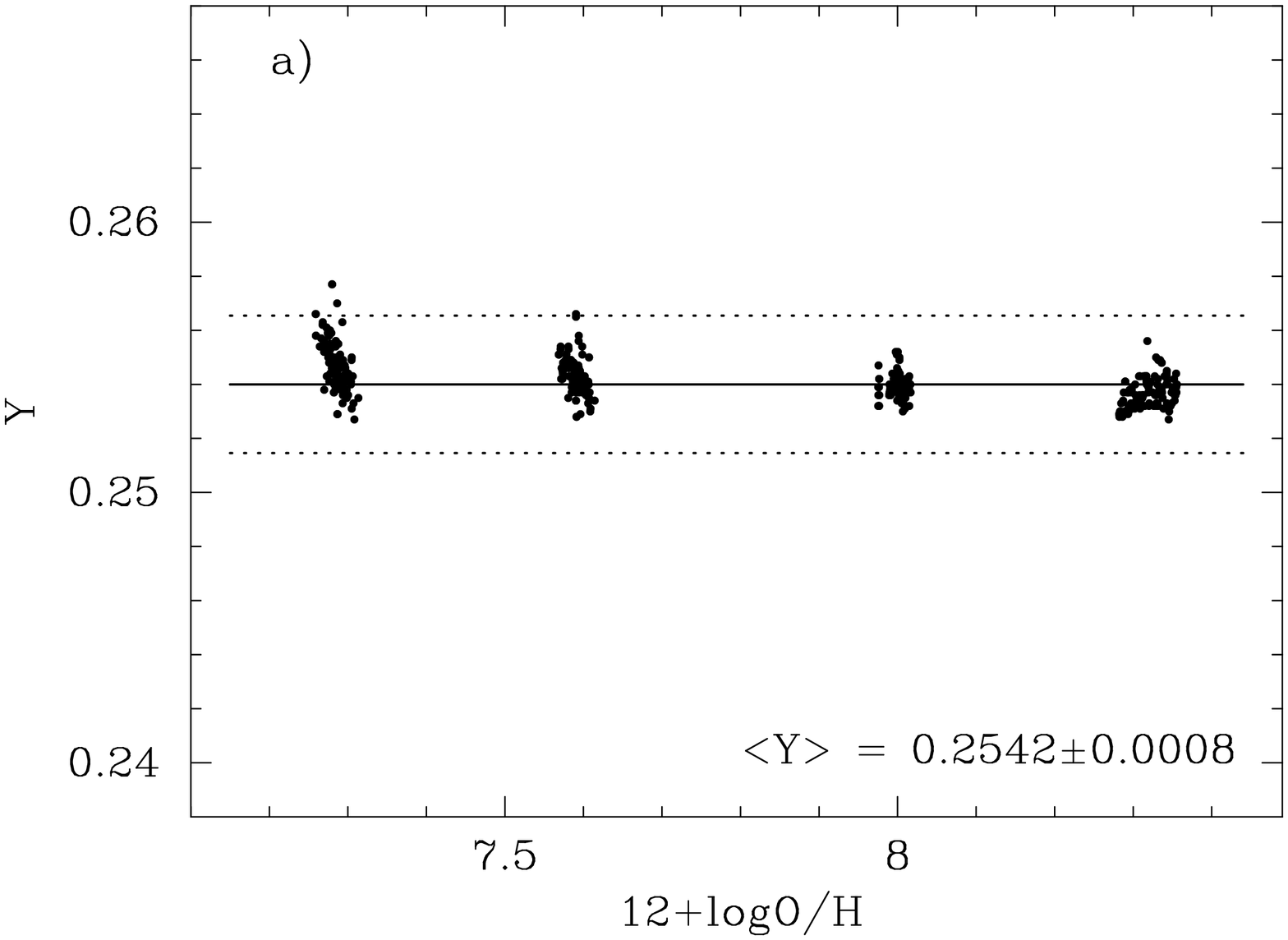}
   \hspace{0.2cm}\includegraphics[width=4cm,angle=-90]{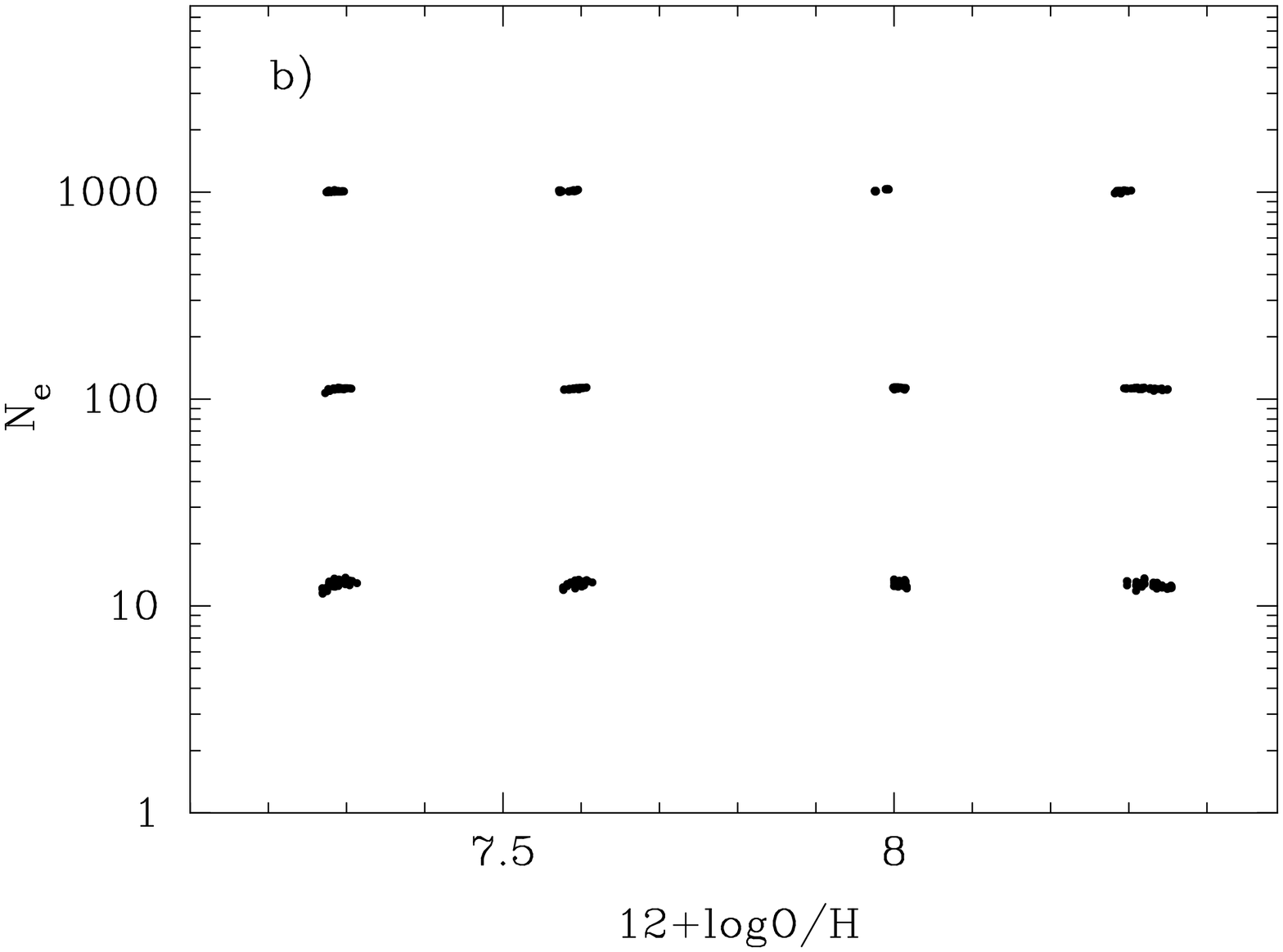}
   \hspace{0.2cm}\includegraphics[width=4cm,angle=-90]{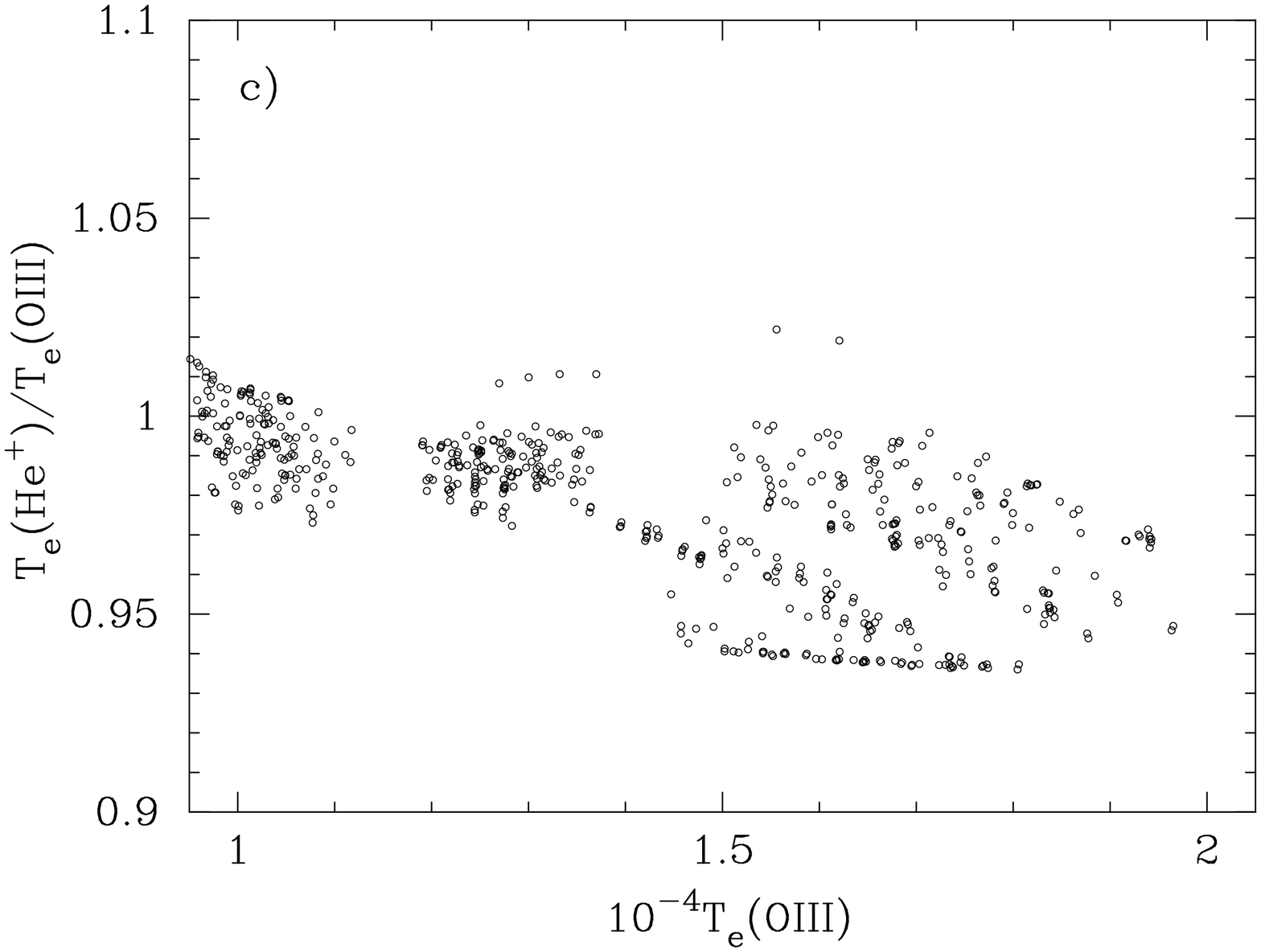}}
\vspace{0.2cm}
  \hbox{ \includegraphics[width=4cm,angle=-90]{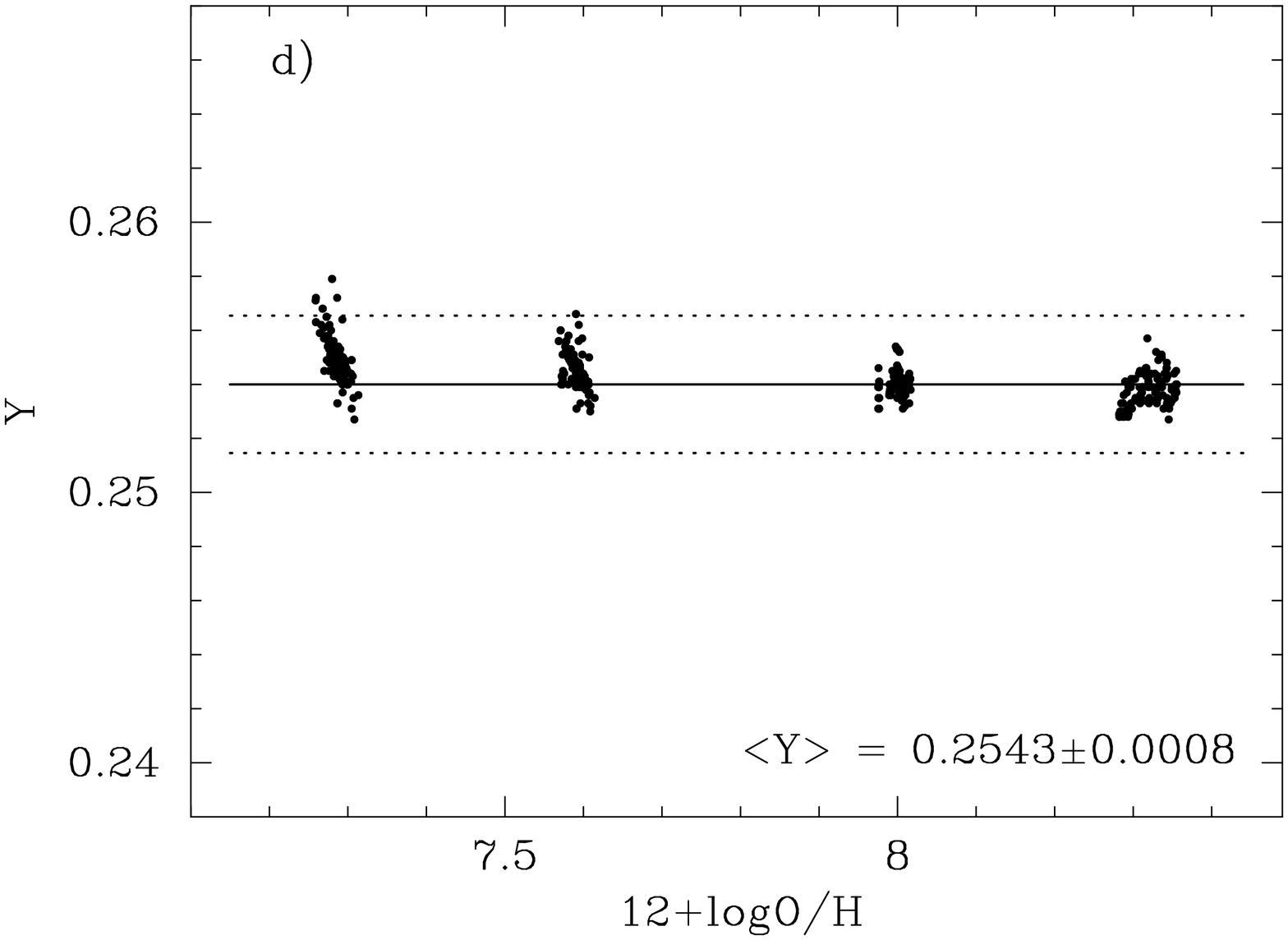}
   \hspace{0.2cm}\includegraphics[width=4cm,angle=-90]{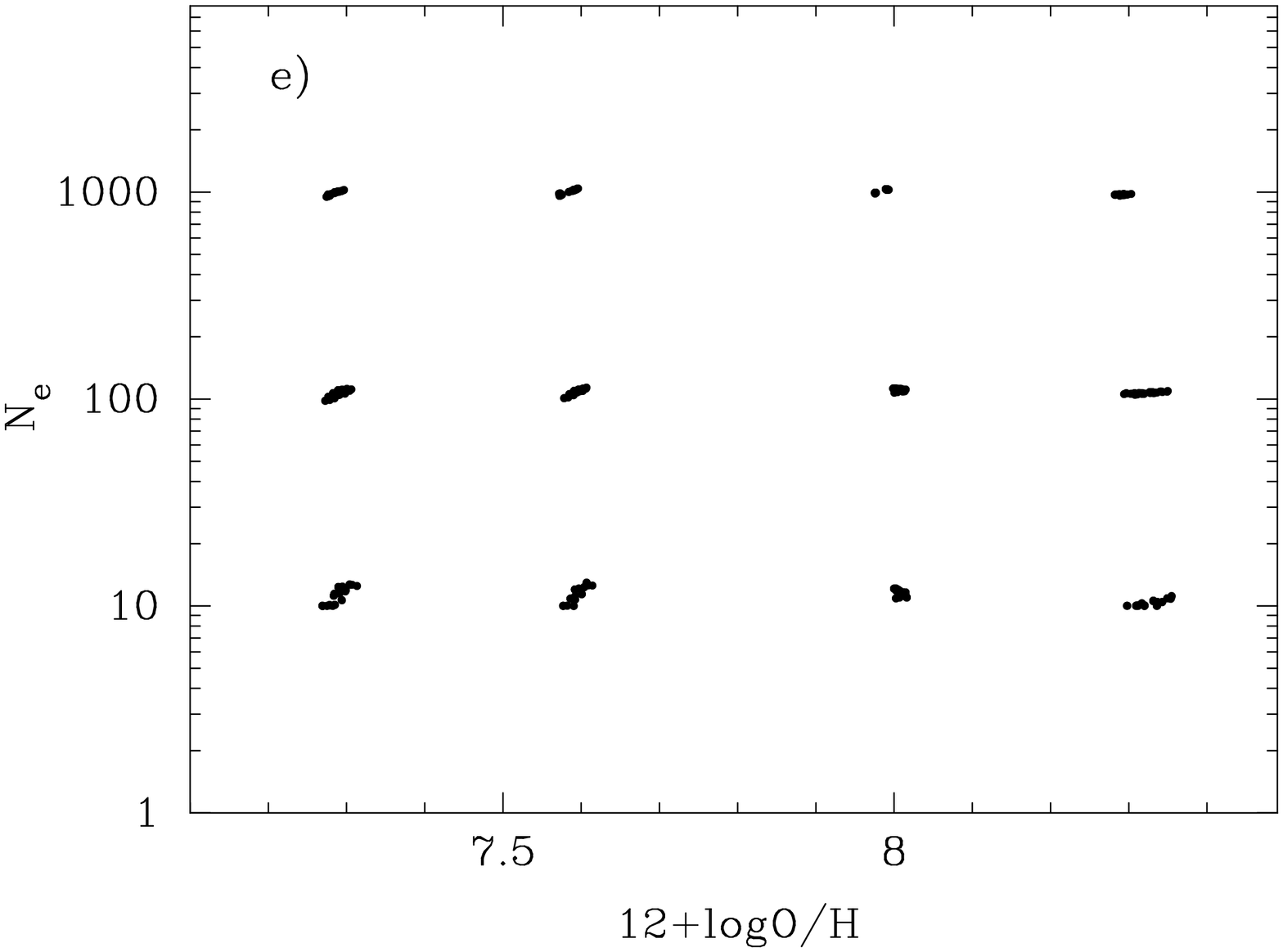}
   \hspace{0.2cm}\includegraphics[width=4cm,angle=-90]{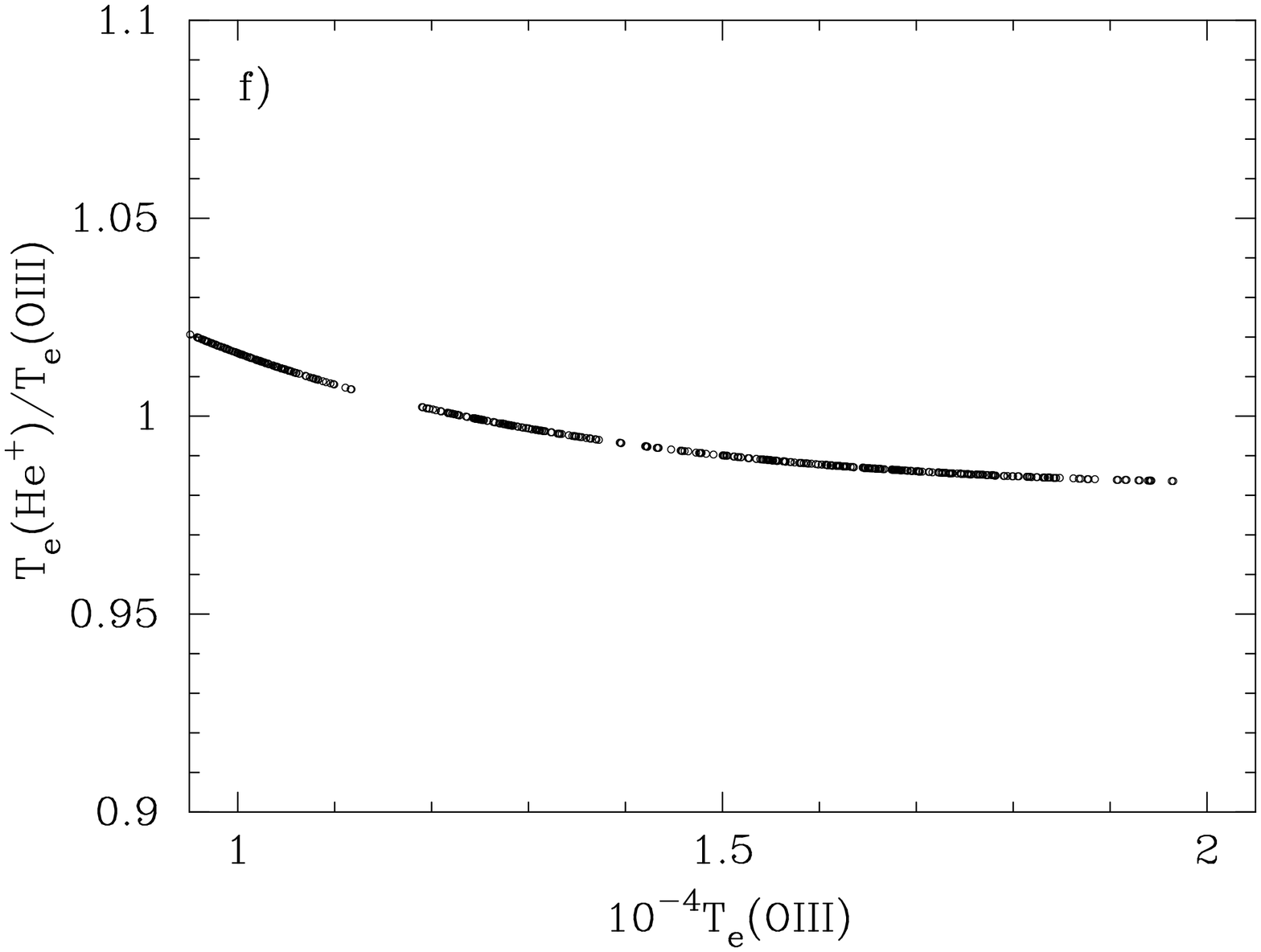}}
\vspace{0.2cm}
 \hbox{  \includegraphics[width=4cm,angle=-90]{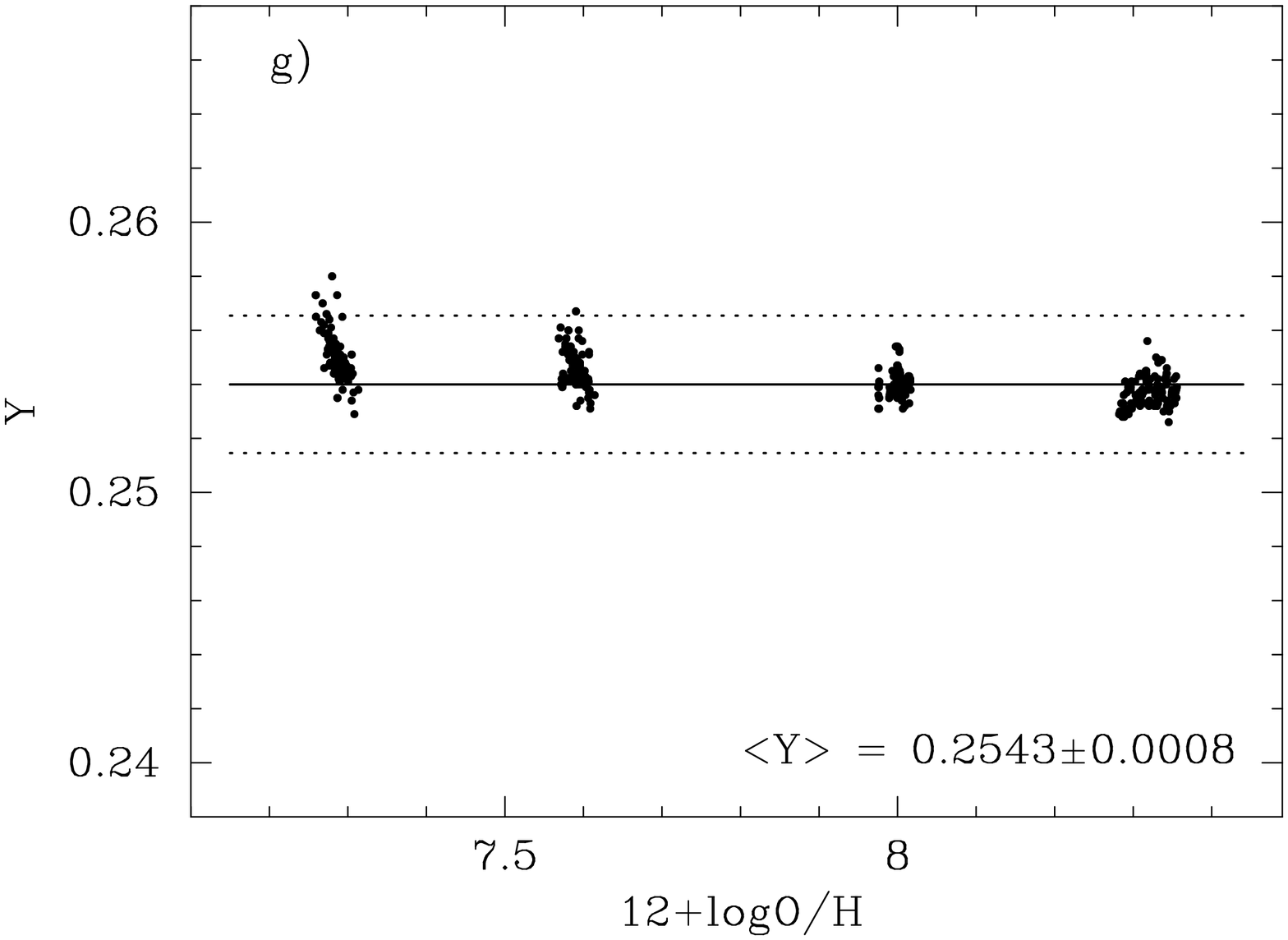}
   \hspace{0.2cm}\includegraphics[width=4cm,angle=-90]{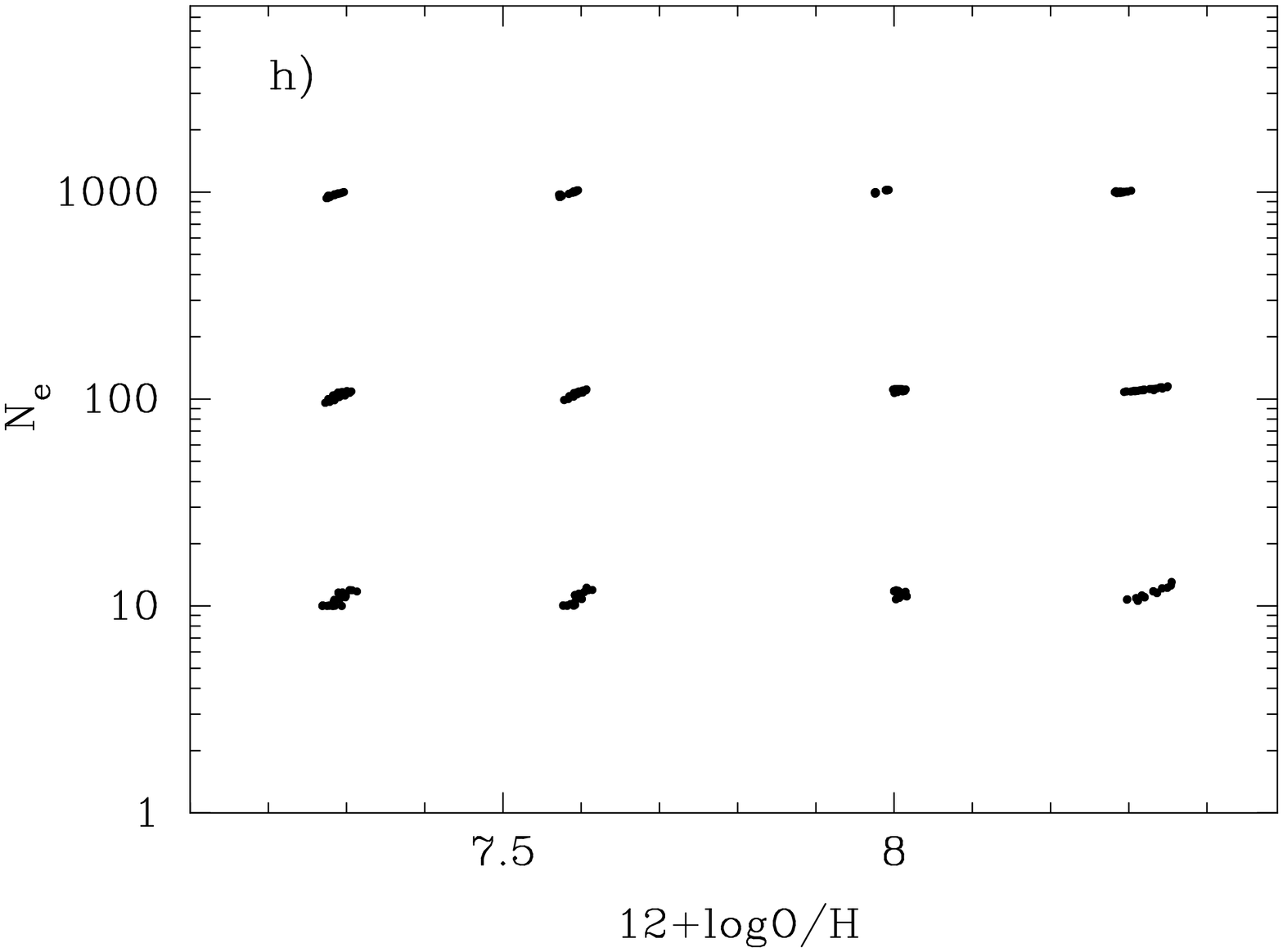}
   \hspace{0.2cm}\includegraphics[width=4cm,angle=-90]{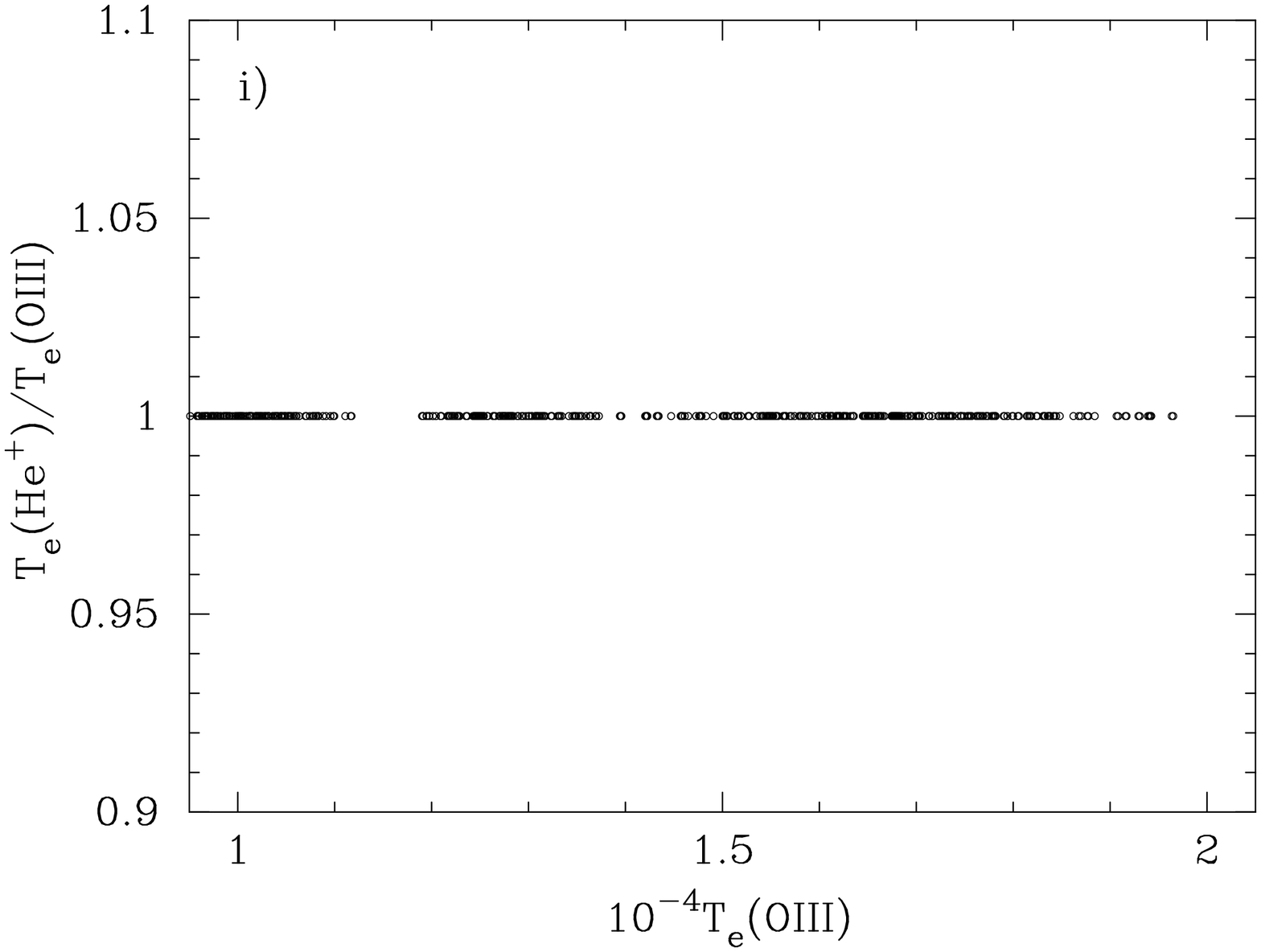}}
      \caption{ Distribution with the oxygen abundance of the 
empirically derived weighted mean $Y$ (a) and $N_{\rm e}$ (b) values calculated with 
the \citet{P13} He~{\sc i} emissivities. Nine He~{\sc i}
$\lambda$3889, $\lambda$4026, $\lambda$4388, $\lambda$4471, $\lambda$4922, 
$\lambda$5876, $\lambda$6678, $\lambda$7065, and $\lambda$10830 emission lines 
are used for $\chi^2$ minimisation and determination of $Y$. In the 
lower parts of the panels showing $Y$ as a function of O/H we indicate the 
mean of all $Y$ values derived from the models, together with the
dispersion. {\bf Top} The electron temperature $T_{\rm e}$(He$^+$) is 
randomly varied in the range 
(0.95 -- 1.05)$\times$ $\widetilde{T}_{\rm e}$(He$^+$)
where $\widetilde{T}_{\rm e}$(He$^+$) is defined by Eq. \ref{tHeOIII},
and the best derived values of $T_{\rm e}$(He$^+$)
for every model are shown in (c). The solid 
line in (a) shows the input CLOUDY $Y$ value of 0.254, the dotted lines are
1\% deviations, and $<$$Y$$>$ is the average value of $Y$s shown by 
filled circles. {\bf Middle} The same as in the top panel, but 
$T_{\rm e}$(He$^+$) = $\widetilde{T}_{\rm e}$(He$^+$). 
{\bf Bottom} The same as in the top 
panel, but $T_{\rm e}$(He$^+$) = $T_{\rm e}$(O~{\sc iii}).}
         \label{fig5}
   \end{figure*}

In Fig. \ref{fig5} we show the empirically derived 
oxygen abundances 12+log O/H, $^4$He mass 
fractions $Y$ and the electron temperatures $T_{\rm e}$(He$^+$)
from the CLOUDY-predicted emission-line intensities 
for all 566 models 
and the electron number densities $N_{\rm e}$ for 363 models with spatially 
constant number density.
Here we adopted the \citet{P13} He~{\sc i} emissivities defined by 
Eqs. \ref{eqa1} and \ref{eqa2} with coefficients from Tables \ref{taba1}
and \ref{taba2}.
We used the nine strongest He~{\sc i} 
$\lambda$3889, $\lambda$4026, $\lambda$4388, $\lambda$4471, $\lambda$4922, 
$\lambda$5876, $\lambda$6678, $\lambda$7065, and $\lambda$10830
emission lines both for $\chi^2$
minimisation and determination of the weighted mean $Y$. 
Equal weights for different lines were adopted. We varied $N_{\rm e}$(He$^+$) 
in our Monte Carlo simulations in the range 10 -- 2$\times$10$^3$ cm$^{-3}$.
The electron temperature $T_{\rm e}$(O~{\sc iii}) was
derived in our empirical routine from the 
[O~{\sc iii}]$\lambda$4363/($\lambda$4959+$\lambda$5007) emission line 
intensity ratio prior to determining $Y$ and is very close to the 
volume-emissivity-averaged value of the electron
temperature in the O$^{2+}$ zone calculated with CLOUDY grid models for the 
whole range of metallicities.
For variations of the electron temperature $T_{\rm e}$(He$^+$) we adopted
three cases: 

1) the electron temperature $T_{\rm e}$(He$^+$) was randomly varied
in the range (0.95 -- 1.05) of the temperature derived from the relation 
between volume-averaged temperatures in
our CLOUDY models, which can be fitted by the expression
\begin{equation}
\widetilde{T}_{\rm e}({\rm He}^+) = 2.51\times 10^{-6} T_{\rm e}({\rm O}~\textsc{iii})+0.8756+
1152/T_{\rm e}({\rm O}~\textsc{iii}).
\label{tHeOIII}
\end{equation}
This relation predicts 
$\widetilde{T}_{\rm e}$(He$^+$) $<$ $T_{\rm e}$(O~{\sc iii}) for hotter H~{\sc ii} regions
and $\widetilde{T}_{\rm e}$(He$^+$) $>$ $T_{\rm e}$(O~{\sc iii}) for cooler H~{\sc ii} 
regions; 

2) $T_{\rm e}$(He$^+$) = $\widetilde{T}_{\rm e}$(He$^+$);

3) $T_{\rm e}$(He$^+$) = $T_{\rm e}$(O~{\sc iii}).

The results of our tests for the above three choices of $T_{\rm e}$(He$^+$)
variations and randomly varying $N_{\rm e}$(He$^+$)
are shown in Figs. \ref{fig5}a - \ref{fig5}c,
\ref{fig5}d - \ref{fig5}e, and \ref{fig5}g - \ref{fig5}i, respectively.
In  panels a, d, and g we indicate the mean of all $Y$ values derived 
from the models, together with the dispersion. It is seen that
empirically derived oxygen abundances 12+log O/H (dots)
agree well with the input CLOUDY values of 7.3, 7.6, 8.0, and
8.3 and deviate from them by not more than $\sim$ 0.03 dex for 12 + log O/H =
7.3 - 8.0 and by $\sim$ 0.06 dex for 12 + log O/H = 8.3.  
It is also seen from Figs. \ref{fig5}a, \ref{fig5}d, and \ref{fig5}g that the 
empirically derived $^4$He mass fractions $Y$ (dots) agree well with the 
input CLOUDY value (solid line) and deviate from it by not more than
at most 0.5\% in most cases and only occasionally by up to 1.5\%. 
The dispersion of the empirically derived $Y$s is
slightly lower when the electron temperature $T_{\rm e}$(He$^+$)
is varied in the range 
(0.95 -- 1.05) $\times$ $\widetilde{T}_{\rm e}$(He$^+$)
(Fig. \ref{fig5}c). The input CLOUDY number densities of 10, 10$^2$, and
10$^3$ cm$^{-3}$ in models with spatially constant $N_{\rm e}$ are also very well 
reproduced (Figs. \ref{fig5}b, \ref{fig5}e, and \ref{fig5}h).
We also note that despite the broad adopted range of $T_{\rm e}$(He$^+$) 
variations of (0.95 -- 1.05) $\times$ $\widetilde{T}_{\rm e}$(He$^+$)
for the models in the 
top panel of Fig. \ref{fig5}, its best derived values behave
in Fig. \ref{fig5}c similar to values calculated with CLOUDY 
(Fig. \ref{fig5}f),
suggesting that empirically derived $T_{\rm e}$(He$^+$) is also reproduced quite
well.

\begin{figure*}
   \centering
\hbox{
    \includegraphics[width=7cm,angle=-90]{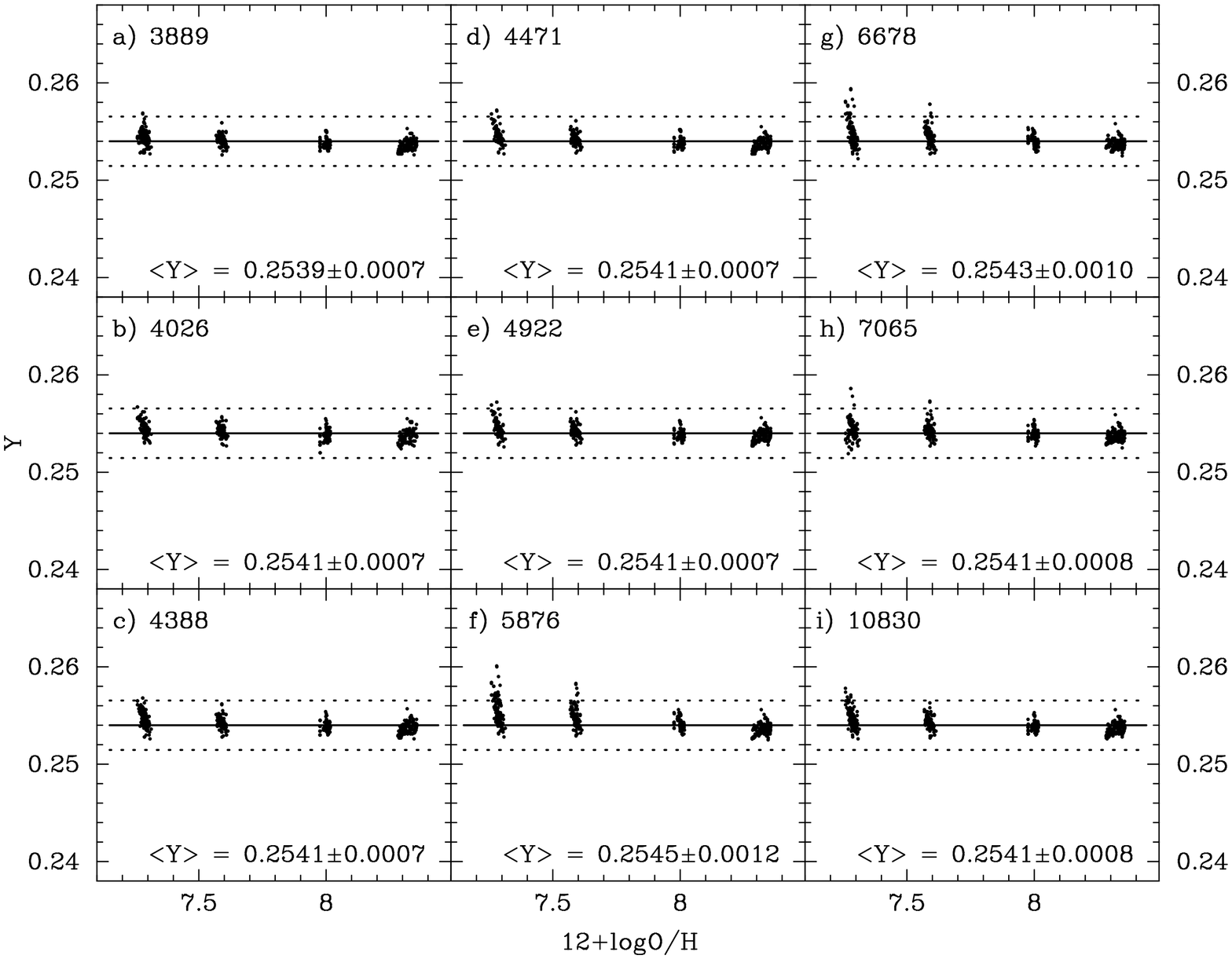}
   \hspace{0.2cm}\includegraphics[width=7cm,angle=-90]{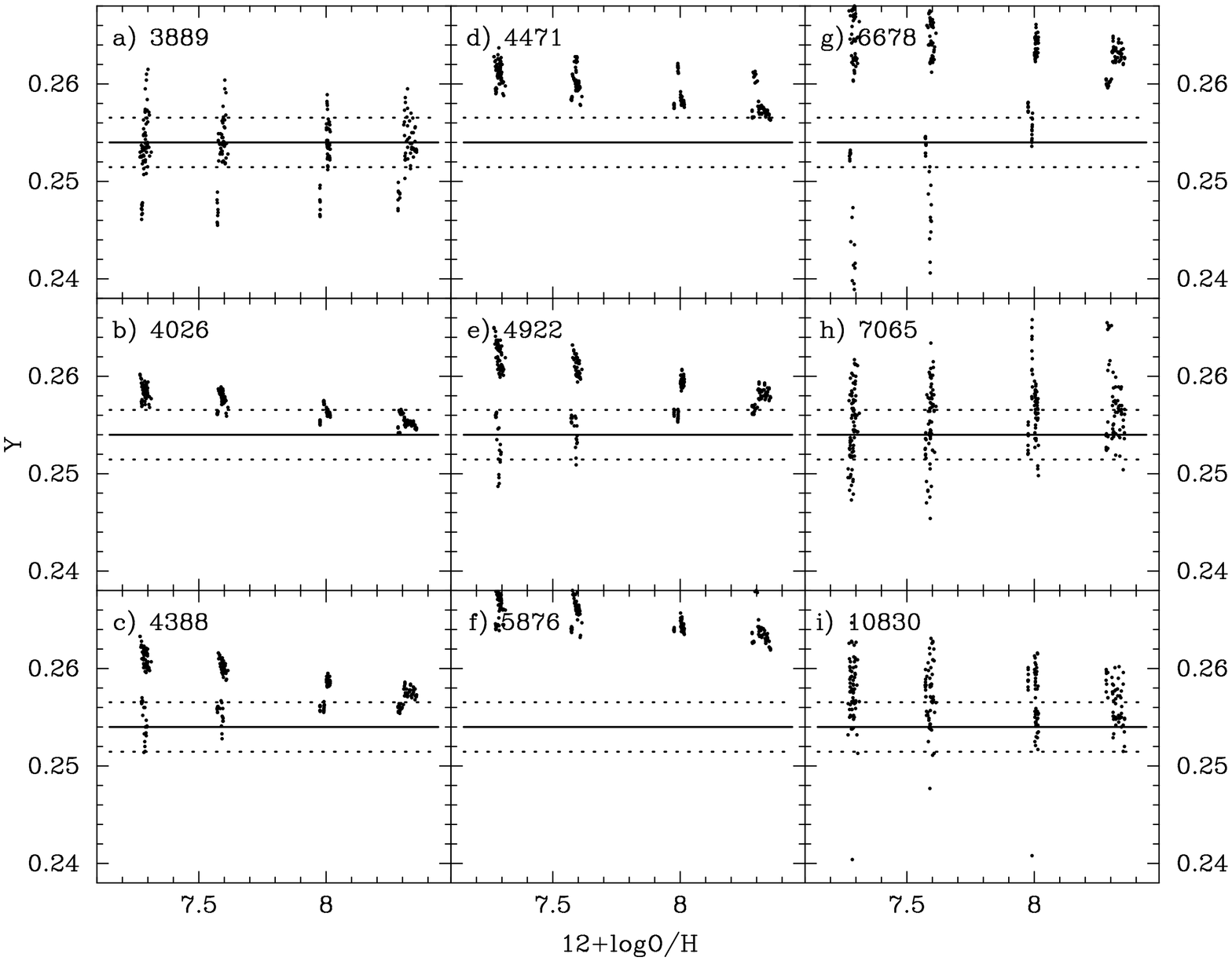}
}
      \caption{{\bf Left} Distribution with the oxygen abundance of the 
empirical $Y$ values calculated 
from the case B intensities for individual He~{\sc i} emission lines.
The best values of $Y$s in every model are derived
by varying $T_{\rm e}$(He$^+$) in the range 
(0.95 -- 1.05)$\times$ $\widetilde{T}_{\rm e}$(He$^+$).
The definition of solid and dotted lines, and $<$$Y$$>$ is the same as in 
Fig. \ref{fig5}. The mean value of $Y$ is indicated at the bottom of each 
plot. {\bf Right:} The same as in the left panel, but empirical $Y$ values 
are calculated from the intensities for individual He~{\sc i} emission lines
with all processes included.}
\label{fig6}
   \end{figure*}

In Fig. \ref{fig6} (left panel) we show the $^4$He mass fractions $Y$ 
empirically derived from the case B intensities of
individual He~{\sc i} emission lines. These $Y$ values
correspond to the same empirical models as those in Fig. \ref{fig5}.
Again, the empirically derived $Y$s (dots) broadly agree
with the input CLOUDY value (solid lines). 
The poorest agreement is for the He {\sc i} $\lambda$5876 and $\lambda$6678
emission lines in models with 12+logO/H = 7.3. 
But even in these cases the empirical $Y$ values
on average do not deviate from the CLOUDY input value by  more than 
1\%. We note that the most deviant points correspond to the models with
Gaussian density distribution. These differences are mainly due 
to the larger uncertainties in the analytical fits of the He {\sc i} 
emissivities at $N_e$ $\sim$ 10$^3$ cm$^{-3}$ 
(dotted lines in Fig. \ref{fig1}).
In any case, it is better to rely on several lines for an accurate derivation 
of $Y$, not only because of possible observational errors in line intensities,
but also because of uncertainties in emissivities, their analytical
fits, etc. For comparison, in Fig. \ref{fig6} (right panel) are shown 
the $^4$He mass fractions $Y$ empirically derived from the CLOUDY intensities 
of individual He~{\sc i} emission lines with all processes included. 
Note that in this case the correction for fluorescent excitation 
of He~{\sc i} emission lines is taken into account using the expressions given 
by \citet{B99,B02}. The
agreement between the CLOUDY input value of $Y$ and empirically derived
values is very poor, especially for He {\sc i} $\lambda$5876 and $\lambda$6678
emission lines. 
This stems from the fact that, as noted in Sect. \ref{problem}, the intensities
of the lines calculated by CLOUDY with all processes included are not 
compatible with what we understand of the emission theory of these lines.

   \begin{figure*}
   \centering
  \hbox{  \includegraphics[width=4cm,angle=-90]{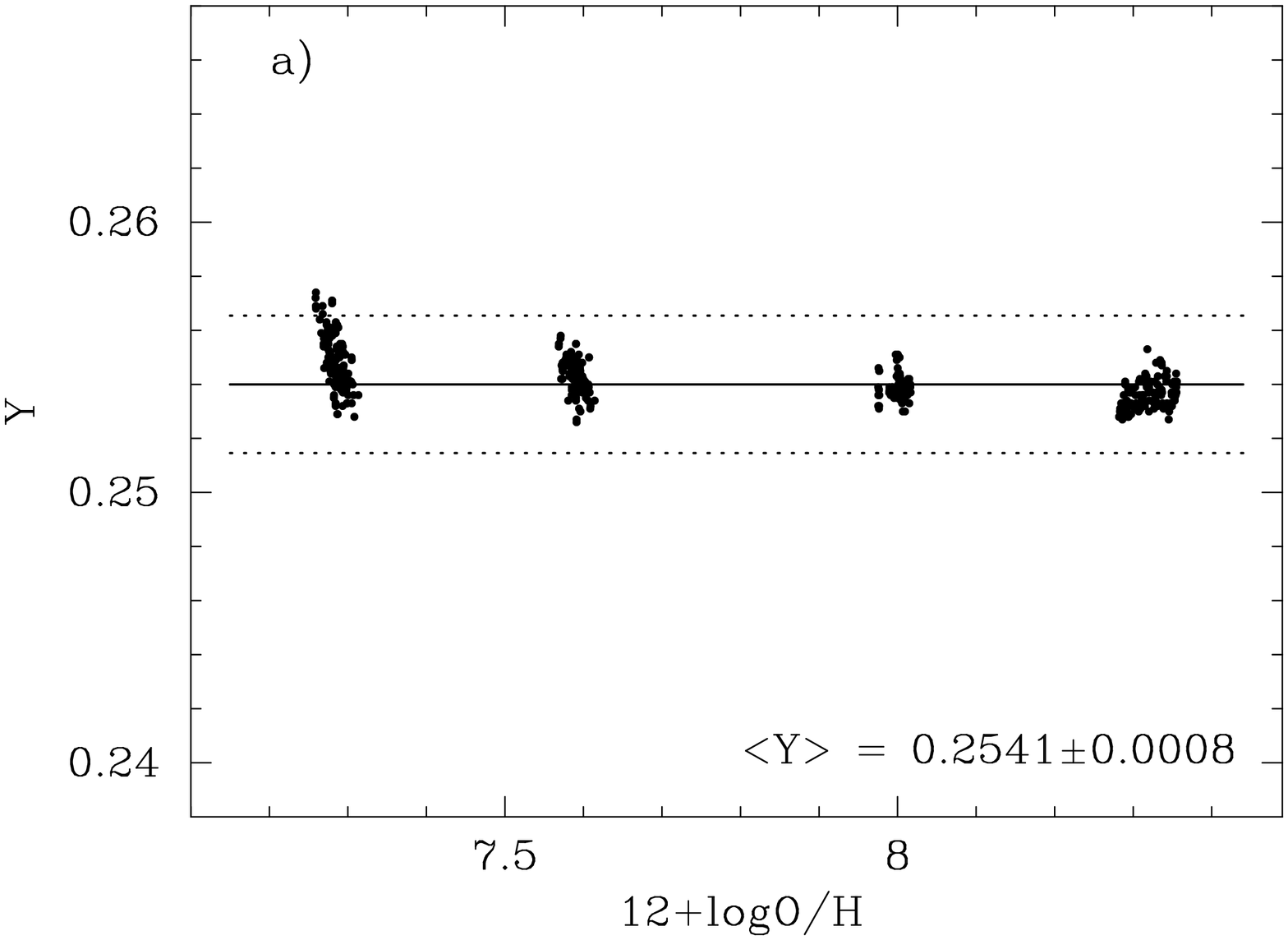}
   \hspace{0.2cm}\includegraphics[width=4cm,angle=-90]{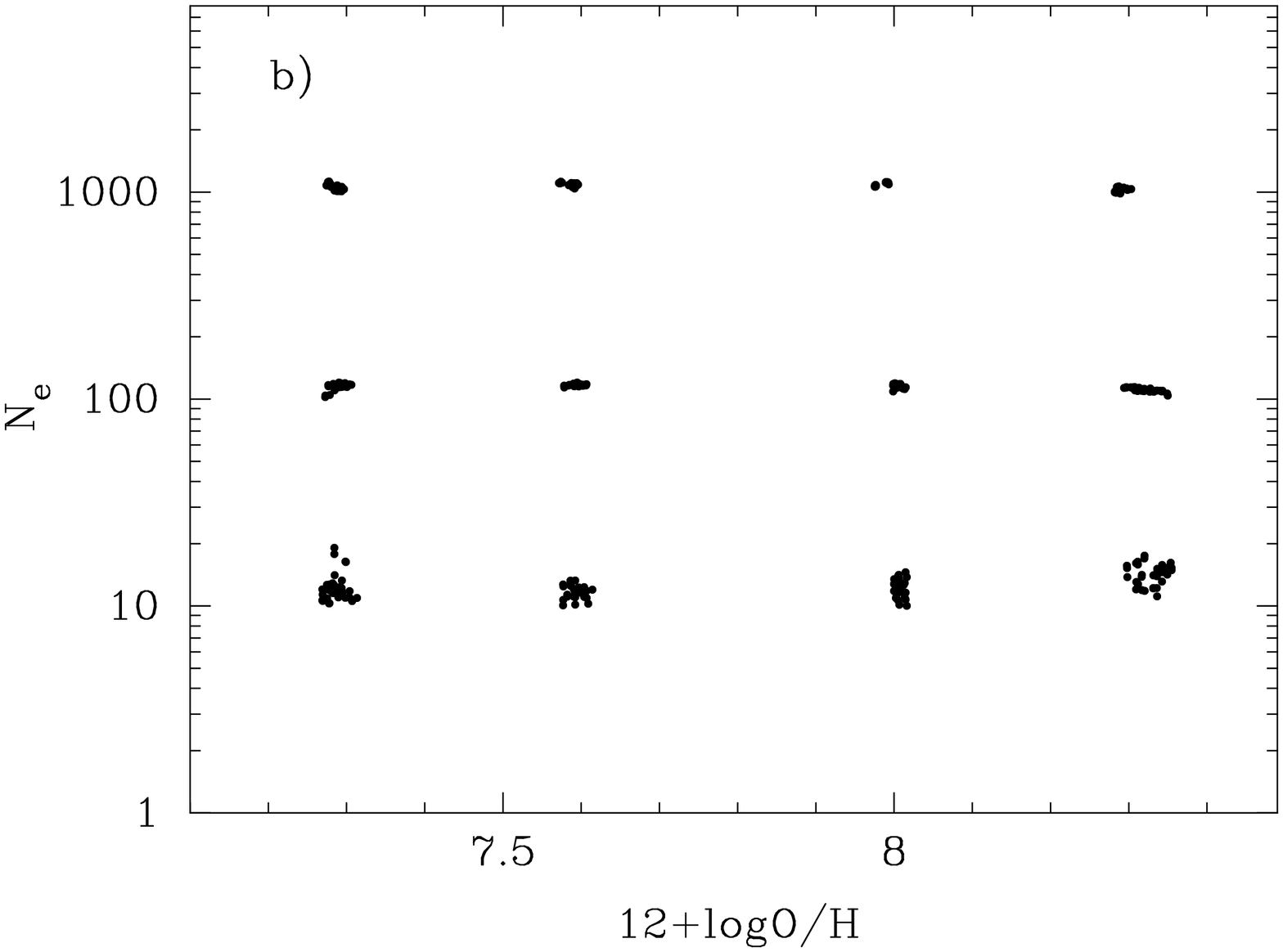}
   \hspace{0.2cm}\includegraphics[width=4cm,angle=-90]{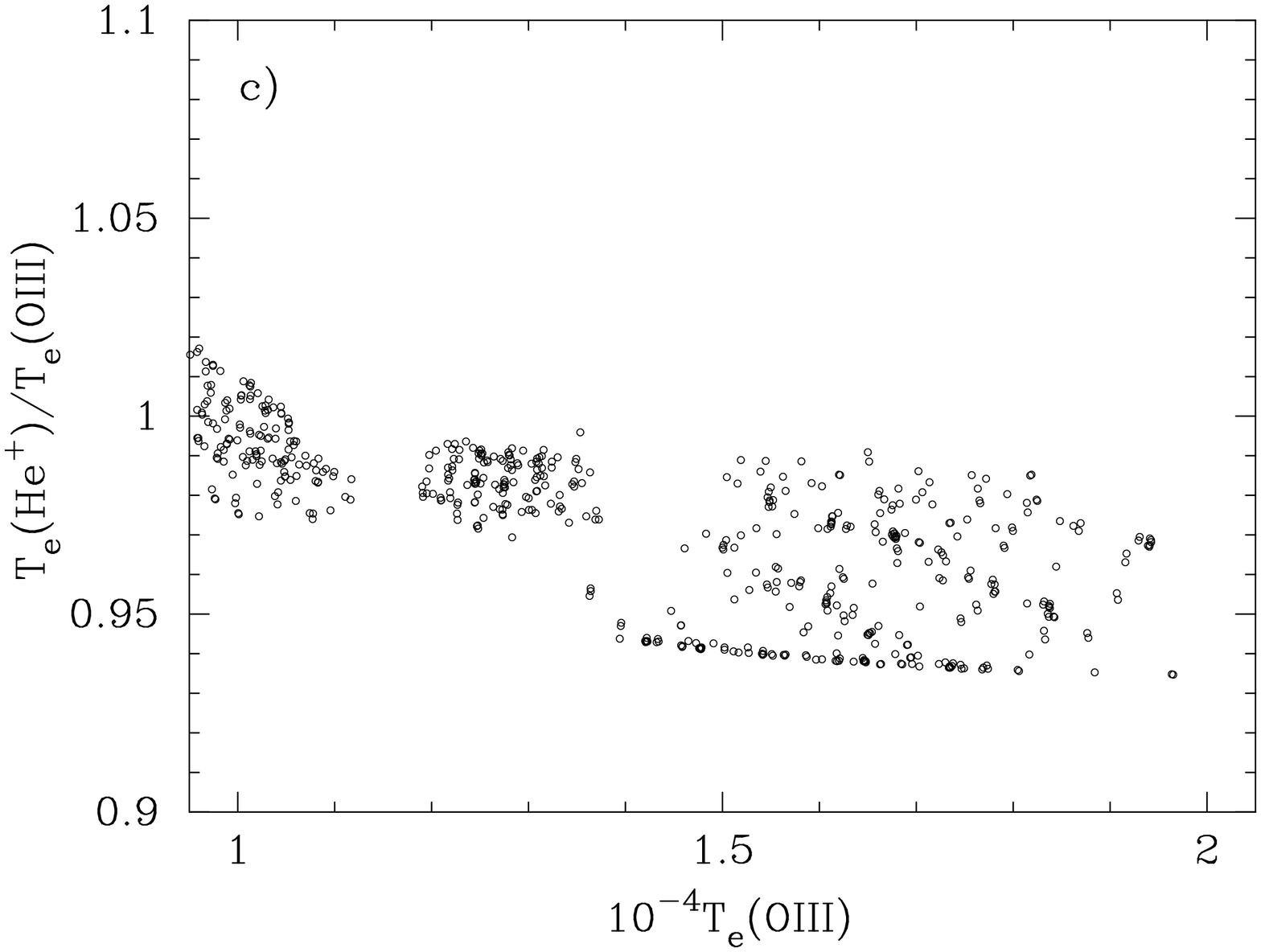}}
\vspace{0.2cm}
  \hbox{  \includegraphics[width=4cm,angle=-90]{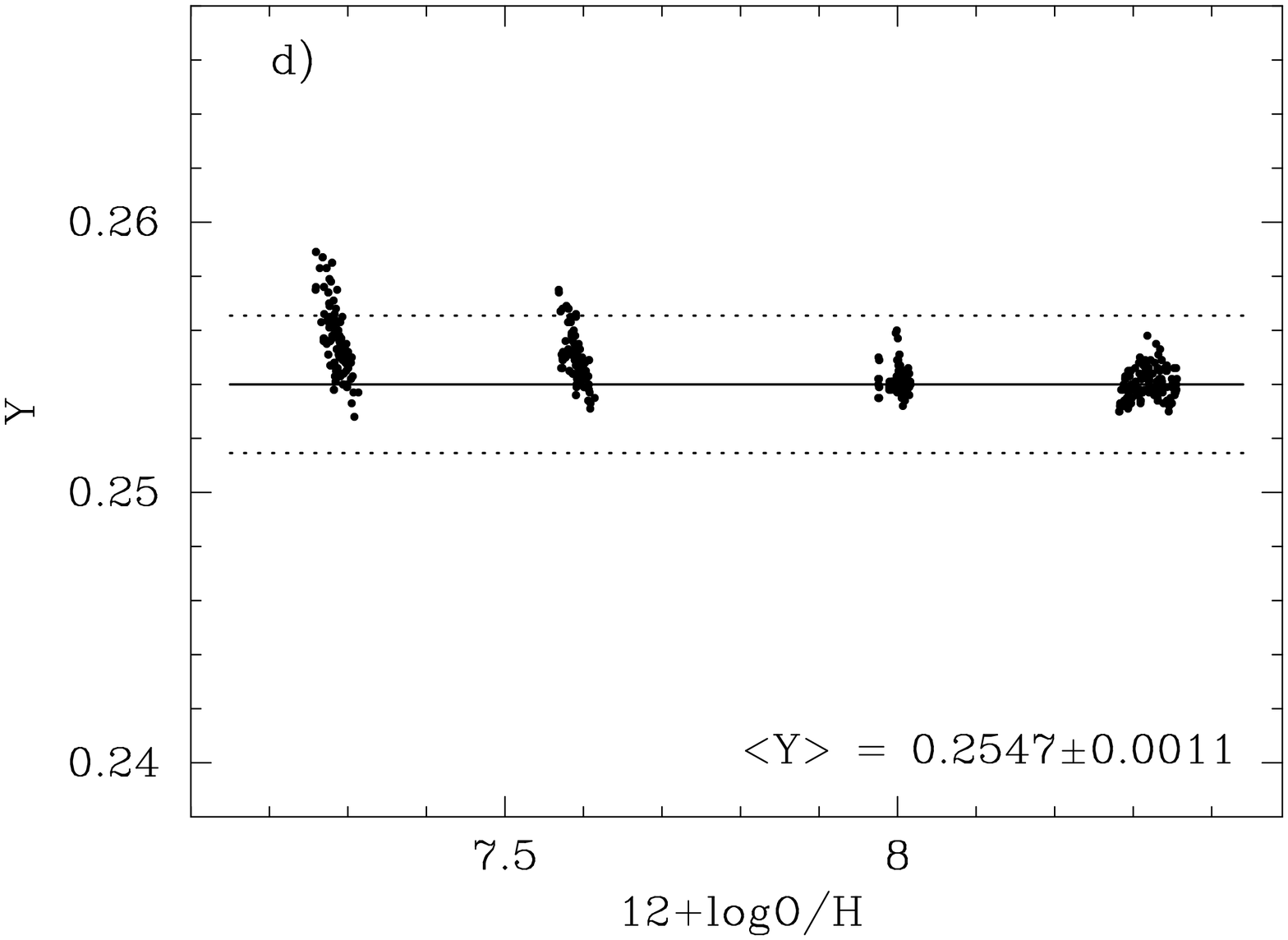}
   \hspace{0.2cm}\includegraphics[width=4cm,angle=-90]{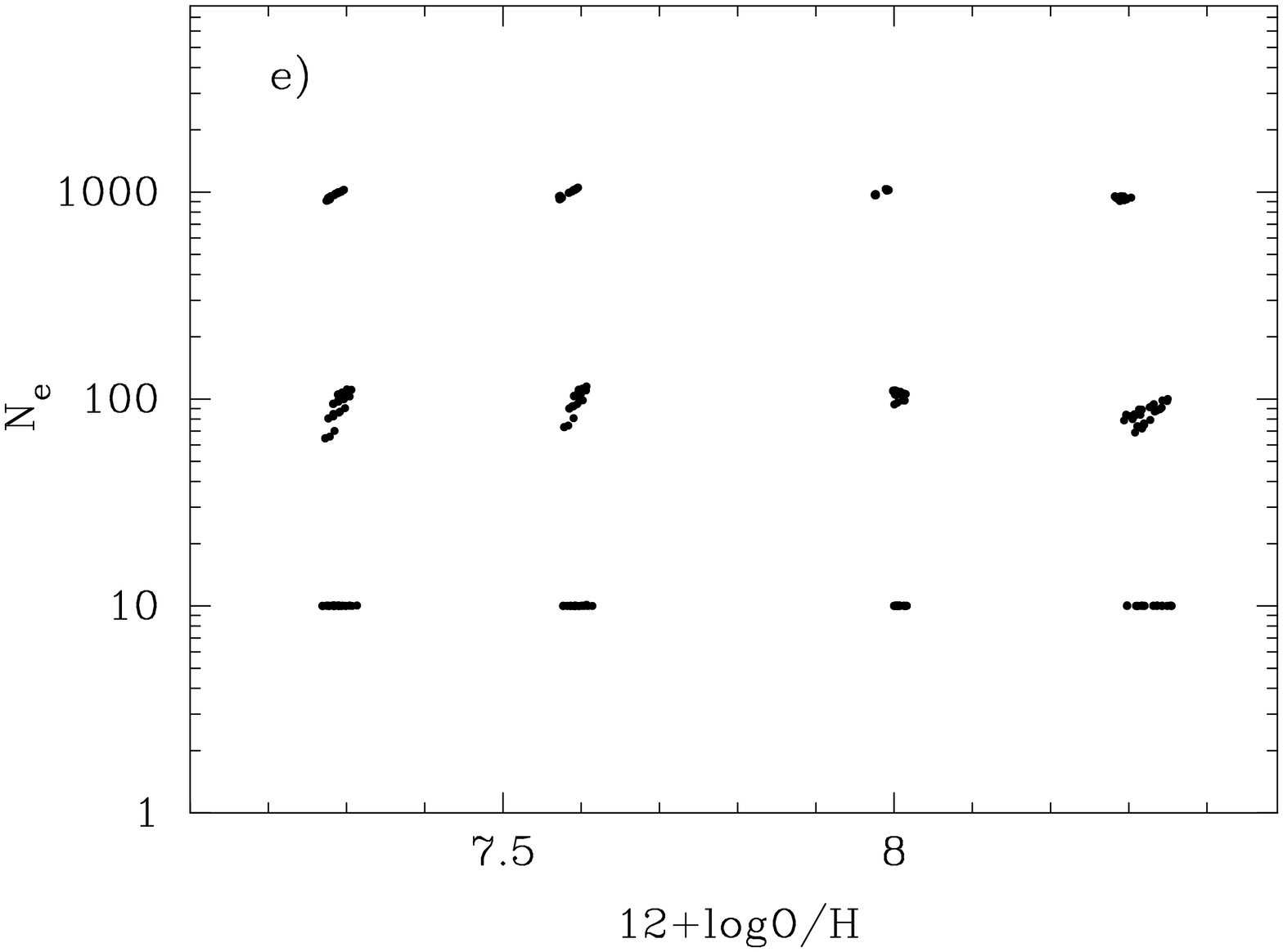}
   \hspace{0.2cm}\includegraphics[width=4cm,angle=-90]{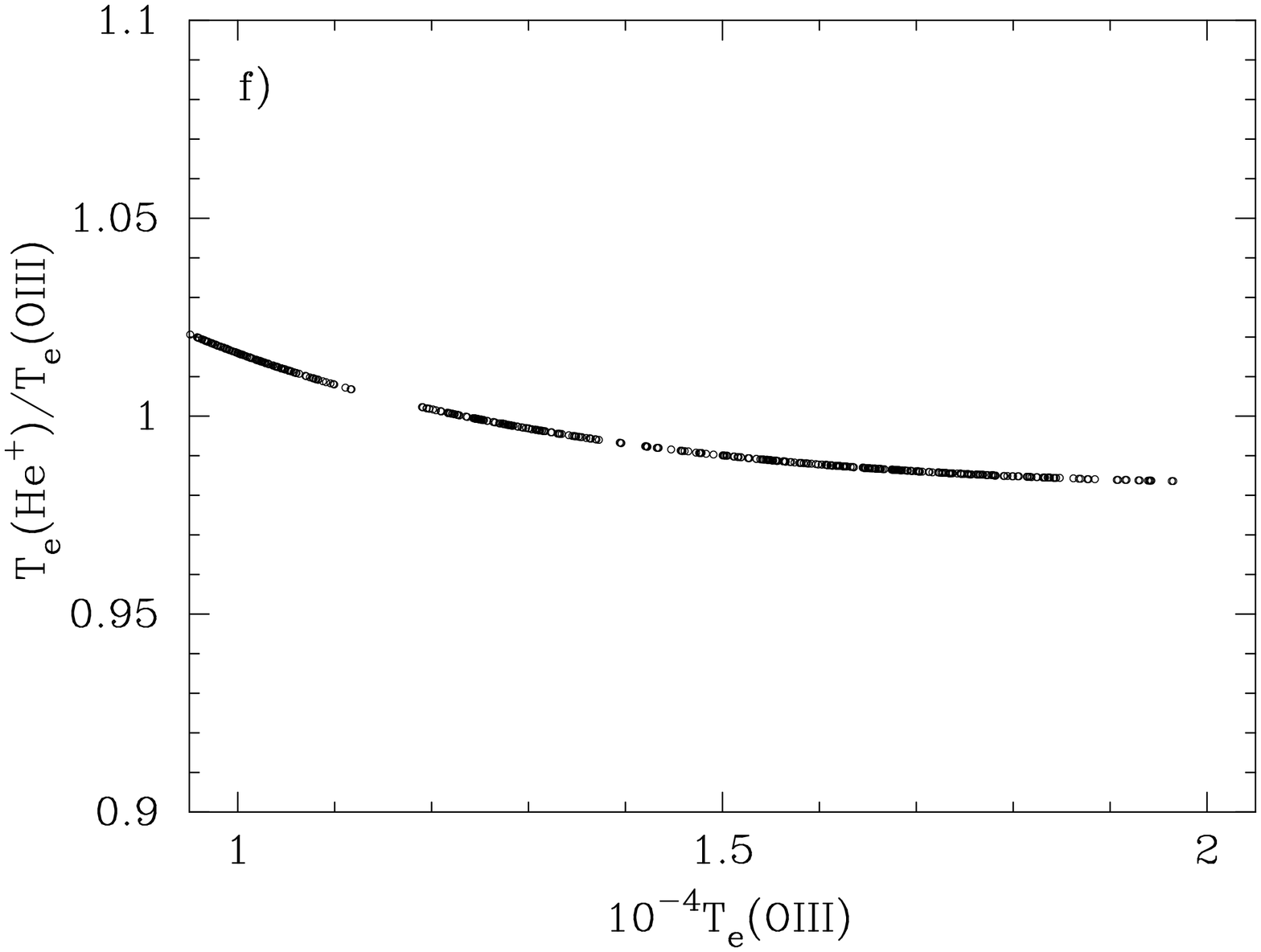}}
\vspace{0.2cm}
  \hbox{  \includegraphics[width=4cm,angle=-90]{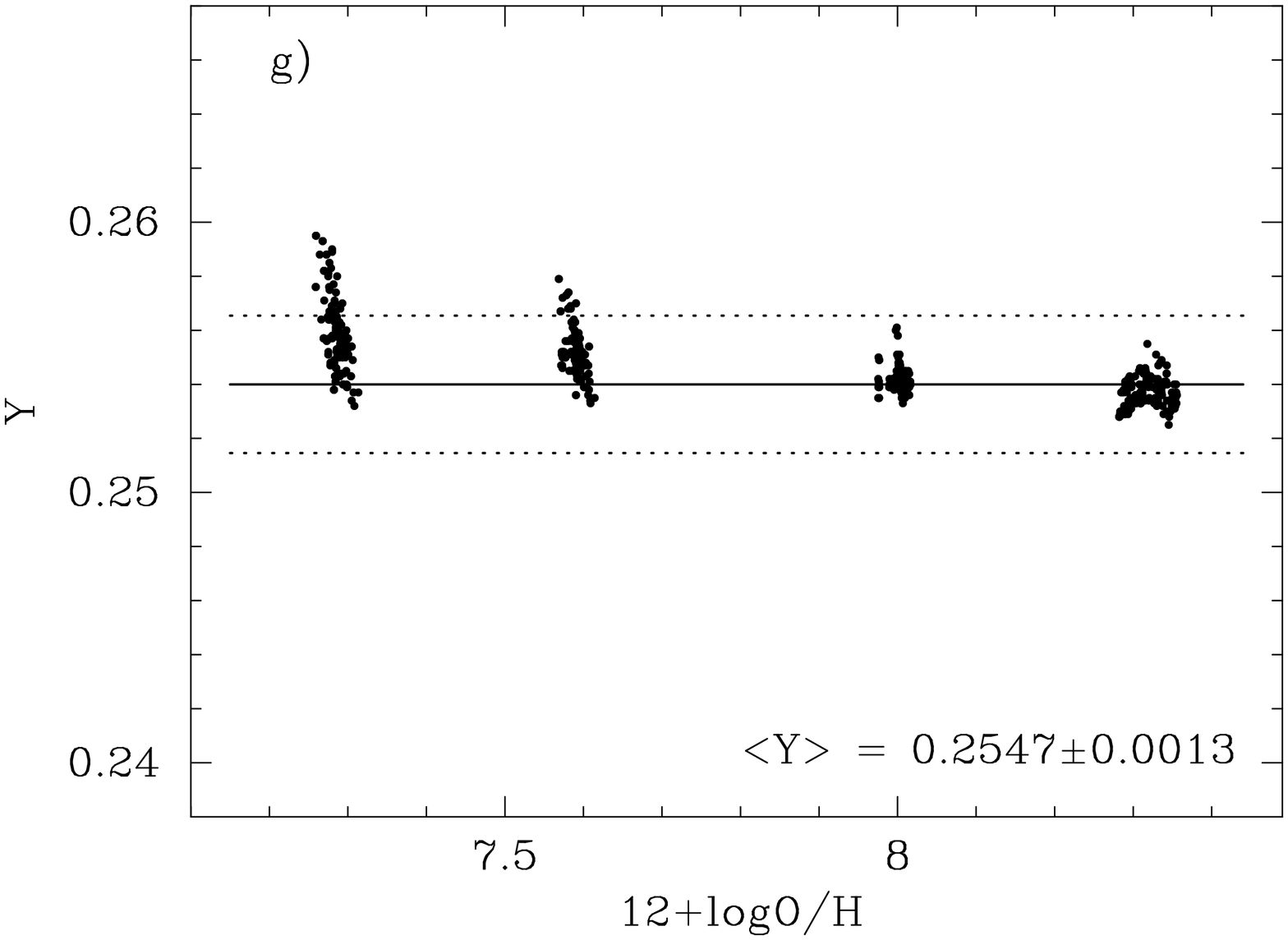}
   \hspace{0.2cm}\includegraphics[width=4cm,angle=-90]{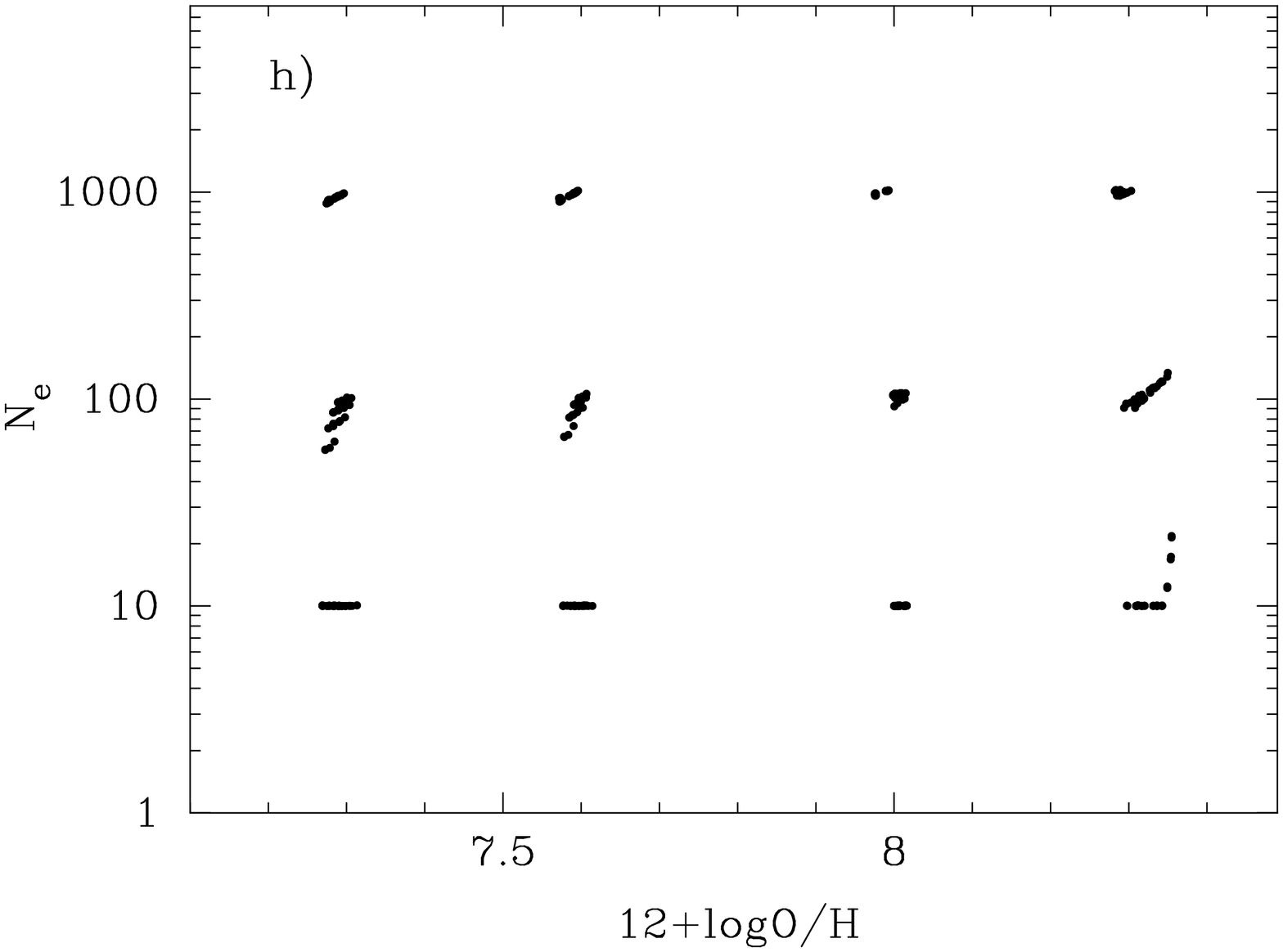}
   \hspace{0.2cm}\includegraphics[width=4cm,angle=-90]{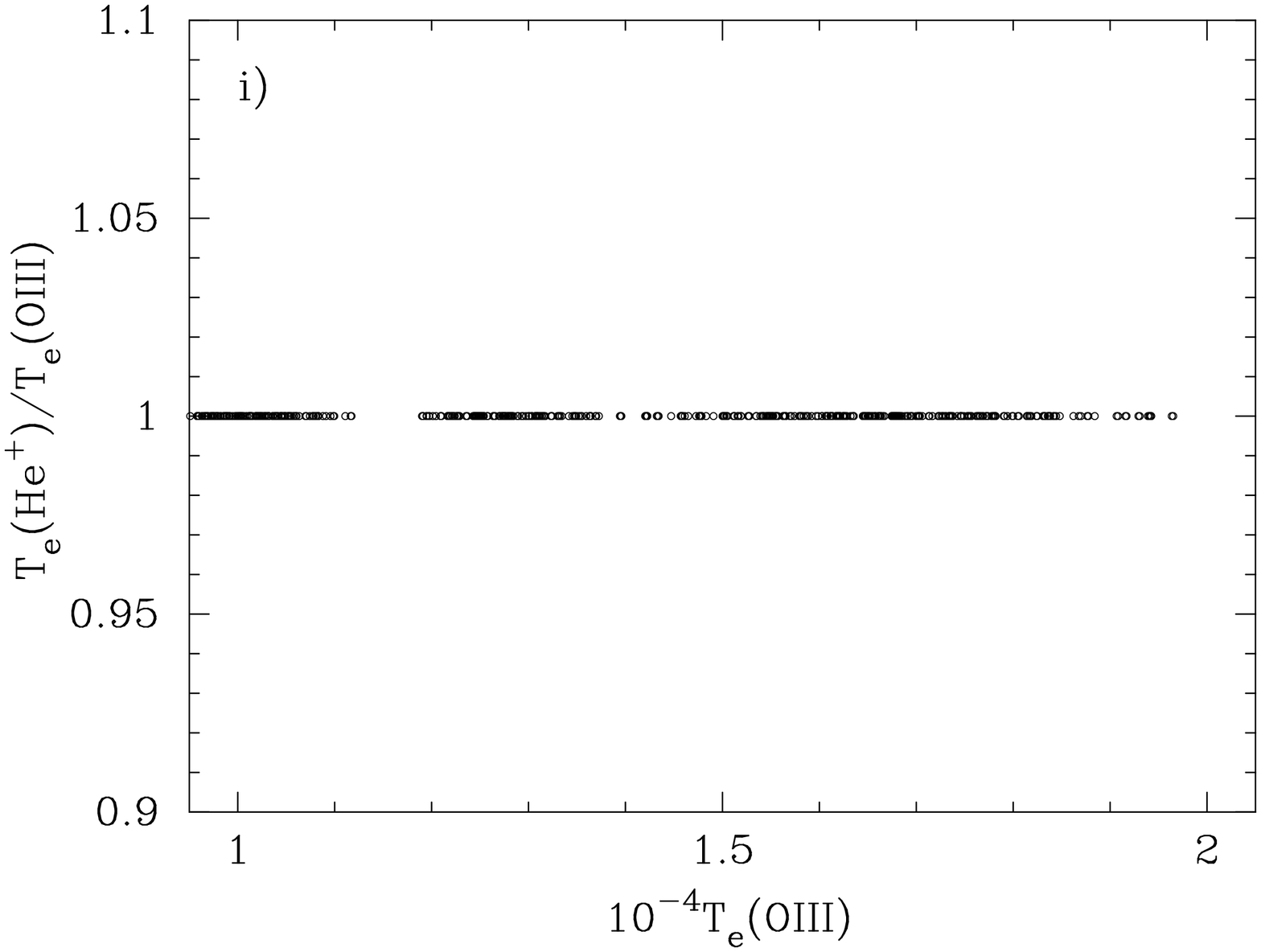}}
      \caption{Same as in Fig. \ref{fig5}, but the five He~{\sc i}
$\lambda$3889, $\lambda$4471, $\lambda$5876, $\lambda$6678, 
and $\lambda$7065 emission lines are used for $\chi^2$ minimisation and 
determination of $Y$.}
         \label{fig7}
   \end{figure*}

The above analysis is based on empirical $Y$ values, derived from the
nine He~{\sc i} emission lines. However, in practice, a smaller number of
He~{\sc i} emission lines is used for determining the $^4$He abundance.
This is because He~{\sc i} emission lines $\lambda$4026, $\lambda$4388,
$\lambda$4922 are several times weaker than the remaining He~{\sc i} emission
lines and therefore their intensities are measured with lower accuracy. Furthermore,
these emission lines are more subject to the uncertainties in correcting for 
underlying stellar He~{\sc i} absorption lines because of their low equivalent 
widths. Additionally, observations of the 
strongest near-infrared He~{\sc i} $\lambda$10830 emission line need 
facilities different from those used for the visible range. Therefore,
this line is rarely observed in the low-metallicity emission-line galaxies.
In particular, \citet{ITL94,ITL97}, \citet{IT98,IT04,IT10}, and
\citet{I07} used only the five He~{\sc i} emission lines
$\lambda$3889, $\lambda$4471, $\lambda$5876, $\lambda$6678, and $\lambda$7065
in the visible range for $\chi^2$ minimisation in Eq. \ref{eq1} 
(i.e. $n$ = 5).

In Figs. \ref{fig7}a, \ref{fig7}d, and \ref{fig7}g 
we compare the empirically derived $Y$s
for three choices of $T_{\rm e}$(He$^+$) variations 
with the input CLOUDY $Y$ value in the case when the
five He~{\sc i} emission lines
$\lambda$3889, $\lambda$4471, $\lambda$5876, $\lambda$6678, and $\lambda$7065
are used for $\chi^2$ minimisation in Eq. \ref{eq1} 
and for determining the weighted mean value of $Y$.
However, at variance with the case of the models with nine lines, we adopted
weights for He~{\sc i} lines proportional to their intensity. This makes our 
comparison closer to real observations when stronger lines
can be measured with better accuracy and therefore should be taken
with heavier weights. Similarly to Fig. \ref{fig5}, we also compared 
the derived values of $N_{\rm e}$ and $T_{\rm e}$(He$^+$).

It is seen from Fig. \ref{fig7}c that the empirically derived
electron temperature $T_{\rm e}$(He$^+$) has the same trend as the value
calculated with CLOUDY (compare with Fig. \ref{fig7}f). 
Similarly, the empirically derived electron number density reproduces 
the CLOUDY input value fairly well (Fig. \ref{fig7}b). As a consequence,
the empirical $Y$s are derived with accuracy similar to that in the case
of nine lines. The accuracy of the $Y$ determination is not as good when
the electron temperature $T_{\rm e}$(He$^+$) is not varied, but is derived
from Eq. \ref{tHeOIII} (middle panel of Fig. \ref{fig7}) or is
equal to $T_{\rm e}$(O~{\sc iii}) (bottom panel of Fig. \ref{fig7}). 
$Y$s in models with 12+log O/H = 7.3 and 7.6 are overestimated and show 
a high dispersion.

The above analysis indicates that five He~{\sc i} emission lines 
are quite enough to reproduce the input CLOUDY $Y$ value with the desired 
accuracy; additional He~{\sc i} emission lines are needed if one wishes to 
reduce the dispersion of $Y$s at low oxygen abundances and to reduce
trends seen in Figs. \ref{fig7}a, \ref{fig7}d, and \ref{fig7}g.
Furthermore, we note that the most problematic emission line in the
He abundance determination is the He {\sc i} $\lambda$3889 line. 
This is because this line is blended with the hydrogen H8 $\lambda$3889
emission line and its intensity is subject to uncertainties of the
subtraction of hydrogen emission and correction for underlying absorption
not only of He {\sc i} line, but also of H8 line. Therefore, 
the use of more He {\sc i}
emission lines would reduce the effect of these uncertainties.
The most promising line for that is the 
strongest line, He~{\sc i} $\lambda$10830. This line is subject to a large
extent to collisional excitation and therefore its intensity is sensitive 
to the electron number density $N_{\rm e}$ and electron temperature,
and consequently, including it would better constrain both 
$T_{\rm e}$ and $N_{\rm e}$.

\begin{figure*}
   \centering
  \hbox{  \includegraphics[width=4cm,angle=-90]{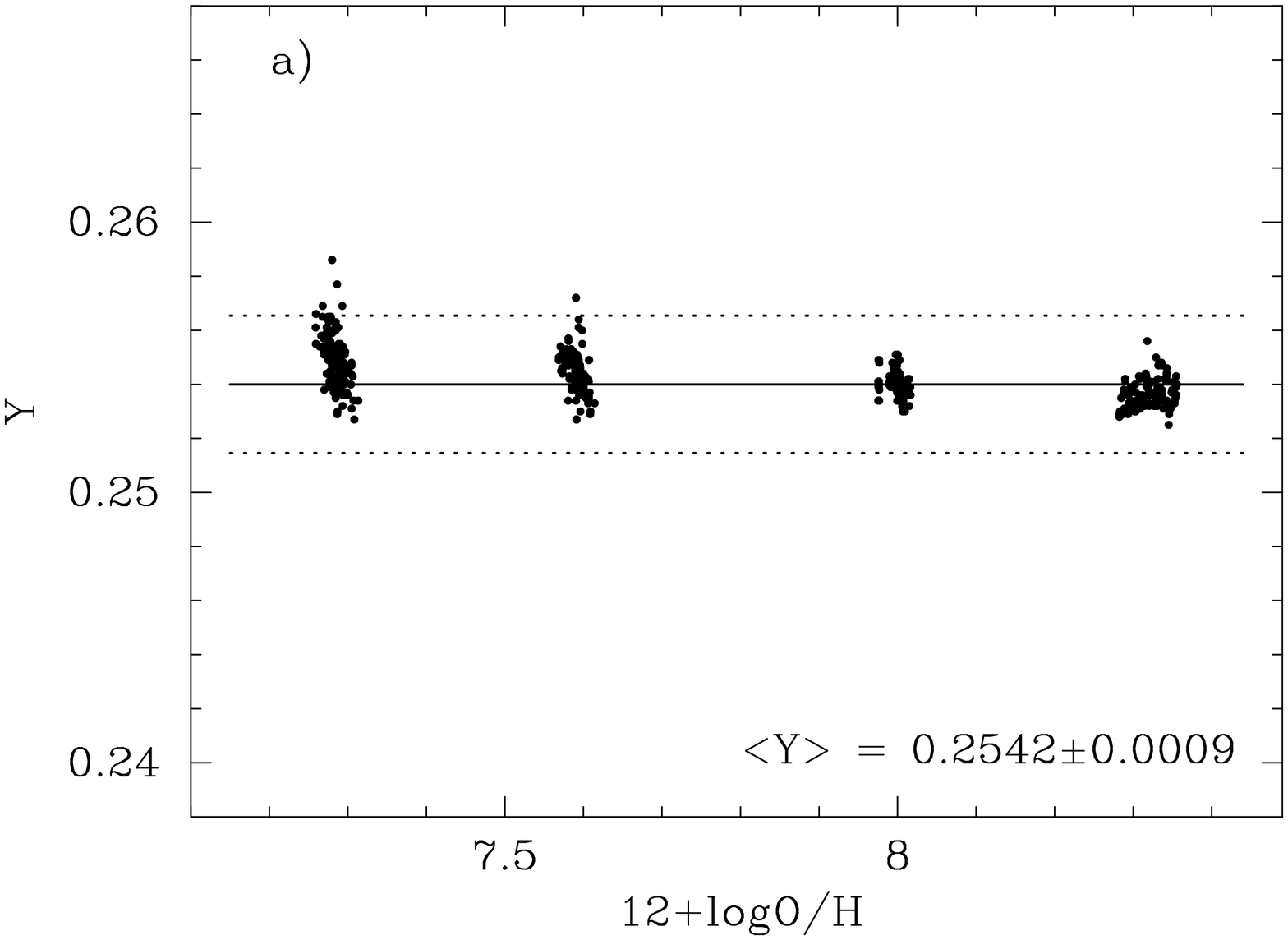}
   \hspace{0.2cm}\includegraphics[width=4cm,angle=-90]{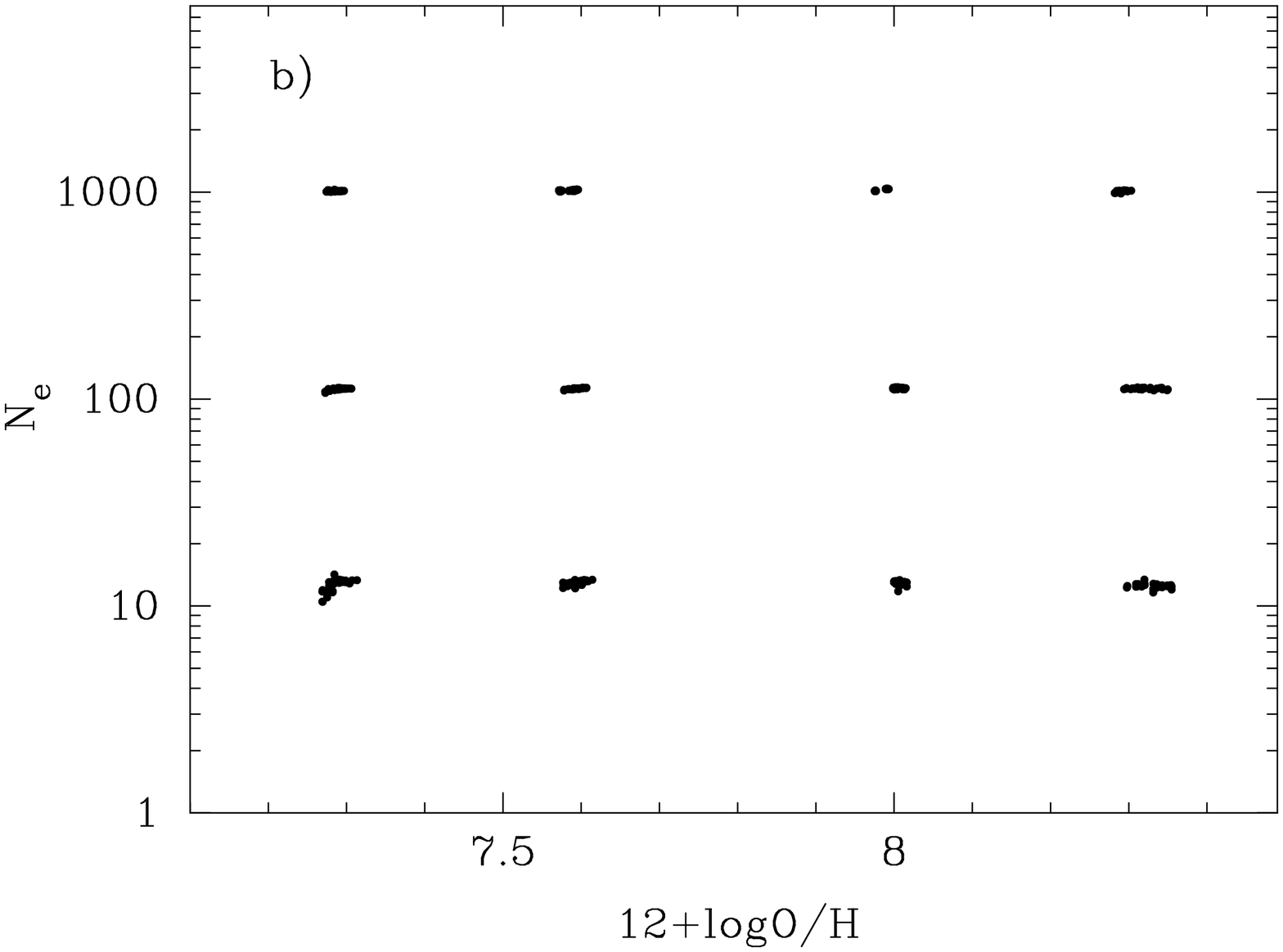}
   \hspace{0.2cm}\includegraphics[width=4cm,angle=-90]{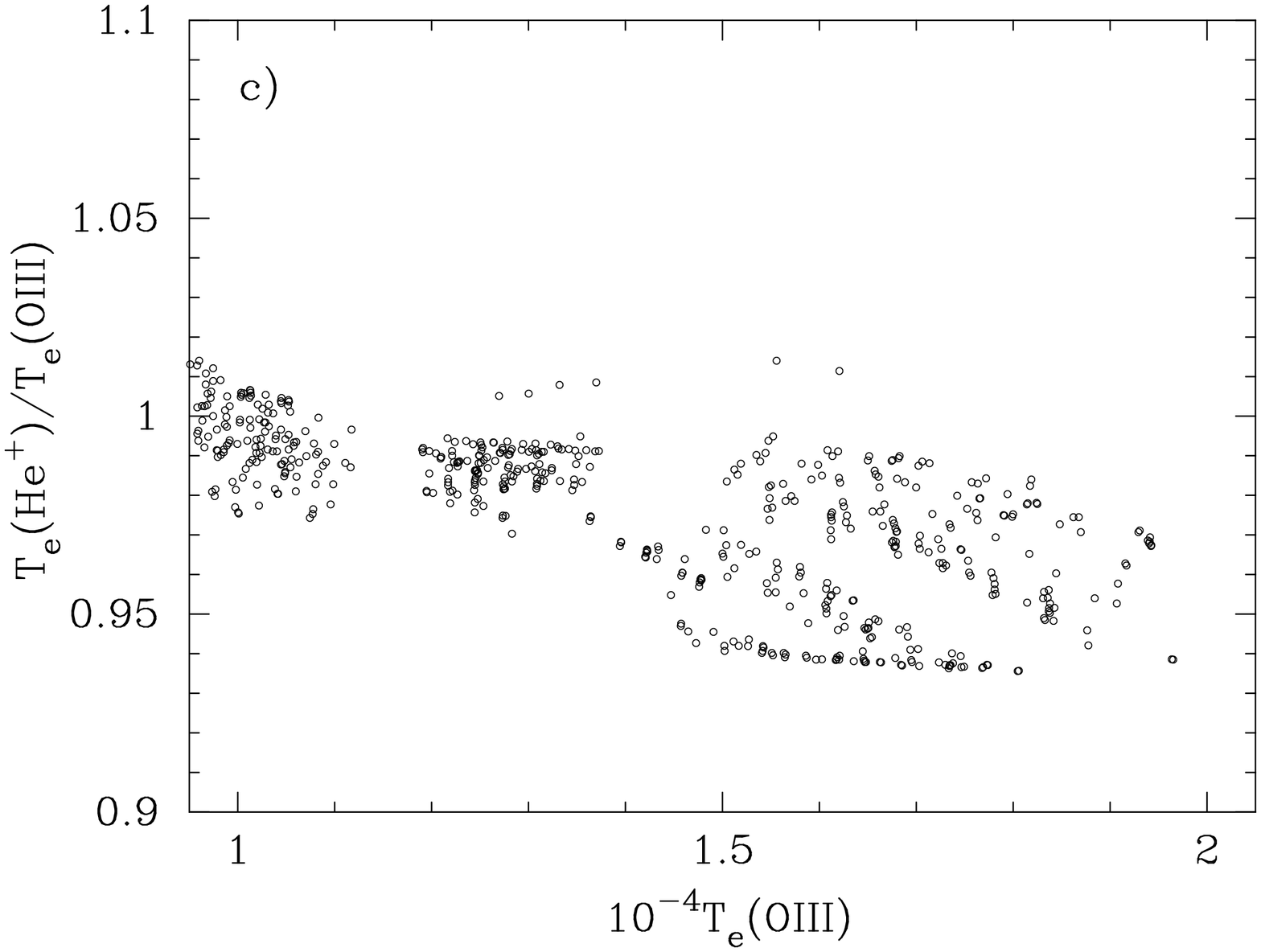}}
\vspace{0.2cm}
  \hbox{  \includegraphics[width=4cm,angle=-90]{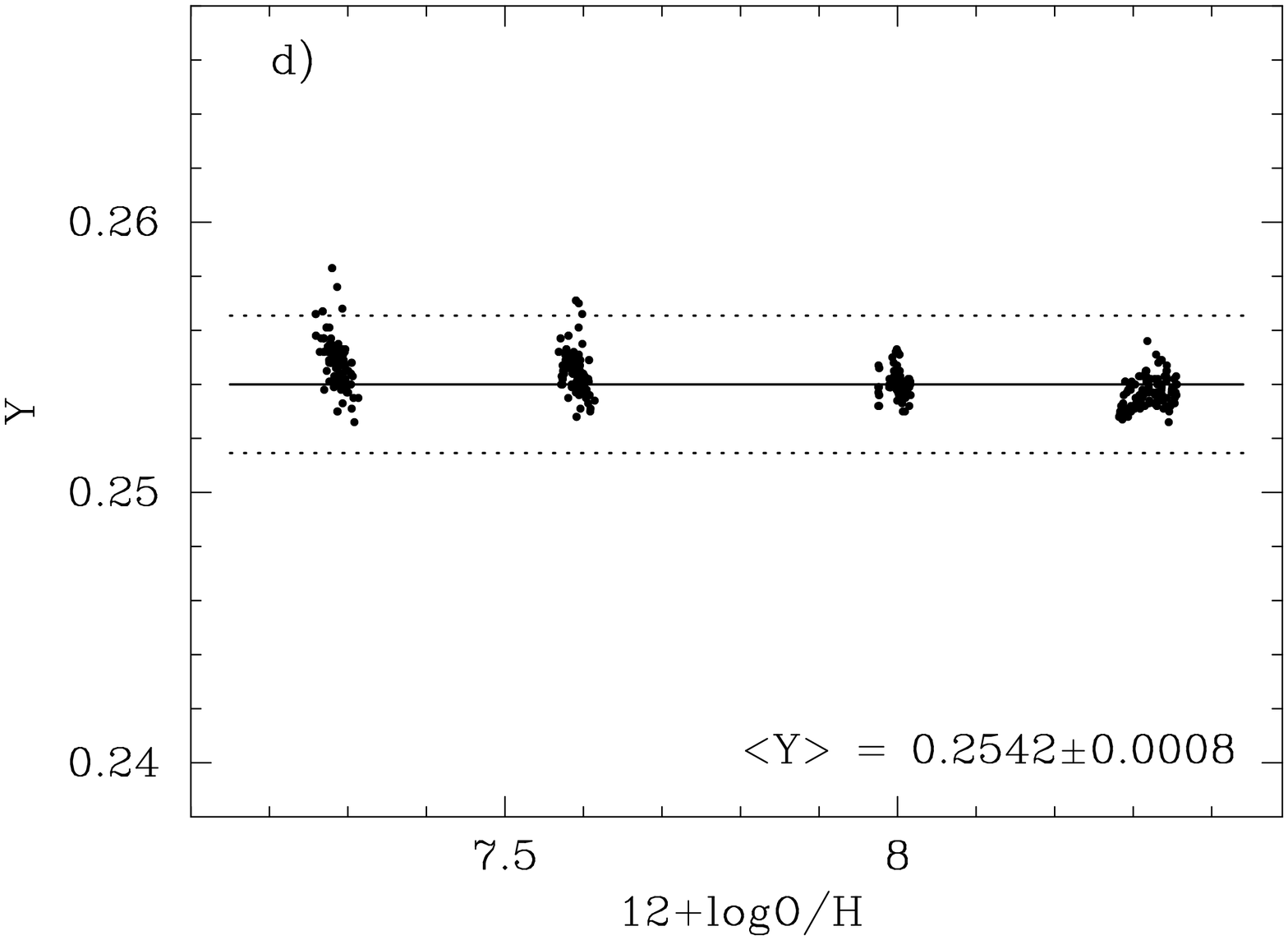}
   \hspace{0.2cm}\includegraphics[width=4cm,angle=-90]{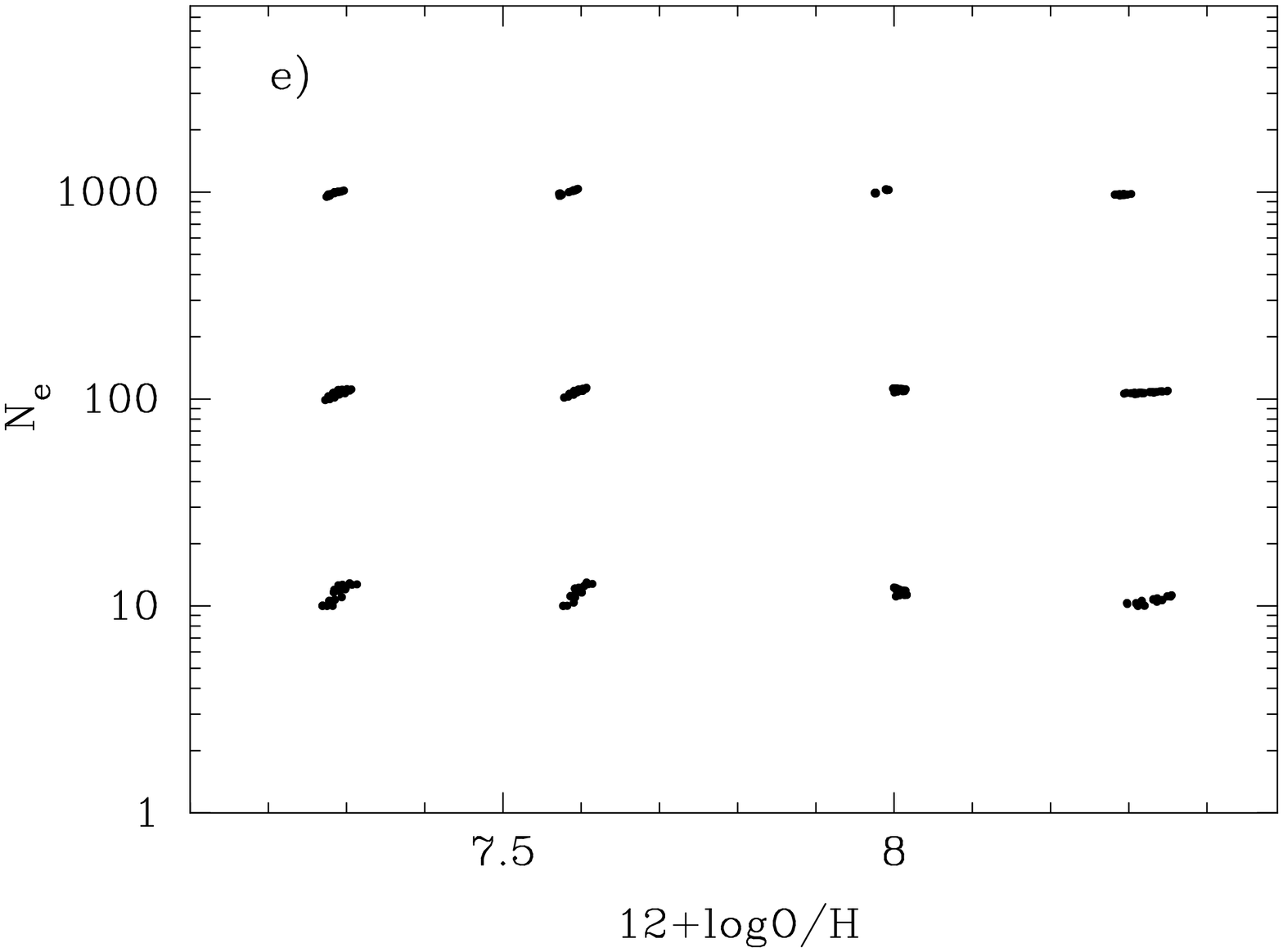}
   \hspace{0.2cm}\includegraphics[width=4cm,angle=-90]{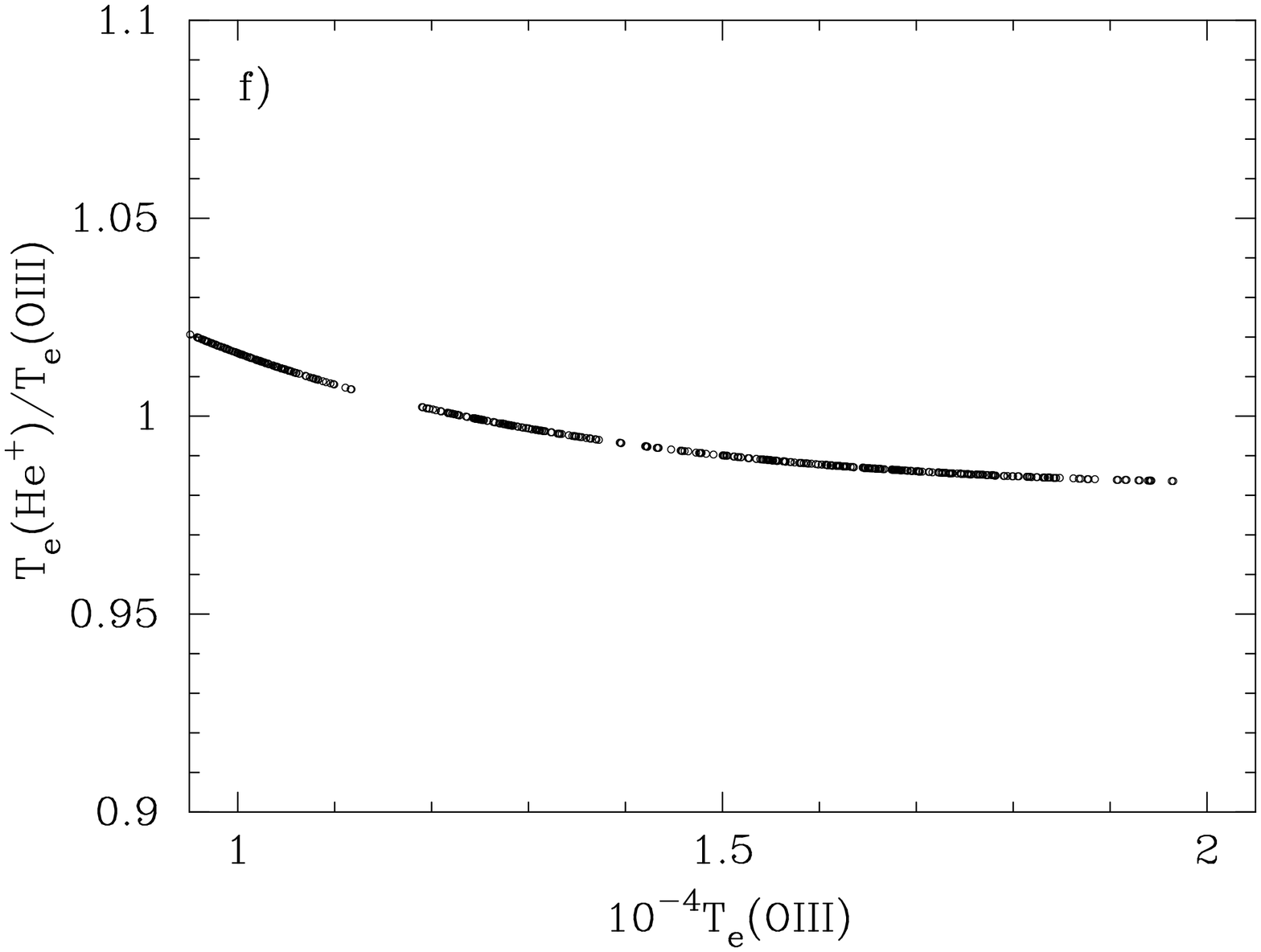}}
\vspace{0.2cm}
  \hbox{  \includegraphics[width=4cm,angle=-90]{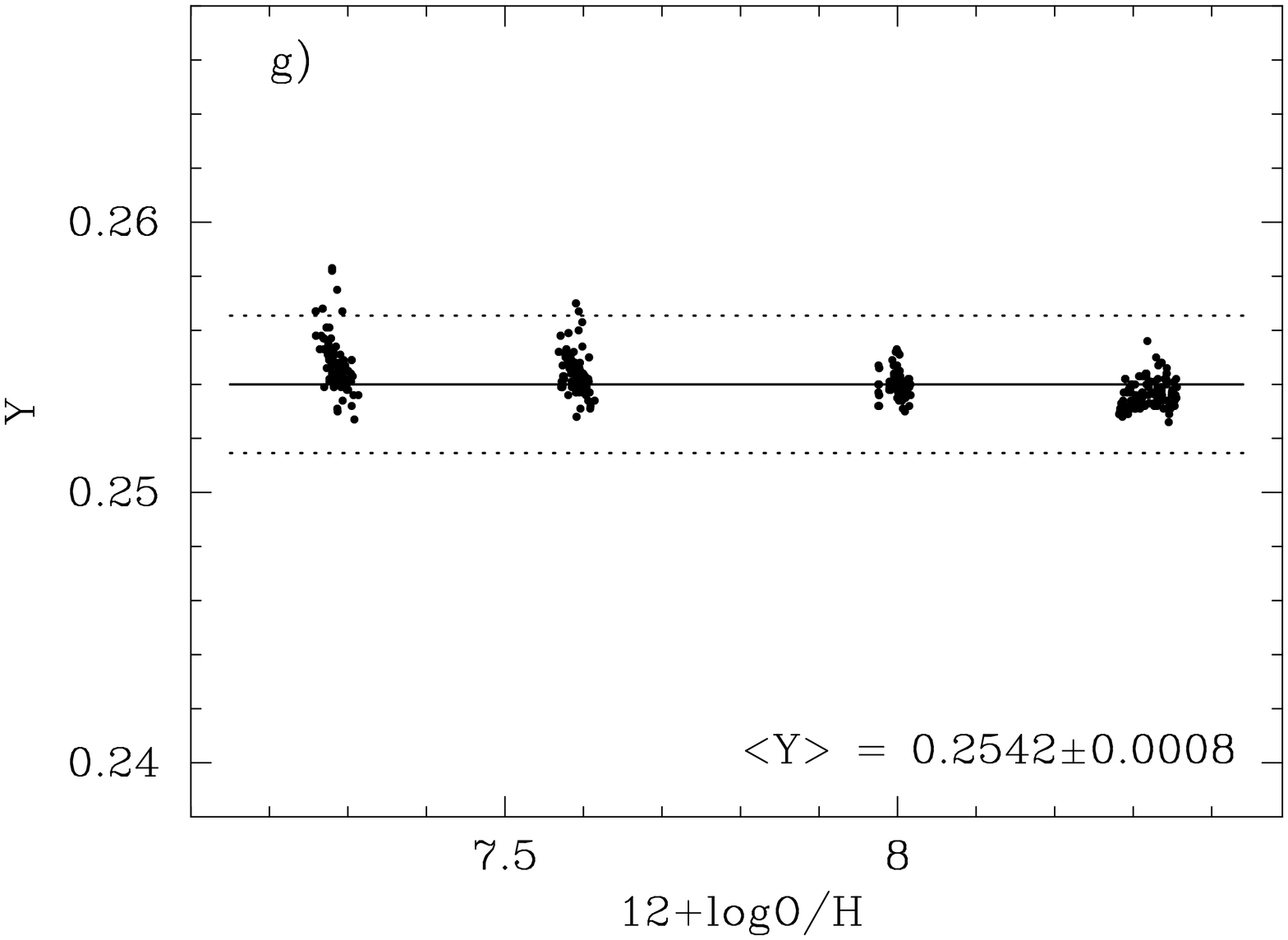}
   \hspace{0.2cm}\includegraphics[width=4cm,angle=-90]{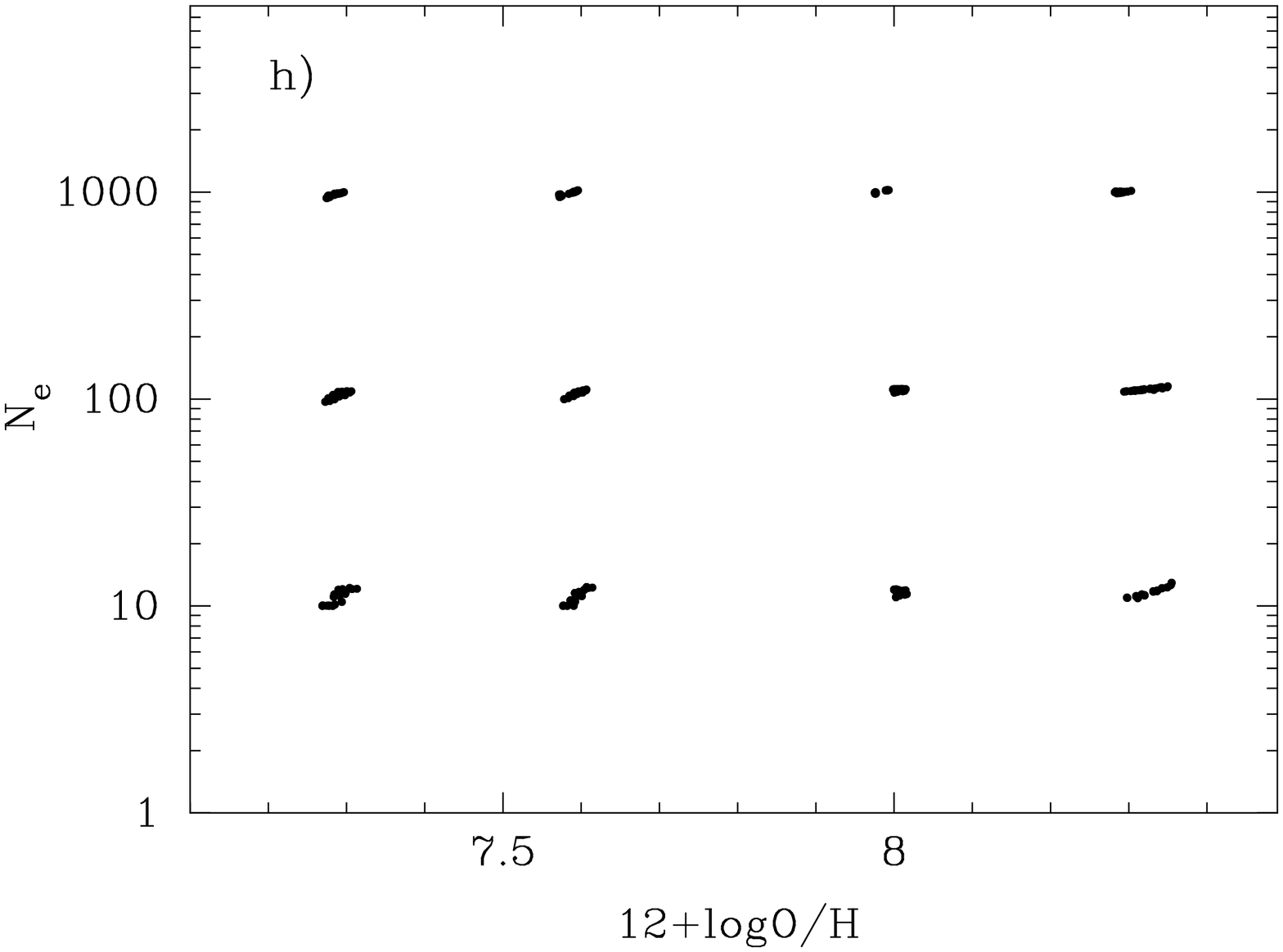}
   \hspace{0.2cm}\includegraphics[width=4cm,angle=-90]{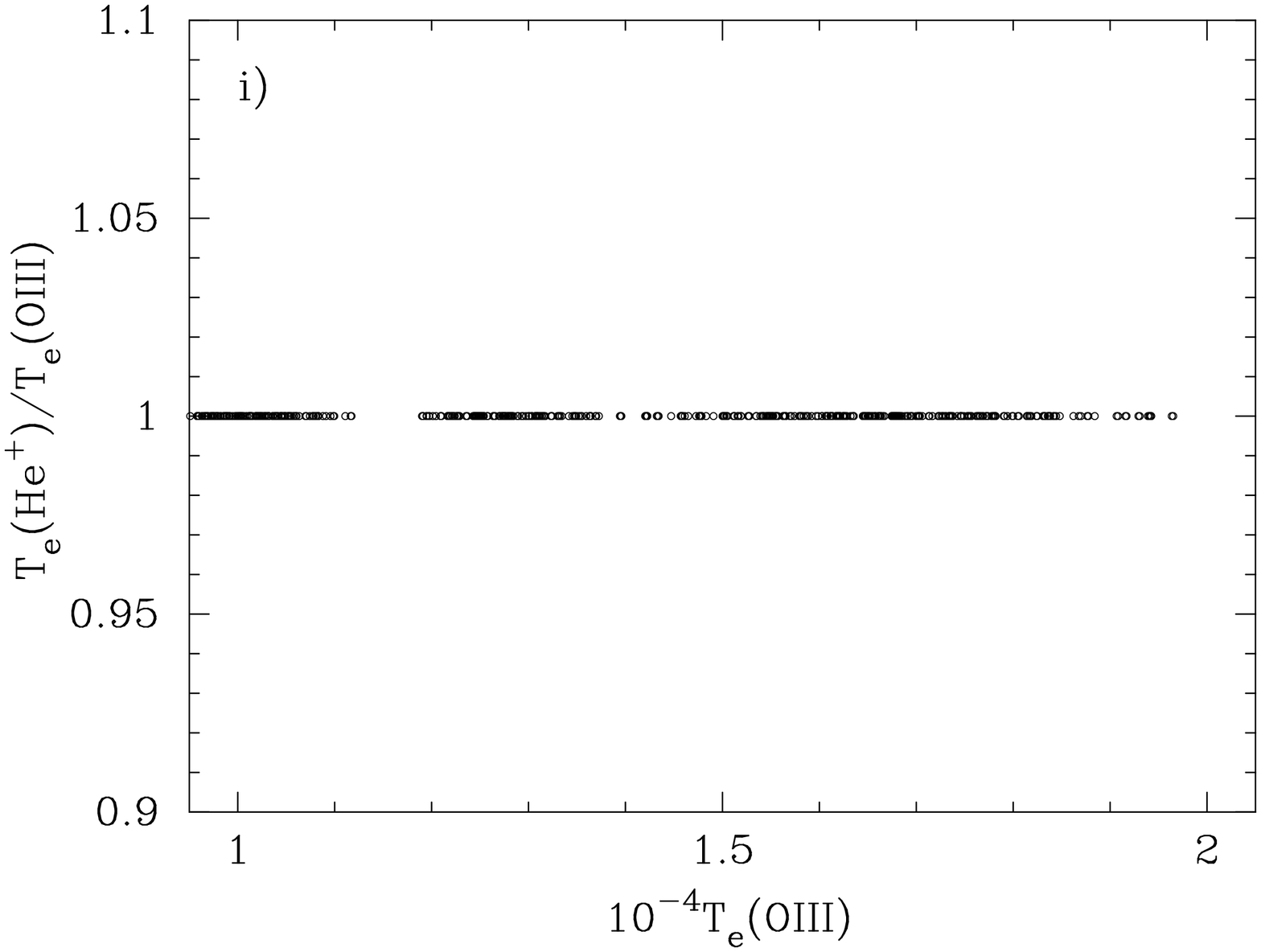}}
      \caption{Same as in Fig. \ref{fig5} but six He~{\sc i}
$\lambda$3889, $\lambda$4471, $\lambda$5876, $\lambda$6678, 
$\lambda$7065, and $\lambda$10830 emission lines 
are used for the $\chi^2$ minimisation and the determination of $Y$.}
         \label{fig8}
   \end{figure*}

In Fig. \ref{fig8}a, \ref{fig8}d, and \ref{fig8}g we show the empirical $Y$s 
derived by minimising $\chi^2$ with the use of the six He~{\sc i}
$\lambda$3889, $\lambda$4471, $\lambda$5876, $\lambda$6678, $\lambda$7065,
and $\lambda$10830 emission lines. All these lines were used to
determine the weighted mean $Y$ with weights proportional to
their intensities. The empirical $Y$ values very well reproduce
the input CLOUDY value, similar to the case of nine emission lines.
The electron number density $N_{\rm e}$ is also reproduced very well 
(Figs. \ref{fig8}b, \ref{fig8}e, and \ref{fig8}h). 
The empirically derived $T_{\rm e}$(He$^+$) (Fig. \ref{fig8}c)
follows the trend with $T_{\rm e}$(O~{\sc iii}), similarly to that 
in the case of nine He~{\sc i} emission lines (Fig. \ref{fig5}c).
Summarising, we conclude that our empirical technique satisfactorily 
reproduces the input CLOUDY $Y$ values and can be used for the $^4$He abundance
determination in real H~{\sc ii} regions. 

\section{Determining the $^4$He abundance in real H~{\sc ii} regions}\label{real}

\subsection{Method for determining the $^4$He abundance in
real H~{\sc ii} regions}\label{ydet}

While testing our empirical method on CLOUDY models calculated with the
most recent v13.01 code, we took into account the 
new set of He~{\sc i} emissivities by \citet{P13},
the collisional excitation of He~{\sc i} emission lines,
the non-recombination contribution to hydrogen emission-line intensities, and 
the correction for the ionisation structure of the H~{\sc ii} region.

To apply our empirical method to real H~{\sc ii} regions several other
effects should be taken into account. First, the Balmer decrement corrected for
the non-recombination contribution was used to simultaneously determine
the dust extinction and equivalent widths of underlying stellar 
hydrogen absorption lines, as described for example by \citet{ITL94}, and to 
correct line intensities for both effects. Second, since the
spectra of the extragalactic H~{\sc ii} regions include both the ionised gas 
and the stellar emission, the underlying
stellar He~{\sc i} absorption lines should be taken into account
\citep[see e.g. ][]{I07}. Third, the He~{\sc i} emission lines
should be corrected for fluorescent excitation that is parametrised 
by the optical depth $\tau$($\lambda$3889) of the He~{\sc i} $\lambda$3889
emission line. We used the correction factors for fluorescent excitation
derived by \citet{B99,B02}. The fluorescent excitation was not
considered in test calculations because of the problems with the CLOUDY 
He~{\sc i} intensities when all processes are included (see Fig. \ref{fig2} and
discussion in Sect. \ref{Heemiss}).

Finally, the oxygen abundance should
be corrected for its fraction locked in dust grains. \citet{I06} found that
the Ne/O abundance ratio in low-metallicity emission-line galaxies
increases with increasing oxygen abundance. They interpreted this
trend by a larger fraction of oxygen locked in dust grains in galaxies
with higher O/H. We used the relation of \citet{I06} between log~Ne/O
and 12 + logO/H to derive the fraction of oxygen confined in dust:

\begin{equation}
\Delta \left(\frac{\rm O}{\rm H}\right)_{\rm dust} =
10^{0.088(12+\log{\rm O/H})-0.616}. \label{dustcorr}
\end{equation}

The equivalent width of the He~{\sc i} $\lambda$4471 absorption
line was chosen to be EW$_{\rm abs}$($\lambda$4471) = 0.4\AA, following
\citet{GD05} and \citet{IT10}. The equivalent widths of the other absorption
lines were fixed according to the ratios 
\begin{eqnarray}
{\rm EW}_{\rm abs}(\lambda 3889)/{\rm EW}_{\rm abs}(\lambda 4471)& = &1.0,\nonumber \\
{\rm EW}_{\rm abs}(\lambda 5876)/{\rm EW}_{\rm abs}(\lambda 4471)& = &0.8,\nonumber \\
{\rm EW}_{\rm abs}(\lambda 6678)/{\rm EW}_{\rm abs}(\lambda 4471)& = &0.4,\nonumber \\
{\rm EW}_{\rm abs}(\lambda 7065)/{\rm EW}_{\rm abs}(\lambda 4471)& = &0.4.\label{ew}
\end{eqnarray}
The EW$_{\rm abs}$($\lambda$5876) / EW$_{\rm abs}$($\lambda$4471)
and EW$_{\rm abs}$($\lambda$6678) / EW$_{\rm abs}$($\lambda$4471) ratios
were set equal to the values 
predicted for these ratios by a Starburst99 \citep{L99} instantaneous
burst model with an age of 3 -- 4 Myr and a heavy-element
mass fraction $Z$ = 0.001 -- 0.008.
These values are significantly higher than the
corresponding ratios of 0.3 and 0.1 adopted by \citet{I07}.
We note that the value chosen for the EW$_{\rm abs}$($\lambda$5876)/
EW$_{\rm abs}$($\lambda$4471) ratio is also consistent with
the one given by \citet{GD05}. Since the
output high-resolution spectra in Starburst99 are calculated
only for wavelengths $<$ 7000 \AA, we do not have a prediction
for the EW$_{\rm abs}$($\lambda$7065)/
EW$_{\rm abs}$($\lambda$4471) ratio. We set it to be
equal to 0.4, the value of the EW$_{\rm abs}$($\lambda$6678)/
EW$_{\rm abs}$($\lambda$4471) ratio. As for He~{\sc i} $\lambda$3889,
this line is blended with the hydrogen H8 $\lambda$3889 line. Therefore,
EW$_{\rm abs}$(He~{\sc i} $\lambda$3889) cannot be estimated from the 
Starburst99 models. We assumed the value shown in Eq. \ref{ew}.

Finally, the age $t_{\rm burst}$ of a starburst should be derived. This is 
because ionisation correction factors $ICF$(He) and the non-recombination 
contribution to hydrogen lines both depend on the starburst age 
(see Sect. \ref{icfHe} and \ref{hydro}). As a first approximation, we
used the relation between $t_{\rm burst}$ and EW(H$\beta$) from the 
Starburst99 instantaneous burst models with a heavy-element mass fraction
$Z$ = 0.004 \citep{L99}. We fitted this relation by the expression
\begin{eqnarray}
 t_{\rm burst}&=&167.6w^3-2296w^2+12603w-35651 \nonumber \\
              &+&54976/w-43884/w^2+14208/w^3, \label{tb}
\end{eqnarray}
where $t_{\rm burst}$ is in Myr, $w$ = log EW(H$\beta$) 
and EW(H$\beta$) is in \AA.  
Eq. \ref{tb} does not take into account the contribution of old stellar 
populations in the underlying galaxy. The effect of an 
underlying galaxy is discussed in Sect. \ref{primo}.
Since the relations for $Z$ = 0.001, $Z$ = 0.004, and 
$Z$ = 0.008 are similar at EW(H$\beta$) $\ga$ 100\AA, corresponding 
to $t_{\rm burst}$ $\la$ 4 Myr, 
we adopted Eq. \ref{tb} for the entire range of oxygen
abundances in our sample galaxies. We also adopted $t_{\rm burst}$ = 1 Myr
and 4 Myr, when the derived starburst age was $<$ 1 Myr or $>$ 4 Myr. For
$t_{\rm burst}$ in the range of 1 -- 4 Myr we used the linear interpolation
to derive $ICF$(He) (Eqs. \ref{icf7.3_2.0} - \ref{icf8.3_4.0})
and the non-recombination contribution to the intensities
of hydrogen lines H$\alpha$, H$\beta$, H$\gamma$, and H$\delta$
(Eqs. \ref{collHa_2.0} - \ref{collHd_4.0}).

One caveat with the ionisation correction factors that we obtained from 
CLOUDY models is that they all have negligible $y^{2+}$. However, in  real 
H~{\sc ii} regions, the He~{\sc ii} $\lambda$4686 line is often measured but 
not explained \citep[see ][]{SI03}, and may give $y^{2+}$ as large as 3\% that
of $y^+$. In particular, the He~{\sc ii} $\lambda$4686 line is measured
in $\sim$58\% of spectra from our sample with an average intensity
of $\sim$1\% that of H$\beta$, corresponding to $\sim$1\% of He in
He$^{2+}$ form. It tends to be stronger in H~{\sc ii} regions with lower
metallicity. \citet{TI05} and \citet{I12} proposed that a small fraction
(a few percent) of ionising radiation could be produced by shocks. This
radiation is harder than the stellar one and is responsible for
He~{\sc ii} $\lambda$4686 emission, but only weakly influences the
He~{\sc i} and H line intensities. However, we cannot quantify 
contribution of shocks using CLOUDY, and to estimate how $ICF$(He)
changes when the small fraction of ionising radiation from shocks
is present. 

On the other hand, we can add some small fraction of harder
non-thermal AGN-like ionising radiation to obtain the intensity
of He~{\sc ii} $\lambda$4686 emission line of $\sim$ 1\% -- 3\% that of 
the H$\beta$ emission line. For this we assumed that the number
of ionising photons due to the harder radiation is 5\% - 10\% of the
stellar number of ionising photons $Q$(H). 
Modelling with CLOUDY shows that the difference
between $ICF$(He)'s with and without harder radiation is very small, not
exceeding 0.2\%.

\subsection{Sample of low-metallicity emission-line galaxies}\label{sample1}

We determined $Y_{\rm p}$ and d$Y$/d(O/H) for a large sample of low-metallicity
emission-line galaxies, consisting of three subsamples.
Most of our galaxies were compact. The spectra of these galaxies are
obtained within apertures that are similar to or larger than the angular 
sizes. Therefore, these spectra are characteristics of the
integrated galaxy properties averaged over their volume, not along the line of 
sight, and the above test analysis
can be applied since it was made for volume-averaged characteristics.

The HeBCD subsample is composed of
93 different observations of 86 H~{\sc ii} regions in 77 galaxies.
The majority of these galaxies are low-metallicity BCD galaxies. This sample
is the same as the one described in \citet{IT04,IT10}.

The VLT subsample \citep{G11} is composed of 75 VLT spectra 
of low-metallicity H~{\sc ii} regions selected from the ESO data archive.

The third subsample is composed of spectra of low-metallicity 
H~{\sc ii} regions selected from the SDSS DR7. 
The SDSS \citep{Y00} offers a gigantic
data base of galaxies with well-defined selection criteria and observed in a
homogeneous way. In addition, the spectral resolution is much better than that
of most previous data bases on emission-line galaxies including all the 
spectra in the HeBCD sample. First, we extracted 
$\sim$ 15000 spectra with strong emission lines from the whole data base of
$\sim$ 800000 galaxy spectra.
We measured the emission line intensities and determined the
element abundances for the SDSS subsample following the same 
procedures as for the HeBCD and VLT subsamples. Then, for the $^4$He 
abundance determination, we selected 1442 spectra (SDSS subsample) in which the
intensity of the temperature-sensitive emission line 
[O~{\sc iii}] $\lambda$4363
was measured with accuracy better than $\sim$ 25\%, allowing a reliable
abundance determination. Part of the SDSS subsample was discussed for instance
by \citet{I06,I11a}. In total, our sample consists
of 1610 spectra, which we used to determine $Y_{\rm p}$.

\subsection{Linear regressions for determining the primordial
$^4$He abundance}\label{method1}

As in previous work \citep[see ][ and references therein]{I07,IT10},
we determined the primordial $^4$He mass fraction
$Y_{\rm p}$ by fitting the data points in the $Y$ -- O/H
plane with a linear regression line of the
form \citep{PTP74,PTP76,P92}
\begin{equation}
Y = Y_p + \frac{{\rm d}Y}{{\rm d}({\rm O/H})} ({\rm O/H}).               \label{eq:YvsO}
\end{equation}

Assuming a linear dependence of $Y$ on O/H appears to be
reasonable because there are no evident non-linear trends
in the distributions of the data points in the $Y$ vs O/H
diagram \citep[e.g., ][]{IT04}. 
The linear regression (Eq. \ref{eq:YvsO})
implies that the initial mass function (IMF) averaged stellar yields for 
different elements do not depend on metallicity. 
It has been suggested in the past \citep[e.g., ][]{B83} that,
at low metallicities, the IMF may be top-heavy, that is, that there are 
relatively more massive stars 
than lower mass stars than at high metallicities. 
If this is the case, the IMF-averaged 
yields would be significantly different for low-metallicity 
stars than those for more 
metal-enriched stars, resulting in a non-linear relationship between $Y$ 
and O/H \citep{SF03}. However, until now, no persuasive evidence has been 
presented for a metallicity dependence of the IMF. Furthermore, 
the properties of extremely metal-deficient stars remain poorly known, 
excluding quantitative estimates
of possible non-linear effects in the $Y$ -- O/H
relation. Therefore, we continue to use the linear regression 
(Eq. \ref{eq:YvsO}) to fit the data in the following analysis.
However, for the sake of comparison we also consider non-linear fits.

To derive the parameters of the linear regressions,
 we used the maximum-likelihood method \citep{Pr92},
 which takes into account the errors in 
$Y$ and O/H for each object.

   \begin{figure*}
   \centering
  \hbox{  \includegraphics[width=6cm,angle=-90]{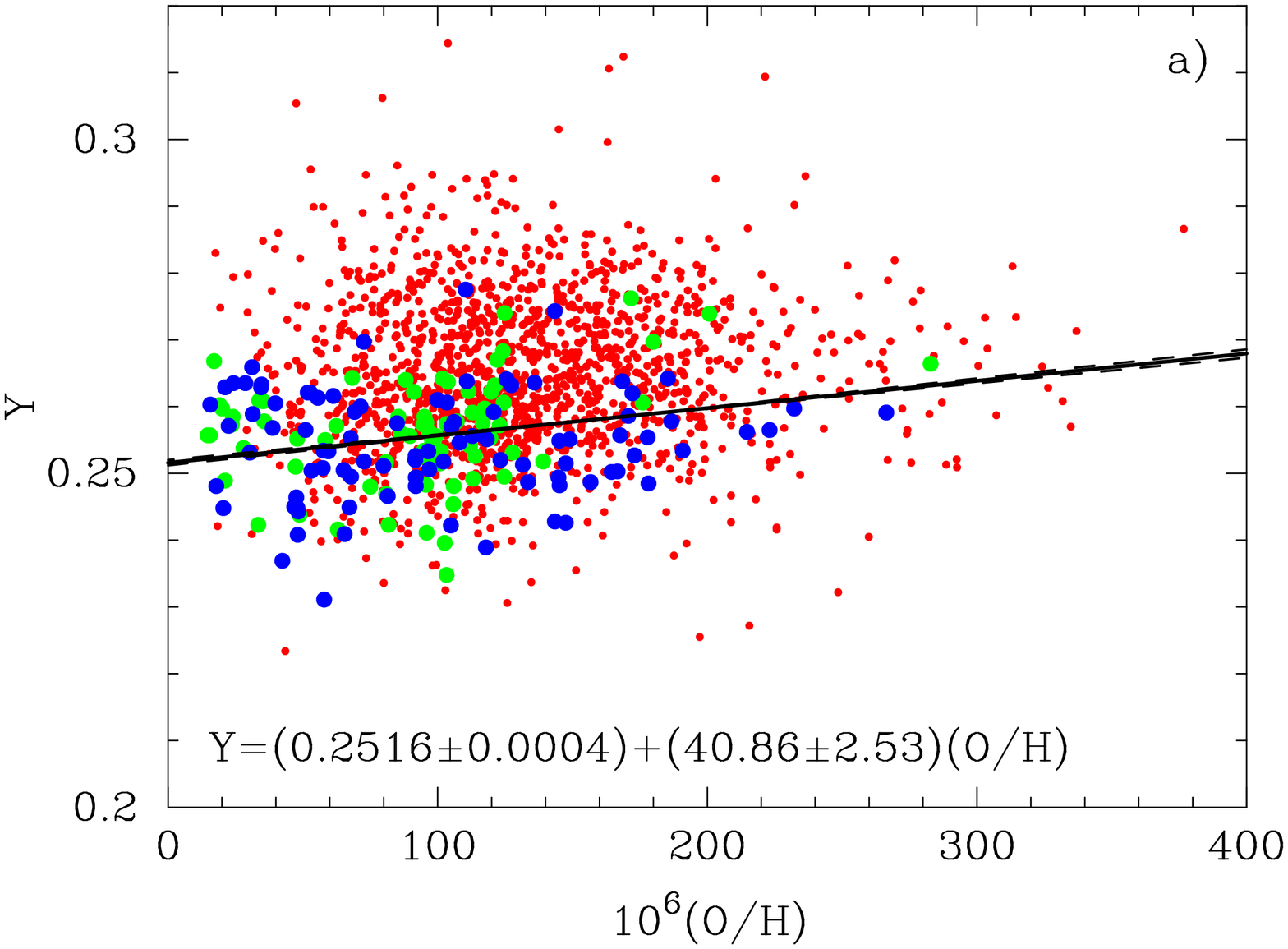}
   \hspace{0.2cm}\includegraphics[width=6cm,angle=-90]{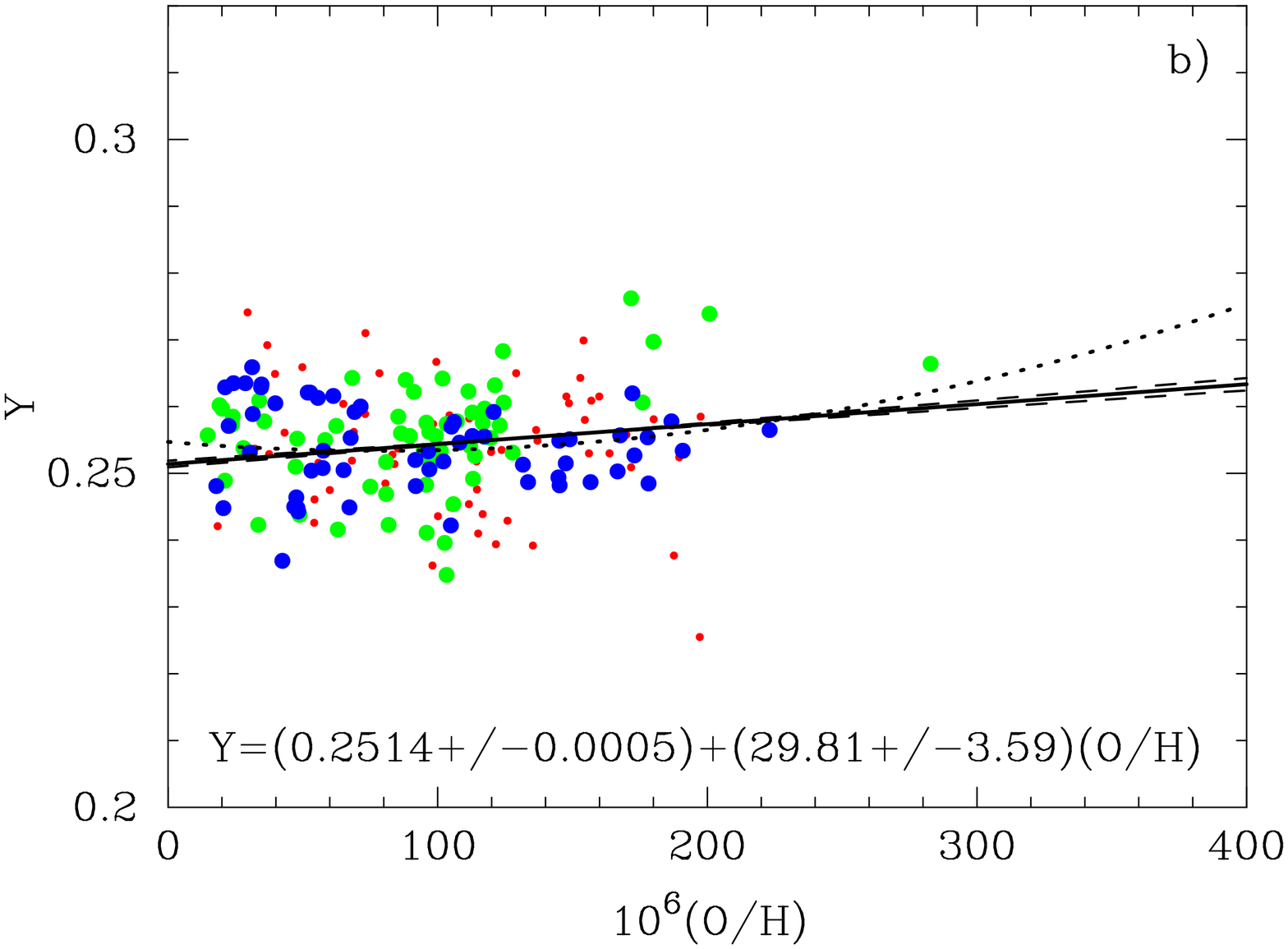}}
\vspace{0.2cm}
  \hbox{  \includegraphics[width=6cm,angle=-90]{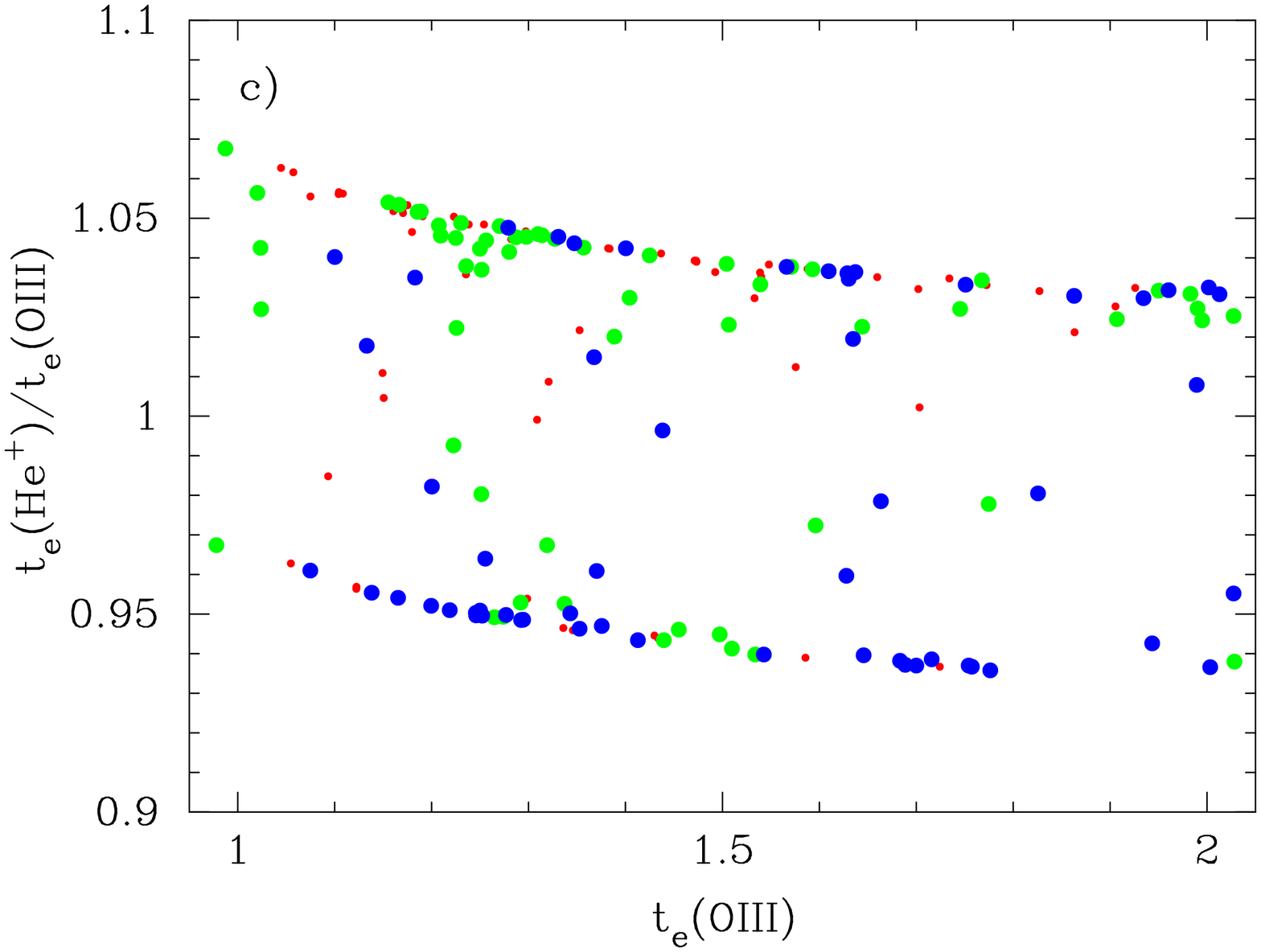}
   \hspace{0.2cm}\includegraphics[width=6cm,angle=-90]{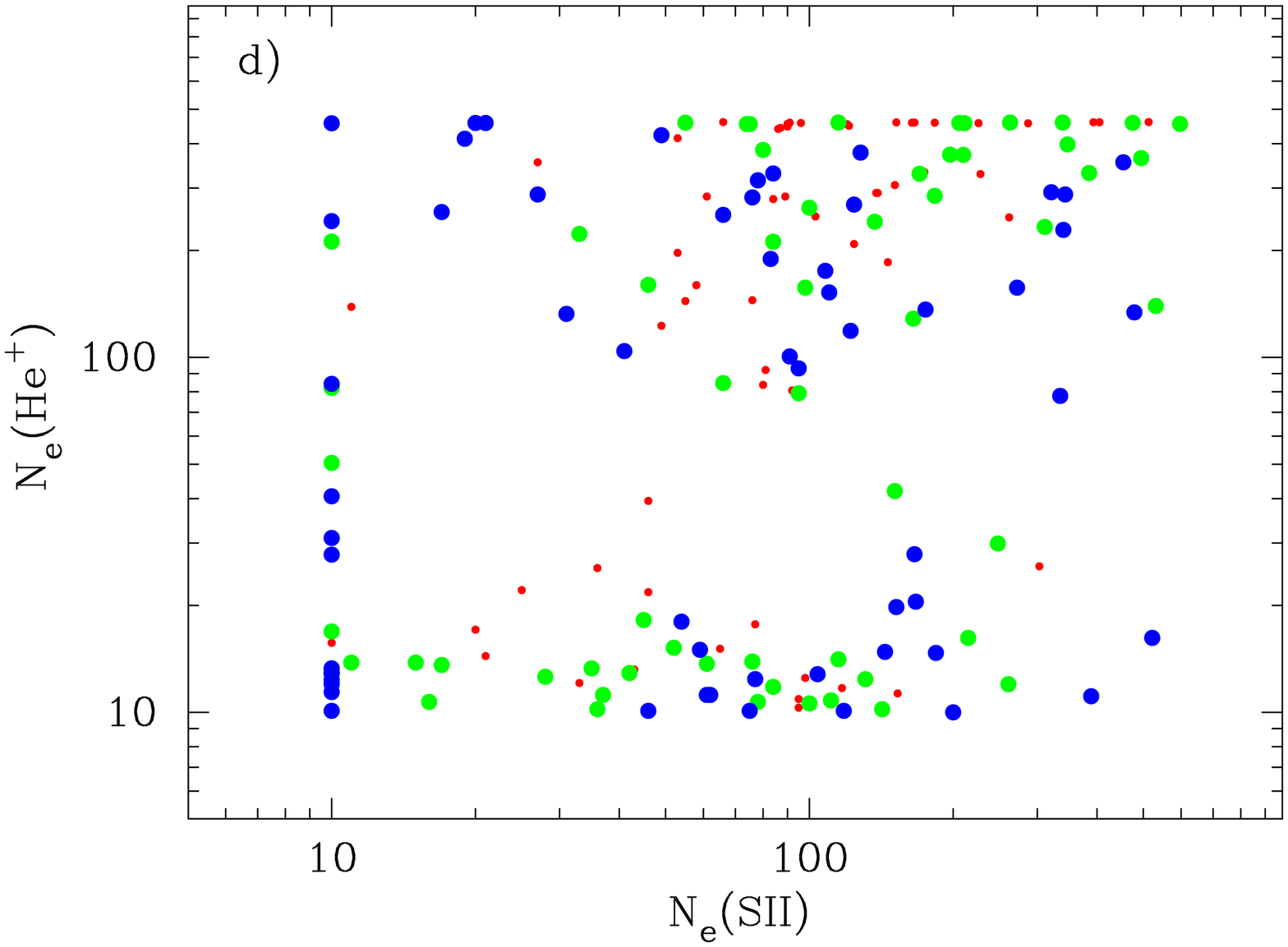}}
      \caption{(a) $Y$ - O/H
for the sample of 1610 H~{\sc ii} regions. The
five He~{\sc i} emission lines $\lambda$3889, $\lambda$4471, $\lambda$5876, 
$\lambda$6678, and $\lambda$7065 are used for the $\chi^2$ minimisation and 
the determination of $Y$.
Large blue and green filled circles are for the HeBCD and VLT samples,
small red filled circles are SDSS galaxies. We chose to let $T_{\rm e}$(He$^+$) 
vary freely 
 in the range 0.95 -- 1.05 of the $\widetilde{T}_{\rm e}$(He$^+$) value. 
The continuous line
represents the linear regression (whose equation is given at the bottom of the 
panel). (b) Same as
(a), but only H~{\sc ii} regions with EW(H$\beta$) $\geq$ 100\AA\ are shown, 
with the excitation parameter $x$ = O$^{2+}$/O $\geq$ 0.7 and with 1$\sigma$ 
error in $Y$ $\leq$3\%. The dotted line is the quadratic maximum-likelihood 
fit to the data. (c) Distribution of the best derived values
$T_{\rm e}$(He$^+$)/$T_{\rm e}$(O~{\sc iii}) versus $T_{\rm e}$(O~{\sc iii})
for the case shown in (b). 
(d) Distribution of the best derived values $N_{\rm e}$(He$^+$) versus  
$N_{\rm e}$(S~{\sc ii}) for the case shown in (b).}
         \label{fig9}
   \end{figure*}

   \begin{figure*}
   \centering
  \includegraphics[width=12cm,angle=-90]{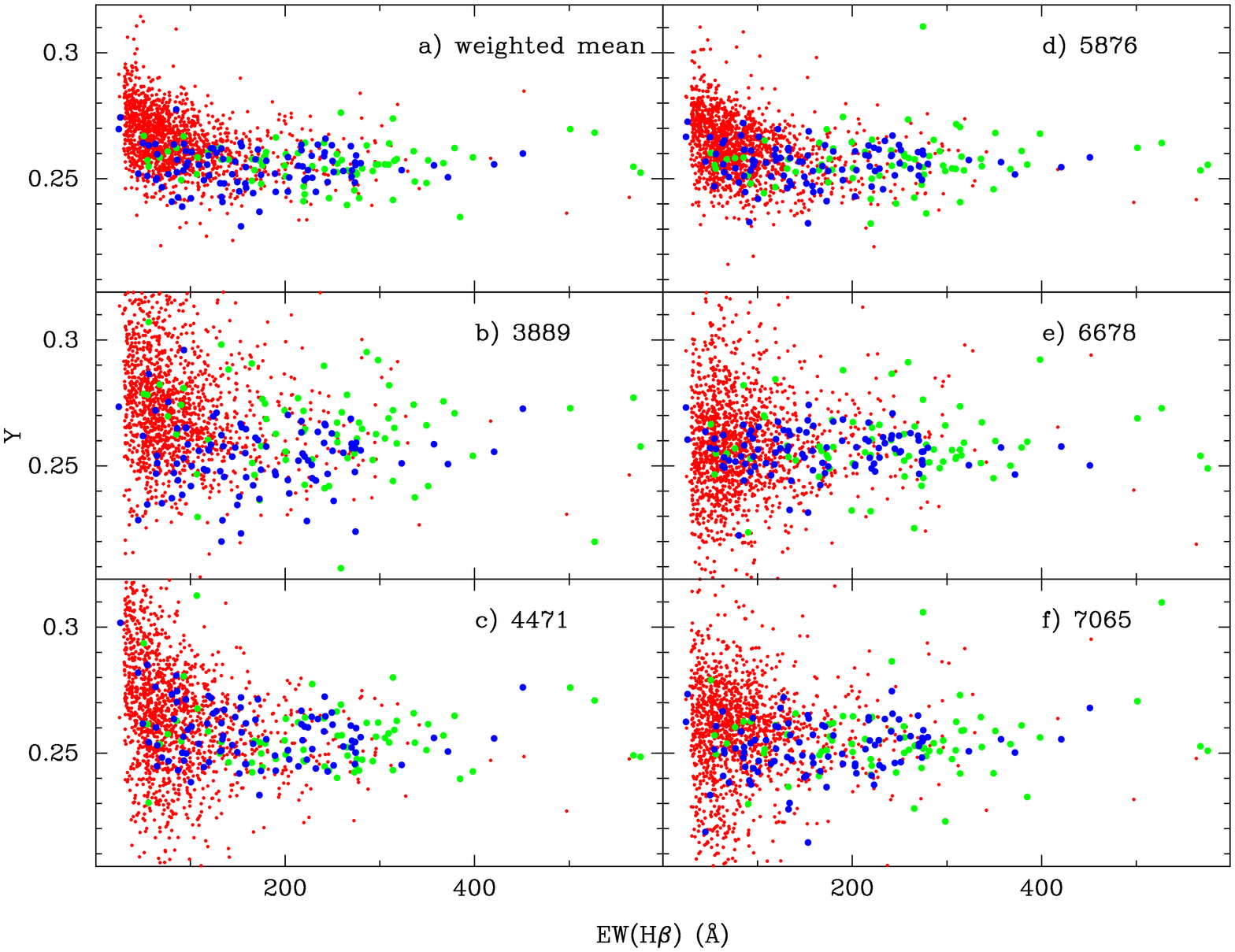}
      \caption{Distribution of helium mass fraction $Y$ with H$\beta$
equivalent width EW(H$\beta$) for the entire sample of 1610 H~{\sc ii}
regions. $Y$ values are the same as in Fig. \ref{fig9}a.}
         \label{fig10}
   \end{figure*}



\section{Primordial $^4$He mass fraction $Y_{\rm p}$}\label{primo}

We considered linear regressions $Y$ -- O/H for the entire 
HeBCD+VLT+SDSS sample, 
adopting that the electron 
temperature $T_{\rm e}$(He$^+$) is randomly varied in the range 
(0.95 -- 1.05)$\widetilde{T}_{\rm e}$(He$^+$) to minimise $\chi^2$,
where $\widetilde{T}_{\rm e}$(He$^+$) is defined by Eq.~\ref{tHeOIII}.
This regression is shown in Fig. \ref{fig9}a.

The most notable feature in the figure is that the SDSS
H~{\sc ii} regions are offset relative to the HeBCD and VLT H~{\sc ii}
regions. 
This suggests that HeBCD+VLT and SDSS H~{\sc ii} samples have different
properties. First, the SDSS spectra are in general of lower quality,
therefore the line intensities have much higher statistical errors.
Second, SDSS galaxies have on average higher oxygen abundances.
The temperature-sensitive [O~{\sc iii}] $\lambda$4363 emission line
is weaker in their spectra and the electron temperature is derived
with larger uncertainties. To study the sources of differences we
show in Fig. \ref{fig10} the dependences of
the weighted mean $^4$He mass fraction $Y$ and $Y$s derived
from individual lines on the equivalent width EW(H$\beta$).
It is clearly seen that the SDSS galaxies have on average lower EW(H$\beta$)
and consequently lower EWs of He~{\sc i} emission lines. Therefore, the
corrections for underlying stellar absorption in these objects are
on average higher than those in HeBCD and VLT H~{\sc ii} regions,
implying larger uncertainties caused by this effect.

We note the broad spread of $Y$s derived from the weakest He~{\sc i} 
$\lambda$4471, 6678, and 7065 emission lines in the SDSS spectra, implying 
that they have a lower quality than HeBCD and VLT spectra. As for the
strongest $\lambda$3889 and $\lambda$5876 lines, the spread of $Y$s
derived from the SDSS spectra is similar to that derived from
the HeBCD and VLT spectra. However, the spread of $Y$s derived from the
$\lambda$3889 emission line is much broader than that derived from
the $\lambda$5876 emission line. Furthermore, it is much broader than that
derived from faintest He~{\sc i} lines in the HeBCD and VLT samples.
This line is blended with the hydrogen H8 line and
therefore the determination of its intensity is more uncertain. This
also implies that including an additional strong He~{\sc i} $\lambda$10830
emission line and its use instead of the He~{\sc i} $\lambda$3889
line is highly important to improve the determination of the He abundance. 

The clear offset of the SDSS galaxies to higher $Y$ values is seen 
in Fig. \ref{fig10}, with the possible exception for the $\lambda$6678 line.
One of the likely reasons for this offset 
is that the SDSS spectra are obtained with a lower 
signal-to-noise ratio (S/N). Measurements with
low S/Ns tend to be dominated by objects whose true value, for instance, 
the flux of the [O {\sc iii}] $\lambda$4363 emission line, is slightly lower 
than the cutoff, and the 
uncertainties push them above the threshold (Malmquist-type bias). 
This effect is stronger for H~{\sc ii} regions with low EW(H$\beta$)
where [O {\sc iii}] $\lambda$4363 emission line
in general is weaker, and it would potentially
overestimate the electron temperature and He abundance.
It can be decreased
by selecting only spectra with highest S/N, for example, by selecting spectra 
whose He abundance is derived with an accuracy of better than 3\%.

\begin{figure}
   \centering
   \includegraphics[width=6cm,angle=-90]{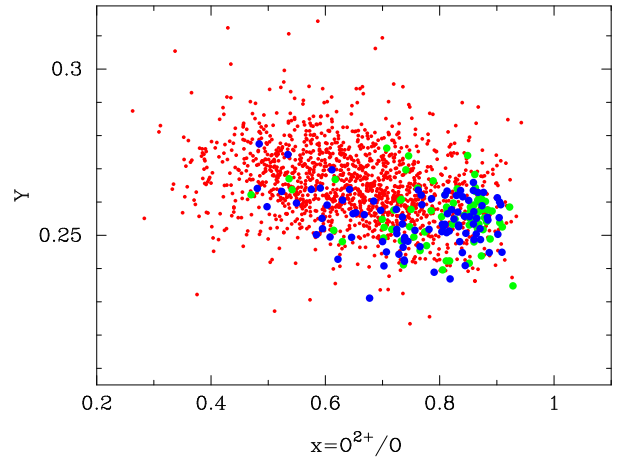}
      \caption{
Distribution of helium mass fraction $Y$ with the excitation
parameter $x$ = O$^{2+}$/O for the entire sample of 1610 H~{\sc ii}
regions. $Y$ values are the same as in Fig. \ref{fig9}a.}
         \label{fig11}
   \end{figure}

A similar offset of the SDSS H~{\sc ii} regions is seen in Fig. \ref{fig11},
where we show the dependence of the weighted mean $Y$ on the excitation 
parameter $x$. The SDSS H~{\sc ii} regions are on average of lower excitation.


Summarising, we conclude that the $^4$He mass fractions $Y$ in 
spectra of H~{\sc ii} 
regions with low EW(H$\beta$), low $x$ and low S/N may be
derived incorrectly.
Therefore, in Fig. \ref{fig9}b we show the linear regression only for 
182 high-excitation H~{\sc ii} regions with EW(H$\beta$) $\geq$ 100\AA, 
$x$ $\geq$ 0.7 and with the 1$\sigma$ error in $Y$ value not exceeding 3\%. 
There is no offset between H~{\sc ii} regions from different 
subsamples. We may further decrease the sample for instance by selecting 
objects with EW(H$\beta$) $>$ 200\AA\ or/and selecting objects with 
$\sigma$($Y$)/$Y$ $<$ 1\%. However, in this case, the sample may become 
small and the $Y_{\rm p}$ value and the slope d$Y$/d(O/H) derived from the 
maximum-likelihood regressions may not be very certain. 
We prefer to use a sample as large as possible.

The best derived values of the electron temperature $T_{\rm e}$(He$^+$)
(Fig. \ref{fig9}c) are distributed within the entire range of 
the Monte Carlo variations, concentrating at the upper and lower values
of 1.05$\widetilde{T}_{\rm e}$(He$^+$) and 
0.95$\widetilde{T}_{\rm e}$(He$^+$), respectively, and 
do not follow the relation Eq. \ref{tHeOIII} between $T_{\rm e}$(He$^+$) and
$T_{\rm e}$(O~{\sc iii}) obtained from our CLOUDY models. This is 
somewhat surprising, and we do not have a good explanation for it. Note, 
however, that the reason cannot be the ``temperature fluctuations'' first 
studied by \citet{P67} and advocated by \citet{Pe07}, since in that case 
$T_{\rm e}$(He$^+$) is expected to be systematically lower than 
$T_{\rm e}$(O~{\sc iii}). Furthermore, uncertainties of the He {\sc i} line 
intensities may play a role.
The best derived electron number density $N_{\rm e}$(He$^+$)  
does not correlate with the number density $N_{\rm e}$(S~{\sc ii}) obtained 
from the [S~{\sc ii}] $\lambda$6717/$\lambda$6731
emission-line ratio (Fig. \ref{fig9}d).
Of course, one does not expect this to be the case if  
the region of He~{\sc i} emission is not 
spatially coextensive with the region of [S~{\sc ii}] emission. However, 
for young and bright H~{\sc ii} regions one instead expects the inner zones to
be denser than the outer ones, and this is not what Fig. \ref{fig9}d shows.
A good test, but requiring very high S/N data, would be to 
compare $N_{\rm e}$(He$^+$) with the density derived from the  [Cl~{\sc iii}] 
doublet.


Because we used a large sample of H~{\sc ii} regions, the statistical errors in
the $Y_{\rm p}$ determination are small (e.g., Fig. \ref{fig9}b). On the other
hand, other uncertainties can be significantly higher. These are 
uncertainties in the He~{\sc i} emissivities and their analytical
fits ($\sim$1\%, Fig. \ref{fig1})\footnote{In particular, \citet{P09} 
and \citet{P12} 
suggested 0.06\% -- 0.8\% standard deviations of 
He~{\sc i} emissivities for extragalactic H~{\sc ii}
regions.}, the uncertainties in $ICF$(He) 
($\sim$ 0.25\%, Fig. \ref{fig3}), and correction for the non-recombination 
contribution to 
hydrogen-line intensities ($\sim$ 0.5\%, Fig. \ref{fig4}).

   \begin{figure*}
   \centering
  \hbox{  \includegraphics[width=4.4cm,angle=-90]{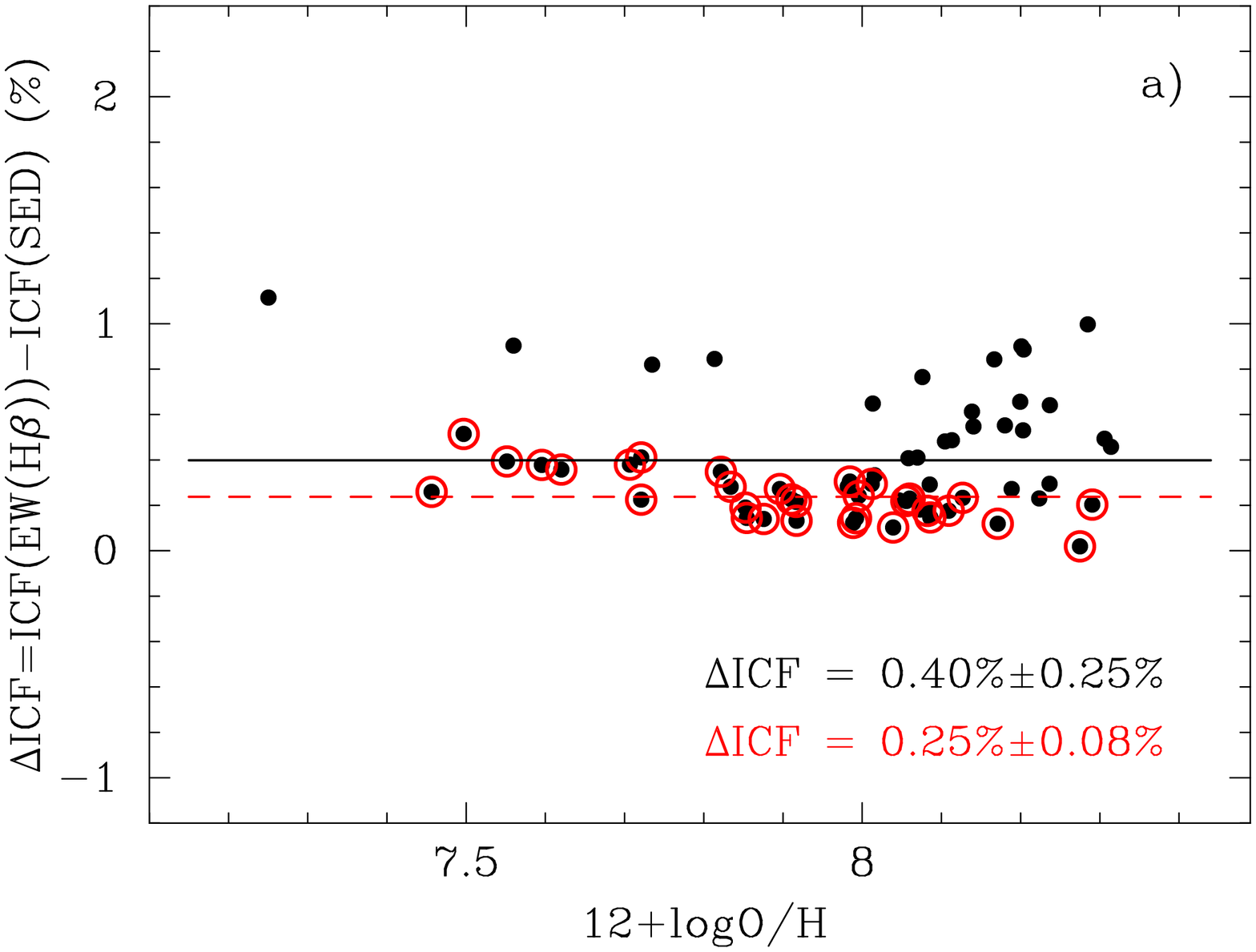}
   \hspace{0.2cm}\includegraphics[width=4.4cm,angle=-90]{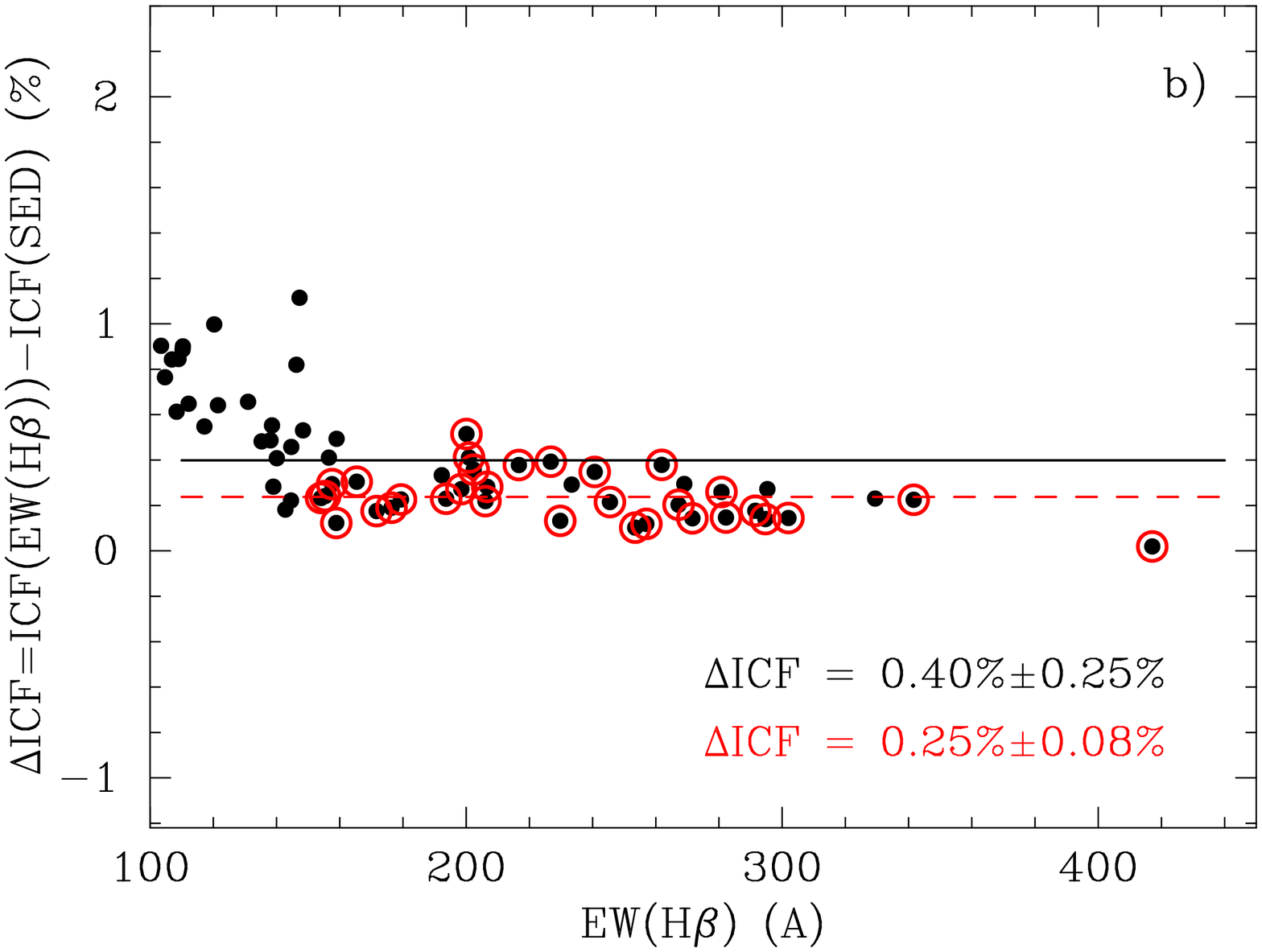}
   \hspace{0.2cm}\includegraphics[width=4.4cm,angle=-90]{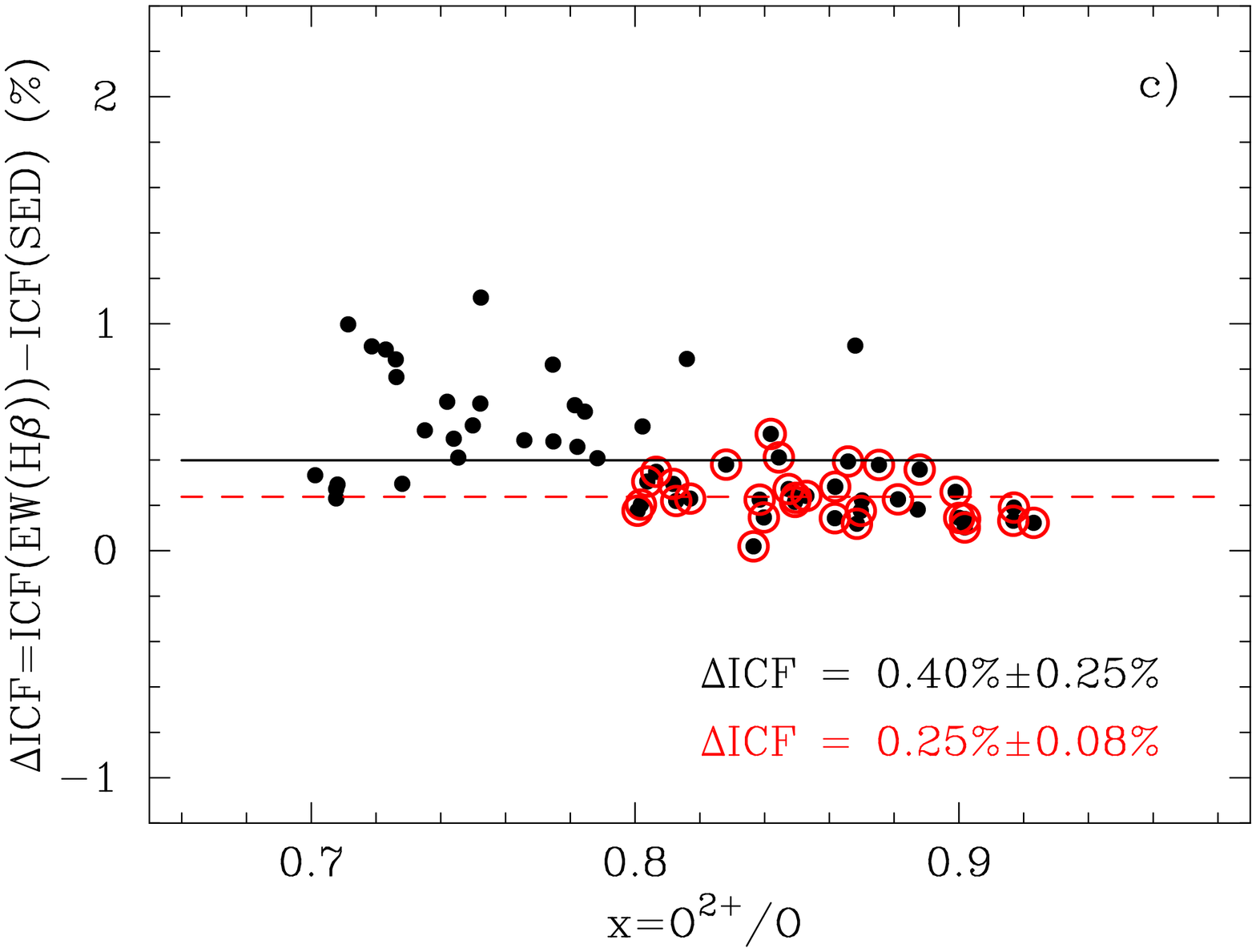}
}
      \caption{ (a) Dependence 
of $\Delta$$ICF$(He) = $ICF$(EW(H$\beta$)) -- $ICF$(SED) on the oxygen 
abundance 12+logO/H for 
64 SDSS galaxies with EW(H$\beta$) $\geq$ 100\AA, 
with the excitation parameter $x$ = O$^{2+}$/O $\geq$ 0.7 and with 1$\sigma$ 
error in $Y$ $\leq$3\% (dots) and for 
32 SDSS galaxies with EW(H$\beta$) $\geq$ 150\AA, 
with the excitation parameter $x$ = O$^{2+}$/O $\geq$ 0.8 and with 1$\sigma$ 
error in $Y$ $\leq$3\% (encircled dots). Here $ICF$(EW(H$\beta$)) is the 
ionisation correction
factor derived for a starburst age estimated from EW(H$\beta$), while
$ICF$(SED) is the ionisation correction factor for the starburst age, which
is derived from the spectral energy distribution (SED) fitting. Horizontal
solid and dashed lines are average values of $\Delta$$ICF$(He) for dots and
encircled dots, respectively. (b) The dependence 
of $\Delta$$ICF$(He) = $ICF$(EW(H$\beta$)) -- $ICF$(SED) on the H$\beta$
equivalent width EW(H$\beta$). Symbols and lines have the same meaning as in 
(a). (c) The dependence 
of $\Delta$$ICF$(He) = $ICF$(EW(H$\beta$)) -- $ICF$(SED) on the excitation
parameter $x$. Symbols and lines have the same meaning as in (a).}
         \label{fig12}
   \end{figure*}

One of the systematic uncertainties is caused by the presence of the underlying
galaxy. Owing to this effect, the EW(H$\beta$) is decreased and the starburst
age is overestimated, resulting in overestimated $ICF$(He) and $Y$. To estimate
the effect of the underlying galaxy we show in Fig. \ref{fig12} the dependence
of the difference $\Delta$$ICF$(He) = $ICF$(EW(H$\beta$)) -- $ICF$(SED) for 
SDSS galaxies from Fig. \ref{fig9}b on the oxygen abundance 12+logO/H,
H$\beta$ equivalent width EW(H$\beta$), and excitation parameter 
$x$ = O$^{2+}$/O. Here, $ICF$(EW(H$\beta$)) is the ionisation correction
factor derived for a starburst age estimated from EW(H$\beta$), while
$ICF$(SED) is the ionisation correction factor for the starburst age, which
was derived from the spectral energy distribution (SED) fitting. To fit the
SED we followed the technique described by \citet{I11a}. Briefly, it takes
into account the emission of the young and old stellar populations as well as
the ionised gas emission. 

It is evident that 
$ICF$(EW(H$\beta$)) $>$ $ICF$(SED) because the starburst age derived from
the SED fitting is always younger than that derived from EW(H$\beta$).
The effect of the underlying galaxy on $ICF$ in galaxies with 
EW(H$\beta$) $\geq$ 100\AA, $x$ $\geq$ 0.7 and $\Delta$$Y$/$Y$ $\leq$ 3\%
(dots in Fig. \ref{fig12}) is small, $\sim$0.4\% on average (horizontal
solid line). We note that there is no clear trend of $\Delta$$ICF$(He) on
12+logO/H (Fig. \ref{fig12}a). On the other hand, there is an increase
of $\Delta$$ICF$(He) with decreasing EW(H$\beta$) (Fig. \ref{fig12}b) 
and $x$ (Fig. \ref{fig12}c), 
suggesting that the effect of an underlying galaxy is 
higher for lower-excitation H {\sc ii} regions that are older.

However, if only H {\sc ii} regions with EW(H$\beta$) $\geq$ 150\AA\ 
and $x$ $\geq$ 0.8 are considered (encircled dots in Fig. \ref{fig12}),
no dependence on EW(H$\beta$) and $x$ is detected with a very low average 
$\Delta$$ICF$(He) = 0.25\% (horizontal dashed line). Thus, from the above
discussion of $Y_{\rm p}$ systematic errors (1\% due to the uncertainties 
in the He {\sc i} emissivities and their analytical
fits, 0.25\% due to the uncertainties in $ICF$ fits, 0.5\% due to the
uncertainties in correction for the non-recombination contribution
to hydrogen-line intensities and 0.25\% due to the contribution of the 
underlying galaxy), we adopted
the value $\sqrt{1^2+0.5^2+0.25^2+0.25^2}$\% = 1.17\% or $\sigma$($Y_{\rm p}$)
=0.003.

As a final sample for determining $Y_{\rm p}$ we adopted 
the sample of 111 H {\sc ii} regions satisfying conditions 
EW(H$\beta$) $\geq$ 150\AA, $x$ $\geq$ 0.8 and $\Delta$$Y$/$Y$ $\leq$ 3\%. 
The linear regression for this sample is shown in Fig. \ref{fig13} with
the primordial value $Y_{\rm p}$ = 0.2542.


Since the statistical error of $Y_{\rm p}$ ($\pm$0.0006) is much lower than the
systematic error, we finally adopted 
\begin{equation}
Y_p=0.254\pm0.003. \label{Yp}
\end{equation}
This new $Y_{\rm p}$ 
agrees with the $Y_{\rm p}$s derived by \citet{IT10} and \citet{A12}. However,
this agreement is in fact accidental because different He~{\sc i}
emissivities and corrections for non-recombination excitation of hydrogen
lines were used in different studies.

The linear dependence $Y$ -- O/H in Fig. \ref{fig13} is
not steep. The weighted mean of all 111 data points in the figure is
$Y$ = 0.255 with $\chi^2$ = 6.13 per degree of freedom. These values are
slightly higher than $Y_{\rm p}$ = 0.254 and $\chi^2$=6.12 per degree of 
freedom
returned by the linear maximum-likelihood technique, implying only weak linear 
dependence. Additionally, we calculated the quadratic maximum-likelihood 
regression (dotted line in Fig. \ref{fig13}) and obtained a primordial value of
0.255 with $\chi^2$ = 5.39 per degree of 
freedom. We preferred to use the $Y_{\rm p}$ value derived from the linear
regression over the weighted mean value because chemical evolution
models predict a general increase of the He abundance with O/H. 
The same chemical evolution argument
does not favour the quadratic regression because the He abundance at low 
oxygen abundances is lower with increasing O/H.


\section{Cosmological implications}\label{cosmo}

\begin{figure}
   \centering
   \includegraphics[width=6cm,angle=-90]{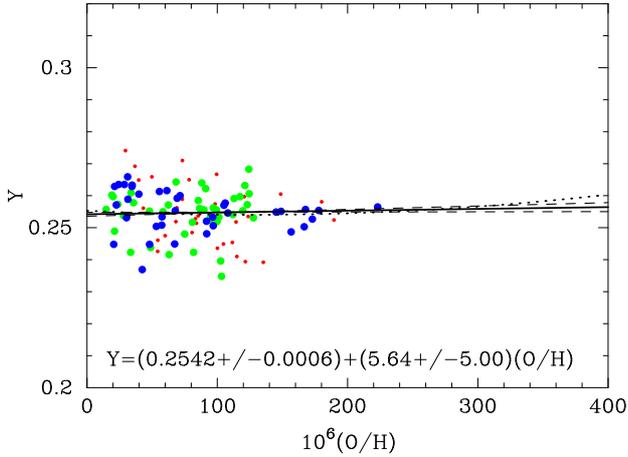}
      \caption{$Y$ - O/H
for the sample of 111 H~{\sc ii} regions with EW(H$\beta$) $\geq$ 150\AA, 
with the excitation parameter $x$ = O$^{2+}$/O $\geq$ 0.8 and with 1$\sigma$ 
error in $Y$ $\leq$3\%. The
five He~{\sc i} emission lines $\lambda$3889, $\lambda$4471, $\lambda$5876, 
$\lambda$6678, and $\lambda$7065 are used for $\chi^2$ minimisation and 
determination of $Y$. Large blue and green filled circles are for HeBCD 
and VLT samples, respectively,
small red filled circles are SDSS galaxies. We chose to let 
$T_{\rm e}$(He$^+$) vary freely 
in the range 0.95 -- 1.05 of the $\widetilde{T}_{\rm e}$(He$^+$) value. 
The continuous line
represents the linear regression (whose equation is given at the bottom of the 
panel) and the dashed lines are 1$\sigma$ alternatives of the linear
regression. The dotted line is the quadratic maximum-likelihood 
fit to the data.}
         \label{fig13}
   \end{figure}

We have derived the primordial $^4$He mass fraction 
$Y_{\rm p}$ = 0.254$\pm$0.003 (Fig. \ref{fig13}), which is higher than
the SBBN value of 0.2477$\pm$0.0001 inferred from the analysis of the 
temperature fluctuations of the microwave background radiation
\citep{A13}, indicating deviations from the standard rate of Hubble 
expansion in the early Universe. These deviations 
can be caused by an extra contribution to the total energy density (for 
example by additional species of neutrinos), which can 
be parameterised by an equivalent number of neutrino species $N_{\rm eff}$. 

\begin{figure}
   \centering
   \includegraphics[width=5.5cm,angle=-90]{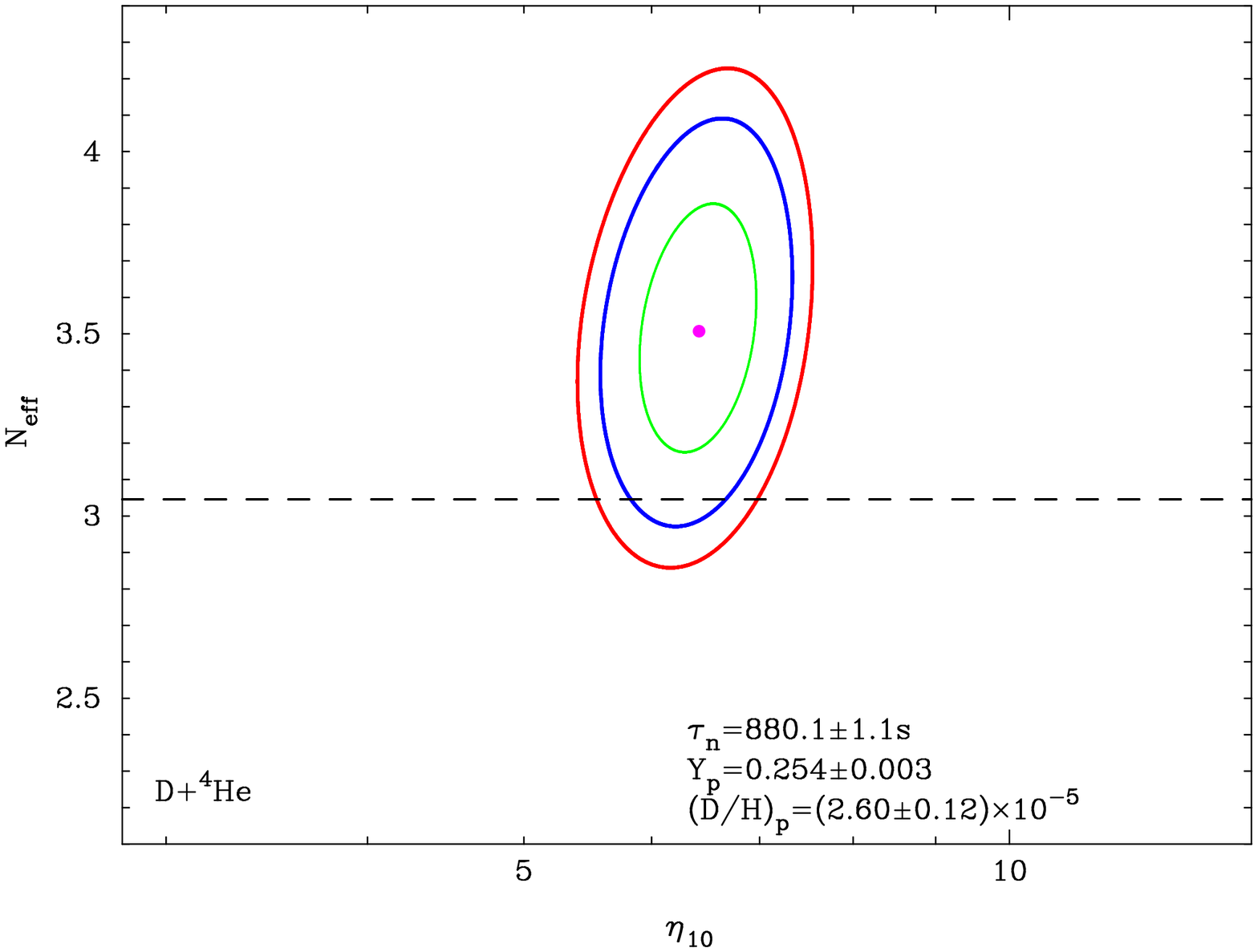}
      \caption{Joint fits to the baryon-to-photon number ratio, 
$\eta_{10}$=10$^{10}$$\eta$, and the equivalent number of light neutrino 
species $N_{\rm eff}$, using a $\chi^2$ analysis with the code developed by 
\citet{Fi98} and \citet{Li99}. The value of the primordial $^4$He abundance has
been set to $Y_{\rm p}$ = 0.254 (Fig. \ref{fig13}) and that of (D/H)$_{\rm p}$ 
is taken from \citet{PC12}. The neutron lifetime of
$\tau_{\rm n}$ = 880.1 $\pm$ 1.1s from \citet{B12}
has been adopted. The filled circle corresponds to $\chi^2$ = 
$\chi^2_{\rm min}$ = 0.
Solid lines from the inside out correspond to confidence levels of 68.3\% 
($\chi^2$ -- $\chi^2_{\rm min}$ = 2.30), 95.4\% 
($\chi^2$ -- $\chi^2_{\rm min}$ = 6.17) and 99.0\% 
($\chi^2$ -- $\chi^2_{\rm min}$ = 9.21), 
respectively. The SBBN value $N_{\rm eff}$ = 3.046
is shown with a dashed line.}
         \label{fig14}
   \end{figure}

To derive $N_{\rm eff}$ we used the statistical $\chi^2$ technique with the code
described by \citet{Fi98} and \citet{Li99} to analyse the constraints 
that the measured $^4$He and D
abundances put on $\eta$ and $N_{\rm eff}$. 
The joint fits of $\eta$ and $N_{\rm eff}$ are shown 
in Figure \ref{fig14}. 
With the two freedoms of degree ($\eta_{10}$ and $N_{\rm eff}$), the 
deviations at the 68.3\% confidence level (CL) corresponding to 
$\chi^2$ -- $\chi^2_{\rm min}$ = 2.30,
at the 95.4\% CL corresponding to $\chi^2$ -- $\chi^2_{\rm min}$ = 6.17, and
at the 99.0\% CL corresponding to $\chi^2$ -- $\chi^2_{\rm min}$ = 9.21 
are shown (from the inside out) by solid lines.

We adopted the most recently published value for neutron lifetimes 
$\tau_n$ = 880.1 $\pm$ 1.1 s \citep{B12}. With $Y_{\rm p}$ = 0.254$\pm$0.003,
(D/H)$_{\rm p}$ = (2.60$\pm$0.12)$\times$10$^{-5}$ \citep{PC12},
the minimum $\chi^2_{\rm min}$ = 0
is obtained for $\eta_{10}$ = 6.42, corresponding to 
$\Omega_{\rm b}h^2$ = 0.0234$\pm$0.0019 and 
$N_{\rm eff}$ = 3.51$\pm$0.35 (68\% CL) (Fig. \ref{fig14}). 
This value of $N_{\rm eff}$ at the 68\% CL
is higher than the SBBN value $N_\nu$ = 3.


We note that the primordial helium abundance sets a tight
constraint on the effective number of neutrino species. These constraints
are similar to or are tighter than those derived using the CMB and 
galaxy clustering
power spectra. For example, using these two sets of data, \citet{Ko11}
derived $N_{\rm eff}$ = 4.34$^{+0.86}_{-0.88}$ at the 68\% confidence level.
On the other hand, \citet{K11} analysed joint {\sl WMAP}7 data and 
South Pole Telescope (SPT)
data, both on the microwave background temperature fluctuations, and
derived 
$N_{\rm eff}$ = 3.85 $\pm$ 0.62 (68\% CL).
Adding low-redshift measurements of the Hubble constant $H_0$ using the Hubble
Space Telescope and the baryon acoustic oscillations (BAO) using
SDSS and 2dFGRS, \citet{K11} obtained
$N_{\rm eff}$ = 3.86 $\pm$ 0.42 (68\% CL).

On the other hand, \citet{A13} using the data of the {\sl Planck} mission
derived $N_{\rm eff}$ = 
3.30 $\pm$ 0.27 (68\% CL).
Thus, there is a general agreement between $N_{\rm eff}$
obtained in this paper and by other researchers with other methods. However, 
uncertainties are too high to make definite conclusions about the deviations 
of the BBN from the standard model. Tighter constraints can be obtained 
by including the additional He~{\sc i} $\lambda$10830 
emission line in the consideration,
which requires new observations. This work is in progress.

Our baryon mass density $\Omega_{\rm b} h^2$ = 
0.0234$\pm$0.0019 (68\% CL) 
(Fig. \ref{fig14}) agrees with the values
of 0.0222$\pm$0.0004 \citep{K11} and 0.0221$\pm$0.0003 \citep{A13} from 
fluctuation studies of the CMB
radiation. \citet{Ar12} developed the most recent code AlterBBN 
for calculating BBN abundances of the elements in alternative cosmologies. 
Adopting a neutron lifetime of $\tau_{\rm n}$ = 880.1 s \citep{B12},
our derived $N_{\rm eff}$ = 3.51 and $\eta_{10}$ = 6.42, it returns the 
predicted primordial abundances $Y_{\rm p}$ = 0.253 and 
(D/H)$_{\rm p}$ = 2.53$\times$10$^{-5}$, which agree well with the
values obtained from observations.

\section{Conclusions}\label{summary}

We have rederived the pregalactic helium abundance, improving 
on several aspects with respect to our previous estimates. First, we used the 
updated He~{\sc i} emissivities published by \citet{P13}, tested our 
overall procedure on a grid of CLOUDY models built
with the most recent version of the code, v13.01 \citep{F13}, using the 
same atomic data. 
Most importantly, we used the largest possible set of suitable observational 
data, which significantly enhance the set used by \citet{IT10}, 
thus reducing the statistical error in determining 
$Y_{\rm p}$ and allowing a more comprehensive analysis of systematic 
effects. 
 
Before proceeding to determine $Y$ in real objects, we produced 
analytical fits to the grid of He~{\sc i} emissivities published by 
\citet{P13}; then we tested and refined our procedure to derive the helium 
mass fraction in H~{\sc ii} regions using an appropriate grid of 
photoionisation models built with CLOUDY.

Finally, we applied our updated empirical code for the 
determination of the primordial $^4$He abundance from the largest sample 
of low-metallicity extragalactic H~{\sc ii} regions ever used (1610 spectra).
It consists of three subsamples: a) the HeBCD subsample of low-metallicity
and high-excitation H~{\sc ii} regions used for instance by \citet{I07} and
\citet{IT10} for the primordial $^4$He abundance determination (93 spectra),
b) the VLT subsample of low-metallicity and high-excitation H~{\sc ii}
regions collected by \citet{G11} and \citet{I09,I11b}
from the European Southern 
Observatory (ESO) archive (75 spectra), and c) the SDSS subsample 
of generally lower-excitation H~{\sc ii} regions (1442 spectra). 
Our main results are summarised below.

1. We fitted the new tabulated He~{\sc i} emissivities by \citet{P13}. Our fits
reproduced the tabulated emissivities of 32 He~{\sc i} emission lines with an
accuracy of better than 1\% in a wide range of electron temperatures 
$T_{\rm e}$ = 10000 -- 20000K and electron number densities
$N_{\rm e}$ = 10 -- 10$^3$ cm$^{-3}$.

2. We obtained ionisation correction factors for He from 
CLOUDY-photoionised H~{\sc ii} region models and produced fits to them as a
function of the starburst age and the excitation parameter $x$ = O$^{2+}$/O, 
where O and O$^{2+}$ are the total oxygen abundance and the abundance of the 
doubly ionised oxygen, respectively. The new ionisation correction factors
agree well with previously published ones, used for instance 
by \citet{I07} and \citet{IT10}.

3. We derived from CLOUDY models the non-recombination contribution 
to the strongest hydrogen Balmer emission line intensities, 
which includes both 
the collisional and fluorescent excitation. For that we compared the
hydrogen line intensities calculated including all processes 
with their case B values. We found that the 
non-recombination contribution to the H$\alpha$ and H$\beta$ line intensities
can be as high as $\sim$ 9\% and $\sim$ 6\%, respectively, in the hot 
lowest-metallicity H~{\sc ii} regions. We produced the fits of these 
contributions as a function of the oxygen abundance  and starburst age.

4. We used the CLOUDY-predicted emission-line intensities to recover the 
physical conditions and chemical composition of the H~{\sc ii} region models
using our updated empirical method. We found that
our method reproduces the electron number density $N_{\rm e}$(He$^+$) and the
emissivity-weighted electron temperature $T_{\rm e}$(He$^+$).
Most importantly, the empirical method 
reproduces the CLOUDY input value of the $^4$He mass fraction with an 
accuracy of
better than 1\%, independently of the number of He~{\sc i} emission lines used 
for the $\chi^2$ minimisation. This gives confidence in our method for 
determining the $^4$He abundance in real H~{\sc ii} regions. 

5. We applied our updated empirical method for determining the $^4$He
mass fraction $Y$ to a large sample of 1610 low-metallicity extragalactic
H~{\sc ii} regions. This is the largest sample used for such a study. 
We found a systematic 
offset between $Y$s derived for the HeBCD+VLT and SDSS subsamples, indicating 
problems in the $Y$ determination in most of SDSS H~{\sc ii} regions.
However, the offset is not present when only 
spectra with the highest S/N of the highest-excitation 
HeBCD+VLT+SDSS H~{\sc ii} regions are used.
Using linear relation $Y$ -- O/H for 111 highest-excitation H~{\sc ii}
regions we derived the primordial $^4$He mass fraction
$Y_{\rm p}$ = 0.254 $\pm$ 0.003, which is higher than the
standard big bang nucleosynthesis (SBBN) value of 0.2477 $\pm$ 0.0001
inferred from the temperature fluctuations of the microwave background 
radiation. This difference possibly indicates the existence of 
additional types of neutrino species.

6. Using the most recently derived primordial abundances of light 
elements D and $^4$He and the $\chi^2$ technique, we found that the best 
agreement between abundances of these light elements is achieved in the BBN 
model with the baryon mass fraction $\Omega_{\rm b} h^2$ = 0.0234$\pm$0.0019
(68\% CL) and the effective number of the neutrino species $N_{\rm eff}$ = 
3.51$\pm$0.35 (68\% CL). Both the $\Omega_{\rm b} h^2$ and 
$N_{\rm eff}$ values nicely agree with those inferred from the 
temperature fluctuations of the microwave background radiation.

\begin{acknowledgements}
Y.I.I. is thankful to the Observatoire de Paris for financing a three-month 
stay during which part of this work was done. 
This paper makes extensive use of the photoionisation code CLOUDY, 
and we are grateful to its authors, in particular G. Ferland. 
Funding for the Sloan Digital Sky Survey (SDSS) and SDSS-II has
been provided by the Alfred P. Sloan Foundation, the Participating
Institutions, the National Science Foundation, the U.S. Department
of Energy, the National Aeronautics and Space Administration, the
Japanese Monbukagakusho, the Max Planck Society, and the
Higher Education Funding Council for England.
\end{acknowledgements}

\Online


\setcounter{table}{1}

\begin{table*}
\centering
\caption{Parameters for fits to He~{\sc i} emissivities
\label{taba1}}
\begin{tabular}{crrrr}
\hline\hline
Wavelength (\AA)&\multicolumn{1}{c}{$a$}&\multicolumn{1}{c}{$b$}
&\multicolumn{1}{c}{$c$}&\multicolumn{1}{c}{$d$} \\ \hline
2945            &$-$1.0931E$+$5&$-$3.9252E$+$2&   1.1960E$+$4&   3.2375E$+$5 \\
3188            &$-$2.1034E$+$5&$-$7.5620E$+$2&   2.3088E$+$4&   6.2101E$+$5 \\
3614            &$-$3.4279E$+$4&$-$1.4370E$+$2&   3.9750E$+$3&   9.7188E$+$4 \\
3889            &$-$4.2407E$+$5&$-$1.5736E$+$3&   4.7538E$+$4&   1.2306E$+$6 \\
3965            &$-$6.3941E$+$4&$-$2.6903E$+$2&   7.4458E$+$3&   1.8049E$+$5 \\
4026            &$-$4.4499E$+$4&$-$3.2773E$+$2&   7.1902E$+$3&   8.2717E$+$4 \\
4121            &   2.0303E$+$4&   1.1227E$+$2&$-$2.5696E$+$3&$-$5.4463E$+$4 \\
4388            &$-$9.4928E$+$3&$-$7.9846E$+$1&   1.6784E$+$3&   1.4537E$+$4 \\
4471            &$-$2.6190E$+$4&$-$4.1379E$+$2&   7.5212E$+$3&$-$1.7296E$+$3 \\
4713            &   4.2872E$+$4&   2.4107E$+$2&$-$5.4550E$+$3&$-$1.1461E$+$5 \\
4922            &$-$2.1899E$+$3&$-$9.6239E$+$1&   1.5392E$+$3&$-$2.0020E$+$4 \\
5016            &$-$1.3257E$+$5&$-$5.7648E$+$2&   1.5752E$+$4&   3.6784E$+$5 \\
5048            &   6.3687E$+$3&   4.3534E$+$2&$-$8.7839E$+$2&$-$1.5904E$+$4 \\
5876            &   2.8913E$+$5&   4.6114E$+$2&$-$2.1977E$+$4&$-$1.0040E$+$6 \\
6678            &   1.0873E$+$5&   2.3872E$+$2&$-$9.2029E$+$3&$-$3.6316E$+$5 \\
7065            &   1.0249E$+$5&   6.1473E$+$2&$-$1.3275E$+$4&$-$2.7178E$+$5 \\
7281            &   1.1356E$+$4&   9.3912E$+$1&$-$1.6876E$+$3&$-$2.6681E$+$4 \\
9464            &$-$6.0453E$+$3&$-$2.1708E$+$1&   6.6141E$+$2&$-$1.7905E$+$4 \\
10830           &$-$5.5328E$+$5&$-$2.4749E$+$3&   6.7852E$+$4&   1.5265E$+$6 \\
11969           &$-$4.4839E$+$3&$-$3.3036E$+$1&   7.2471E$+$2&   8.3299E$+$3 \\
12527           &$-$6.7370E$+$3&$-$2.4221E$+$1&   7.3950E$+$2&   1.9891E$+$4 \\
12785           &   3.9452E$+$4&   1.0969E$+$2&$-$3.6899E$+$3&$-$1.2739E$+$5 \\
12790           &   1.4119E$+$4&   4.0926E$+$1&$-$1.3426E$+$3&$-$4.5241E$+$4 \\
12968           &$-$1.2751E$+$3&$-$1.0404E$+$1&   2.2055E$+$2&   2.0622E$+$3 \\
15084           &$-$3.5046E$+$3&$-$1.4759E$-$1&   4.0785E$+$2&   9.9031E$+$3 \\
17002           &$-$1.8707E$+$3&$-$2.9335E$+$1&   5.3390E$+$2&$-$1.1660E$+$3 \\
18685           &   9.8299E$+$4&   3.2753E$+$2&$-$1.0026E$+$4&$-$2.9046E$+$5 \\
18697           &   3.9198E$+$4&   1.3734E$+$2&$-$4.0786E$+$3&$-$1.1555E$+$5 \\
19089           &$-$2.0601E$+$2&$-$8.9043E$+$0&   1.4258E$+$2&$-$1.8379E$+$3 \\
19543           &$-$3.9284E$+$3&$-$1.4127E$+$1&   4.3125E$+$2&   1.1597E$+$4 \\
20580           &$-$5.7214E$+$4&$-$3.0719E$-$2&   7.7903E$+$3&   1.4348E$+$5 \\
21118           &   6.5439E$+$3&   3.6798E$+$1&$-$8.3265E$+$2&$-$1.7494E$+$4 \\ \hline
\end{tabular}
\end{table*}


\longtab{3}{
\begin{longtable}{ccrrr}
\caption{Fit parameters for the relative collisional contributions, $C/R$, to the
 He~{\sc i} line emissivities \label{taba2}} \\
\hline\hline
Wavelength (\AA)&$i$&\multicolumn{1}{c}{$a_i$}&\multicolumn{1}{c}{$b_i$}
&\multicolumn{1}{c}{$c_i$} \\
\hline
\endfirsthead
\caption{continued.}\\
\hline\hline
Wavelength (\AA)&$i$&\multicolumn{1}{c}{$a_i$}&\multicolumn{1}{c}{$b_i$}
&\multicolumn{1}{c}{$c_i$} \\
\hline
\endhead
\hline
\endfoot
 2945 & 1 &      0.237599&      1.283664&   $-$0.492275\\
      & 2 &      0.018249&   $-$2.537098&   $-$0.280238\\
      & 3 &   $-$0.074595&   $-$0.100902&   $-$0.275109\\
      & 4 &   $-$0.034749&   $-$0.129213&   $-$0.575391\\
      & 5 &   $-$0.012696&      0.187311&   $-$0.703736\\
      & 6 &      0.000781&      3.290271&   $-$3.038961\\
      & 7 &   $-$0.000120&      7.670289&   $-$0.484230\\
      & 8 &   $-$0.000038&   $-$5.633697&      0.244857\\
      & 9 &   $-$0.000010&      5.791438&   $-$0.890583\\ \\
 3188 & 1 &      0.728281&      0.981482&   $-$1.474760\\
      & 2 &      0.003620&   $-$2.737075&      0.014362\\
      & 3 &   $-$0.050690&      0.501912&   $-$0.272824\\
      & 4 &   $-$0.005385&      4.458482&      0.095771\\
      & 5 &      0.003062&   $-$1.251756&      0.159231\\
      & 6 &   $-$0.000188&      3.382599&   $-$1.159856\\
      & 7 &   $-$0.000077&      3.574946&   $-$0.579868\\
      & 8 &   $-$0.000055&      0.044420&   $-$0.322935\\
      & 9 &   $-$0.000002&      5.191432&   $-$1.828550\\ \\
 3614 & 1 &      0.376504&      1.443660&   $-$0.852351\\
      & 2 &      0.014185&   $-$2.418831&   $-$0.167465\\
      & 3 &   $-$0.072682&      0.327648&   $-$0.180120\\
      & 4 &   $-$0.017533&      0.670556&   $-$0.492422\\
      & 5 &   $-$0.010484&      0.418788&   $-$0.789364\\
      & 6 &      0.001705&   $-$2.704035&   $-$0.622900\\
      & 7 &   $-$0.000176&      7.902544&      0.151241\\
      & 8 &      0.000064&      6.131473&      3.450867\\
      & 9 &   $-$0.000037&      5.368607&   $-$5.579913\\ \\
 3889 & 1 &      2.256542&      0.511307&   $-$1.742707\\
      & 2 &   $-$0.034717&      2.949781&      1.041676\\
      & 3 &      0.010993&   $-$2.484143&   $-$0.173011\\
      & 4 &   $-$0.045348&      0.647284&   $-$0.155018\\
      & 5 &      0.002098&   $-$1.517937&      0.058988\\
      & 6 &      0.000026&   $-$2.918819&   $-$0.503535\\
      & 7 &   $-$0.000005&      2.675449&      0.913041\\
      & 8 &      0.000002&      8.872791&      0.077254\\
      & 9 &   $-$0.000001&   $-$8.450243&   $-$1.355353\\ \\
 3965 & 1 &      0.456770&      1.654674&   $-$1.158531\\
      & 2 &      0.015196&   $-$2.045893&   $-$0.252967\\
      & 3 &   $-$0.067550&      0.652662&   $-$0.248281\\
      & 4 &   $-$0.007843&      4.496742&   $-$0.262929\\
      & 5 &   $-$0.000781&      4.890201&      1.464717\\
      & 6 &      0.002827&   $-$1.450538&   $-$0.069850\\
      & 7 &      0.001110&   $-$2.004124&   $-$0.325267\\
      & 8 &      0.000194&   $-$2.716613&   $-$0.724557\\
      & 9 &   $-$0.000004&      4.371841&      0.663073\\ \\
 4026 & 1 &      0.185064&      1.943657&   $-$0.648145\\
      & 2 &      0.008751&   $-$2.724192&   $-$0.097890\\
      & 3 &   $-$0.045073&      0.510649&   $-$0.102190\\
      & 4 &   $-$0.008680&      0.471088&   $-$0.270234\\
      & 5 &   $-$0.002879&      0.286899&   $-$0.525890\\
      & 6 &   $-$0.000016&  $-$11.687314&   $-$0.818941\\
      & 7 &   $-$0.000117&      7.491152&      1.297731\\
      & 8 &   $-$0.000010&      6.109163&      0.682963\\
      & 9 &      0.000002&   $-$3.088925&      0.277057\\ \\
 4121 & 1 &     12.796716&   $-$0.490913&   $-$3.214825\\
      & 2 &   $-$0.512153&      1.066279&   $-$0.652840\\
      & 3 &      0.057334&   $-$1.384667&   $-$0.408928\\
      & 4 &      0.013453&   $-$3.178046&   $-$0.847815\\
      & 5 &   $-$0.003661&   $-$2.663920&      0.154191\\
      & 6 &      0.000411&   $-$1.371718&      0.083031\\
      & 7 &      0.000241&   $-$2.062889&   $-$0.282061\\
      & 8 &      0.000176&   $-$1.443159&      0.046880\\
      & 9 &      0.000126&   $-$0.715603&      0.040848\\ \\
 4388 & 1 &      0.592415&      1.479889&   $-$1.575477\\
      & 2 &      0.009681&   $-$2.014916&   $-$0.241845\\
      & 3 &   $-$0.048354&      0.783922&   $-$0.135038\\
      & 4 &   $-$0.012120&      0.153999&   $-$0.391060\\
      & 5 &      0.004329&   $-$2.227663&   $-$0.307074\\
      & 6 &      0.003904&   $-$2.413184&   $-$0.373145\\
      & 7 &   $-$0.000044&      7.482763&      5.477369\\
      & 8 &      0.002504&   $-$1.897203&   $-$0.129707\\
      & 9 &      0.000659&   $-$2.299944&   $-$0.179990\\ \\
 4471 & 1 &      2.074758&      0.871883&   $-$2.773291\\
      & 2 &      0.000427&   $-$2.791796&   $-$0.254427\\
      & 3 &   $-$0.038291&      2.548854&      0.190174\\
      & 4 &      0.001975&   $-$2.134139&      0.291631\\
      & 5 &   $-$0.006131&      3.477982&      0.217477\\
      & 6 &      0.000150&      1.070168&   $-$0.221179\\
      & 7 &   $-$0.000673&      0.587419&      0.291429\\
      & 8 &   $-$0.000052&   $-$2.436785&   $-$0.073254\\
      & 9 &      0.000008&      2.435703&   $-$0.125751\\ \\
 4713 & 1 &      3.052470&      0.998766&   $-$1.464257\\
      & 2 &   $-$0.128327&      2.958524&      0.620627\\
      & 3 &   $-$0.001934&      3.647304&      2.052811\\
      & 4 &      0.003193&   $-$3.733573&      0.215077\\
      & 5 &   $-$0.049904&      0.299277&   $-$0.126395\\
      & 6 &   $-$0.012788&      0.169174&   $-$0.273934\\
      & 7 &   $-$0.002769&      2.825847&      0.529843\\
      & 8 &   $-$0.000770&   $-$0.102838&   $-$0.729419\\
      & 9 &      0.000005&   $-$1.845712&      0.144187\\ \\
 4922 & 1 &      2.073767&      0.939984&   $-$2.661237\\
      & 2 &      0.000474&   $-$3.111335&   $-$0.032107\\
      & 3 &   $-$0.040763&      1.978845&      0.118798\\
      & 4 &      0.002338&   $-$1.921872&      0.245616\\
      & 5 &   $-$0.003455&      4.486010&      0.486895\\
      & 6 &   $-$0.000375&      5.351461&      2.114852\\
      & 7 &      0.000760&   $-$1.654804&      0.082870\\
      & 8 &      0.000080&   $-$1.016341&      0.093812\\
      & 9 &      0.000004&   $-$2.045578&      0.247970\\ \\
 5016 & 1 &      0.702161&      1.240150&   $-$1.231985\\
      & 2 &      0.015156&   $-$2.139900&   $-$0.193653\\
      & 3 &   $-$0.078849&      0.478302&   $-$0.287219\\
      & 4 &   $-$0.032381&   $-$0.224821&   $-$0.744900\\
      & 5 &   $-$0.003822&      4.922654&      0.471553\\
      & 6 &      0.003009&   $-$1.216575&      0.104354\\
      & 7 &      0.003016&   $-$1.564262&   $-$0.030637\\
      & 8 &      0.001328&   $-$2.090390&   $-$0.294729\\
      & 9 &      0.000080&   $-$1.088301&   $-$0.425798\\ \\
 5048 & 1 &      3.867902&      0.451594&   $-$2.209071\\
      & 2 &      0.005663&   $-$2.715737&   $-$0.232676\\
      & 3 &   $-$0.124564&      2.431123&      0.149969\\
      & 4 &   $-$0.020914&      2.037531&      0.386945\\
      & 5 &      0.006407&   $-$2.191700&      0.008098\\
      & 6 &   $-$0.004000&      2.140291&      0.555353\\
      & 7 &      0.001913&   $-$2.719687&   $-$0.279718\\
      & 8 &   $-$0.002026&      0.764487&   $-$0.136685\\
      & 9 &      0.000031&   $-$1.340548&      0.068556\\ \\
 5876 & 1 &      0.765684&      2.031295&   $-$0.800207\\
      & 2 &   $-$0.120951&      0.665641&   $-$0.115064\\
      & 3 &      0.002838&   $-$2.899645&   $-$0.128681\\
      & 4 &   $-$0.020210&      1.163158&   $-$0.280858\\
      & 5 &   $-$0.011974&      0.567921&   $-$0.702161\\
      & 6 &      0.000941&   $-$1.814439&      0.143205\\
      & 7 &   $-$0.000084&      8.290851&      4.956775\\
      & 8 &      0.001916&   $-$1.626788&      0.081847\\
      & 9 &      0.000330&   $-$0.574927&      0.049323\\ \\
 6678 & 1 &   $-$0.142778&      0.002216&   $-$0.345017\\
      & 2 &      0.194029&      2.957312&      1.012852\\
      & 3 &   $-$0.161264&      0.049074&   $-$0.242204\\
      & 4 &   $-$0.111315&      0.027608&   $-$0.434800\\
      & 5 &   $-$0.071360&      0.104060&   $-$0.491709\\
      & 6 &   $-$0.028272&   $-$0.212975&   $-$0.564984\\
      & 7 &   $-$0.001836&      7.009674&      0.788004\\
      & 8 &   $-$0.000203&   $-$4.712255&      0.361104\\
      & 9 &   $-$0.000004&   $-$1.776468&   $-$0.004734\\ \\
 7065 & 1 &      7.407207&      0.039764&   $-$1.194215\\
      & 2 &   $-$0.832559&   $-$0.209824&   $-$0.057477\\
      & 3 &      0.014316&   $-$3.462613&      0.299767\\
      & 4 &   $-$0.098030&   $-$0.380680&   $-$0.821184\\
      & 5 &   $-$0.027901&   $-$0.193571&   $-$0.691501\\
      & 6 &   $-$0.000820&      6.151002&      2.800949\\
      & 7 &      0.010872&   $-$2.082732&      0.174198\\
      & 8 &      0.004095&   $-$2.451380&      0.145223\\
      & 9 &      0.003886&   $-$1.095082&      0.105919\\ \\
 7281 & 1 &      4.009749&      0.349820&   $-$1.559434\\
      & 2 &   $-$0.169962&      2.247969&      0.103872\\
      & 3 &   $-$0.207715&   $-$1.420517&   $-$1.142405\\
      & 4 &      0.002292&   $-$2.893220&      0.460167\\
      & 5 &   $-$0.047401&   $-$0.341306&   $-$0.537341\\
      & 6 &   $-$0.078960&   $-$1.463006&   $-$1.127866\\
      & 7 &   $-$0.012757&      2.051007&   $-$0.007594\\
      & 8 &      0.011432&   $-$1.438304&      0.173748\\
      & 9 &      0.000589&   $-$0.996816&      0.128812\\ \\
 9464 & 1 &      0.065617&      1.801141&      1.222155\\
      & 2 &   $-$0.211779&   $-$0.373572&   $-$0.349116\\
      & 3 &      0.025911&      2.079864&      0.079155\\
      & 4 &   $-$0.002505&   $-$1.090156&      0.819827\\
      & 5 &   $-$0.019108&   $-$0.117144&   $-$0.799836\\
      & 6 &   $-$0.005019&      0.120752&   $-$0.712156\\
      & 7 &   $-$0.000013&      9.761829&      0.920806\\
      & 8 &   $-$0.002696&   $-$0.979003&   $-$0.735566\\
      & 9 &      0.000375&   $-$2.324543&   $-$0.000464\\ \\
10830 & 1 &     39.752766&   $-$0.298983&   $-$1.655744\\
      & 2 &      0.160380&   $-$2.503047&      0.999384\\
      & 3 &   $-$0.343303&   $-$2.567812&      0.383924\\
      & 4 &   $-$0.179840&   $-$4.593626&   $-$0.093260\\
      & 5 &      0.376525&   $-$2.098506&   $-$0.275385\\
      & 6 &      0.229901&   $-$1.132494&      0.071783\\
      & 7 &      0.120023&   $-$1.277745&      0.063845\\
      & 8 &      0.085351&   $-$2.359180&   $-$0.635813\\
      & 9 &      0.013747&   $-$0.348160&      0.010280\\ \\
11969 & 1 &      0.169269&      2.142598&   $-$0.481203\\
      & 2 &      0.010005&   $-$2.379077&   $-$0.229726\\
      & 3 &   $-$0.051812&      0.593835&   $-$0.113468\\
      & 4 &   $-$0.014826&      0.334608&   $-$0.308218\\
      & 5 &      0.001950&   $-$1.694416&   $-$0.008706\\
      & 6 &   $-$0.000103&      7.060107&      3.687421\\
      & 7 &      0.001868&   $-$2.097287&   $-$0.190160\\
      & 8 &      0.000478&   $-$2.327992&   $-$0.222428\\
      & 9 &   $-$0.000017&      5.173059&      0.222140\\ \\
12527 & 1 &      0.682182&      0.898860&   $-$1.200630\\
      & 2 &      0.018834&   $-$2.134930&   $-$0.211498\\
      & 3 &   $-$0.085191&      0.177796&   $-$0.377210\\
      & 4 &   $-$0.046766&   $-$0.434677&   $-$0.889978\\
      & 5 &   $-$0.027952&   $-$0.977738&   $-$1.300571\\
      & 6 &   $-$0.000271&      6.298621&      2.435816\\
      & 7 &      0.003779&   $-$2.256166&   $-$0.401394\\
      & 8 &      0.001858&   $-$2.210664&   $-$0.382732\\
      & 9 &      0.001313&   $-$1.753759&   $-$0.049094\\ \\
12785 & 1 &   $-$0.017363&      0.732726&      0.414940\\
      & 2 &      0.051895&      3.220740&      0.976371\\
      & 3 &   $-$0.047197&      0.036961&   $-$0.018628\\
      & 4 &   $-$0.016052&   $-$0.065403&   $-$0.125076\\
      & 5 &   $-$0.008674&   $-$0.448546&   $-$0.227839\\
      & 6 &   $-$0.004627&   $-$0.189011&   $-$0.147199\\
      & 7 &   $-$0.000752&      0.008037&   $-$0.125717\\
      & 8 &   $-$0.000015&      6.817896&      0.285434\\
      & 9 &   $-$0.000001&      5.052590&   $-$0.254754\\ \\
12790 & 1 &      0.034784&      4.153957&      1.861917\\
      & 2 &   $-$0.034996&      0.291587&      0.200122\\
      & 3 &   $-$0.024853&      0.472719&      0.306223\\
      & 4 &   $-$0.029037&   $-$0.016732&   $-$0.028320\\
      & 5 &   $-$0.016854&      0.008265&   $-$0.080113\\
      & 6 &   $-$0.012192&      0.006807&   $-$0.139773\\
      & 7 &   $-$0.007873&   $-$0.413797&   $-$0.273093\\
      & 8 &   $-$0.005301&      0.032505&   $-$0.204200\\
      & 9 &   $-$0.001220&      9.140628&   $-$2.456817\\ \\
12968 & 1 &      0.469864&      1.857833&   $-$1.238283\\
      & 2 &      0.014293&   $-$2.206397&   $-$0.234544\\
      & 3 &   $-$0.063890&      0.715108&   $-$0.133062\\
      & 4 &   $-$0.007740&      4.548513&      0.702111\\
      & 5 &      0.003757&   $-$1.363295&      0.021378\\
      & 6 &   $-$0.000520&      0.354133&   $-$0.411930\\
      & 7 &   $-$0.000007&      0.867989&   $-$0.382254\\
      & 8 &      0.000000&   $-$0.312655&   $-$1.229920\\
      & 9 &      0.000000&   $-$1.321470&      0.001483\\ \\
15084 & 1 &      0.370078&      1.313424&   $-$0.211504\\
      & 2 &      0.121869&   $-$2.898463&   $-$0.760698\\
      & 3 &   $-$0.175005&   $-$0.277672&   $-$0.356154\\
      & 4 &   $-$0.231416&   $-$0.688087&   $-$0.921995\\
      & 5 &   $-$0.035201&      0.001357&   $-$0.092017\\
      & 6 &   $-$0.001294&   $-$3.457917&      0.600536\\
      & 7 &   $-$0.000318&      6.265059&      1.835816\\
      & 8 &   $-$0.000655&      0.001848&   $-$0.253604\\
      & 9 &   $-$0.000015&      4.188363&   $-$0.168791\\ \\
17002 & 1 &      2.084653&      0.874461&   $-$2.772676\\
      & 2 &      0.000431&   $-$2.798307&   $-$0.252522\\
      & 3 &   $-$0.038638&      2.593246&      0.193422\\
      & 4 &      0.001978&   $-$2.132807&      0.291096\\
      & 5 &   $-$0.006182&      3.386596&      0.229511\\
      & 6 &      0.000117&      1.312846&   $-$0.134606\\
      & 7 &   $-$0.000664&      0.329870&      0.245240\\
      & 8 &   $-$0.000057&   $-$2.384770&   $-$0.048070\\
      & 9 &   $-$0.000003&   $-$1.271340&      0.074656\\ \\
18685 & 1 &   $-$0.664639&      0.446721&   $-$0.336584\\
      & 2 &      0.321034&      2.361666&      0.631961\\
      & 3 &   $-$0.048314&   $-$0.032944&      0.471416\\
      & 4 &   $-$0.007661&      0.822514&   $-$0.347149\\
      & 5 &   $-$0.000196&      8.385674&      1.098974\\
      & 6 &      0.000501&   $-$1.010528&   $-$0.159621\\
      & 7 &      0.000028&      3.757360&      0.211017\\
      & 8 &      0.000005&      4.581320&   $-$0.254349\\
      & 9 &      0.000000&      1.087205&   $-$0.289278\\ \\
18697 & 1 &   $-$0.479959&   $-$0.397439&   $-$0.551253\\
      & 2 &      0.287592&      2.875301&      0.778442\\
      & 3 &   $-$0.142753&      0.215541&   $-$0.072482\\
      & 4 &   $-$0.054167&   $-$0.107174&   $-$0.312201\\
      & 5 &   $-$0.010114&      0.237588&   $-$0.170092\\
      & 6 &   $-$0.000601&   $-$0.069410&   $-$0.212051\\
      & 7 &   $-$0.000044&      1.161615&   $-$0.595412\\
      & 8 &      0.000002&   $-$7.642257&      1.130464\\
      & 9 &   $-$0.000002&     13.444530&      0.322168\\ \\
19089 & 1 &      2.067268&      0.938927&   $-$2.659532\\
      & 2 &      0.000467&   $-$3.128493&   $-$0.028876\\
      & 3 &   $-$0.040632&      1.966949&      0.117632\\
      & 4 &      0.002336&   $-$1.925862&      0.246575\\
      & 5 &   $-$0.003439&      4.477894&      0.477374\\
      & 6 &   $-$0.000371&      5.378613&      2.119313\\
      & 7 &      0.000746&   $-$1.660655&      0.083380\\
      & 8 &      0.000077&   $-$1.020796&      0.094426\\
      & 9 &      0.000004&   $-$2.059014&      0.250287\\ \\
19543 & 1 &      0.693183&      0.929526&   $-$1.195240\\
      & 2 &      0.020944&   $-$2.058396&   $-$0.255067\\
      & 3 &   $-$0.084119&      0.174134&   $-$0.420265\\
      & 4 &   $-$0.051351&   $-$0.608595&   $-$0.995168\\
      & 5 &   $-$0.059494&   $-$1.207560&   $-$1.453542\\
      & 6 &   $-$0.000305&      5.981650&      3.333056\\
      & 7 &      0.005404&   $-$1.748600&   $-$0.203026\\
      & 8 &      0.002988&   $-$1.787902&   $-$0.251321\\
      & 9 &      0.000925&   $-$1.827335&   $-$0.035773\\ \\
20580 & 1 &      2.893238&      0.202500&   $-$1.175161\\
      & 2 &   $-$0.005668&      4.825179&   $-$4.401486\\
      & 3 &      0.005762&   $-$0.564630&      0.608410\\
      & 4 &   $-$0.011124&   $-$0.689134&      0.077485\\
      & 5 &   $-$0.000787&   $-$1.690374&   $-$0.029622\\
      & 6 &   $-$0.000428&   $-$2.496137&   $-$0.278228\\
      & 7 &   $-$0.000150&   $-$3.959460&   $-$0.871990\\
      & 8 &   $-$0.000007&      4.354130&   $-$2.055856\\
      & 9 &      0.000000&   $-$4.576949&      0.433552\\ \\
21118 & 1 &      4.140184&      0.730139&   $-$1.671074\\
      & 2 &   $-$0.155361&      2.629962&      0.543211\\
      & 3 &   $-$0.008976&      3.021516&      1.896211\\
      & 4 &      0.015029&   $-$2.092910&      0.202714\\
      & 5 &   $-$0.088831&      0.684846&   $-$0.150858\\
      & 6 &      0.004902&   $-$1.603460&      0.130640\\
      & 7 &      0.000382&   $-$2.561481&   $-$0.197627\\
      & 8 &      0.000048&   $-$2.856426&   $-$0.128358\\
      & 9 &   $-$0.000002&   $-$5.382045&      0.626686\\ \\

\hline
\end{longtable}
}


\end{document}